\documentclass[reprint,aps,pre,twocolumn,groupedaddress,nobibnotes]{revtex4-1}
\usepackage{amsmath,amsthm,amssymb}
\usepackage[caption=false]{subfig}
\usepackage{mathtools}
\usepackage{newfloat,algcompatible}
\usepackage{etoolbox}
\usepackage[
  unicode=true,
  bookmarks=false,
  pdfpagelabels=false,
  hyperfootnotes=false,
  hyperindex=false,
  pageanchor=false,
  colorlinks=false,
]{hyperref}
\usepackage{cleveref}
\usepackage{sidecap}
\usepackage{xcolor}

\AtBeginEnvironment{algorithm}{\noindent\hrulefill\par\nobreak\vskip-5pt}
\usepackage{newfloat}

\DeclareFloatingEnvironment[
    fileext=loa,
    listname=List of Algorithms,
    name=ALGORITHM,
    placement=tbhp,
]{algorithm}
\DeclareCaptionFormat{algorithms}{\vskip-15pt\hrulefill\par#1#2#3\vskip-6pt\hrulefill}
\newcommand{\env}[1]{\texttt{#1}}
\DeclareMathOperator{\sech}{sech}
\DeclareMathOperator{\Tr}{Tr}

\DeclareMathOperator{\vol}{vol}
\DeclareMathOperator{\cut}{cut}
\DeclareMathOperator{\NCut}{NCut}
\theoremstyle{definition}
\newtheorem{ax}{Principle}
\newtheorem{problem}{Problem}
\newtheorem{rem}{Remark}
\begin{document}

\title{A Spectral Clustering Approach to\\
Lagrangian Vortex Detection\\
\textcolor{blue}{\small{(to appear in Physical Review E, 2016})}}

\author{Alireza Hadjighasem}
\email{Email address for correspondence: alirezah@ethz.ch}
\altaffiliation{Institute of Mechanical Systems, Department of Mechanical and Process Engineering, ETH Z\"{u}rich, Leonhardstrasse 21, 8092 Z\"{u}rich, Switzerland}

\author{Daniel Karrasch}
\email{karrasch@ma.tum.de}
\altaffiliation{Institute of Mechanical Systems, Department of Mechanical and Process Engineering, ETH Z\"{u}rich, Leonhardstrasse 21, 8092 Z\"{u}rich, Switzerland. Current address: Technische Universit\"{a}t M\"{u}nchen, Zentrum Mathematik M3, Boltzmannstr. 3, 85748 Garching, Germany}

\author{Hiroshi Teramoto}
\email{teramoto@es.hokudai.ac.jp}
\altaffiliation{Molecule \& Life Nonlinear Sciences Laboratory, Research Institute for Electronic Science, Hokkaido University, Kita 20 Nishi 10, Kita-ku, Sapporo 001-0020, Japan}

\author{George Haller}
\email{georgehaller@ethz.ch}
\altaffiliation{Institute of Mechanical Systems, Department of Mechanical and Process Engineering, ETH Z\"{u}rich, Leonhardstrasse 21, 8092 Z\"{u}rich, Switzerland}

\begin{abstract}
One of the ubiquitous features of real-life turbulent flows is the existence and persistence of coherent vortices. Here we show that such coherent vortices can be extracted as clusters of Lagrangian trajectories. We carry out the clustering on a weighted graph, with the weights measuring pairwise distances of fluid trajectories in the extended phase space of positions and time. We then extract coherent vortices from the graph using tools from spectral graph theory. Our method locates all coherent vortices in the flow simultaneously, thereby showing high potential for automated vortex tracking. We illustrate the performance of this technique by identifying coherent Lagrangian vortices in several two- and three-dimensional flows.
\end{abstract}

\maketitle

\section{Introduction}\label{section:introduction}

It has long been recognized that even unsteady flows with aperiodic
time dependence admit persistent patterns that govern the transport
of passive tracers \cite{Mcwilliams84,Provenzale99,Haller15}. Generally
referred to as coherent structures, these patterns are often vortex-type
spatial features that remain recognizable over times exceeding typical
time scales in the flow. Our goal here is to systematically decompose
trajectories in such a general flow into coherent and incoherent families,
providing a conceptual simplification of the underlying dynamical
system.

The majority of coherent structure identification methods used in fluid dynamics continues to be Eulerian (see, e.g., \cite{Zhang14,Dong14,Wilczek15,Oneill15} for recent examples), concerned with features of the instantaneous velocity
field driving the flow \cite{Hunt88,Zhou99}. The resulting Eulerian
coherent structure criteria have been broadly used in flow structure
identification, although none has emerged as a definitive tool of
choice. By their focus on the velocity field, these Eulerian criteria
inherently depend on the reference frame in which they are applied
\cite{Haller05}.

By contrast, Lagrangian methods identify vortical flow structures
based on the properties of fluid particle trajectories \cite{Provenzale99,Boffetta01,Weiss08,Peacock10,Haller15}. Several of these methods are frame-invariant and hence the structures
they locate (or miss) are the same in all frames that translate and
rotate relative to each other. This invariance is especially important
for geophysical flows which are invariably defined in the rotating
frame of the earth. In such flows, long lived coherent vortices may
transport fluid over great distances, surrounded by strongly mixing background
turbulence \cite{Provenzale99,Haller13}.

Lagrangian vortex detection approaches either seek a coherent material
boundary to the vortex, or aim to identify a coherent interior of
a vortex. Coherent material vortex boundaries are special cases of
Lagrangian coherent structures (LCSs), the most influential material
surfaces in the flow \cite{Haller15}. Within this class, Lagrangian
vortex boundaries can either be defined as outermost non-filamenting,
closed material surfaces (elliptic LCSs \cite{Haller13,Blazevski14}),
or as outermost, closed material surfaces of equal material rotation \cite{Farazmand15,Haller16}. Other approaches target Lagrangian vortex boundaries as locations of minimal curvature change \cite{Ma14} or as curves that maximize the volume to boundary size ratio throughout advection \cite{Froyland15_2}.

Approaches seeking the interior of Lagrangian vortices have mostly been 
probabilistic in nature. Early techniques relied on the diagnostic
use of relative and absolute dispersion \cite{Provenzale99}. Later
mathematical approaches offer a bipartition of phase space into minimally
diffusive regions by delineating the density evolution that can be
characterized by the Perron-Frobenius or transfer operator \cite{Froyland10,Froyland13,Froyland14}.
Further diagnostic approaches have also been influenced by techniques
for ergodic dynamical systems, such as trajectory complexity and long-term
averages along trajectories \cite{Mezic10,Rypina11,Budisic12_1}.

The clustering approach developed here falls in the second category, focusing on the identification of the interiors of coherent Lagrangian vortices. Our method is unconcerned with the deformation of the boundary, requiring only a bulk coherence for the interior of the material vortex instead. We build on techniques developed over the past few decades in computer science for data clustering \cite{Everitt11}. While clustering methods have already been used in coherent structure detection in fluid flows \cite{Ser-Giacomi15_1,Froyland15}, here we apply spectral clustering to a graph describing the spatio-temporal evolution of a fluid. This approach identifies coherent vortices as clusters of Lagrangian trajectories remaining close over a finite-time interval. As we show, our proposed method detects coherent vortices in two- and three-dimensional flows, and can be extended to higher dimensional problems as well. Its main advantage is that it requires a relatively low number of Lagrangian trajectories as an input, making it suitable for the analysis of low-resolution trajectory data sets (see also \cite{Froyland15,Williams15,Froyland15_2} for methods with a similar capability). During the peer-review process of this manuscript, we were learned about the more recent preprint \cite{Banisch2016}, which applies a similar spectral clustering approach to the transfer operator framework.

Prior definitions of coherence are tied to specific geometrical requirements such as convexity \cite{Pratt14,Haller16}, lack of filamentation \cite{Haller13}, or shape coherence \cite{Ma14} of the vortex boundary. In contrast, our approach does not pose any geometrical constraint on the vortex boundary, which helps us to identify coherent vortices that may have non-convex or deformable boundaries. Unlike most other Lagrangian methods \cite{Haller13,Ma14,Mezic10,Froyland10}, which rely only on initial and final positions of particles, our method makes use of intermediate particle location information (as does \cite{Froyland15}). Another important feature is the ability to extract the a priori unknown number of coherent structures from the trajectory data set together with their simultaneous detection. This is an important prerequisite for automatic vortex tracking in large-scale data sets (see also \cite{Karrasch15}).

Our approach is based on three basic principles:

\begin{ax}{[}Coherence indicator{]}\label{axm:distance} The \emph{dynamical
distance} between two Lagrangian particles is the distance between their
corresponding trajectories in space-time over a finite time interval
$[t_{0},T]$ of interest. \end{ax}

\begin{ax}{[}Coherent structure{]}\label{axm:coherence} A \emph{coherent structure} is a distinguished set of Lagrangian particles which have
mutually short dynamical distances relative to the distances to particles from its complement.
\end{ax}

This definition adopts the notion of coherence from \emph{spatio-temporal clustering algorithms} \cite{Kisilevich10} to coherence in fluid flows, in a fashion similar to \cite{Froyland15}. A typical unsteady fluid, however, is not a union of coherent
structures. Rather, it is composed of coherent sets and their surrounding
incoherent background turbulence \cite{Mcwilliams84,Provenzale99}.
Our third principle makes this explicit as follows.

\begin{ax}{[}Coherence
vs.~incoherence{]}\label{axm:turbulence} Coherent structures are
surrounded by an incoherent background of particles.
\end{ax}

Our \Cref{axm:turbulence} underlines the impossibility
of a simple clustering of a general fluid flow into coherent structures. Instead,
we formulate the following main objective.

\begin{problem}\label{prob:MainProblem}
Given a fluid domain, possibly sampled discretely, and a finite time interval
$[t_{0},T]$ of interest, find a partition of the fluid domain into coherent
structures surrounded by an incoherent background.
\end{problem}

The rest of the paper is organized as follows. \Cref{section:method}
presents our method for identifying coherent vortices. \Cref{section:previouswork}
describes the relationship of our method with previous methods, namely the transfer operator
approach \cite{Froyland10,Froyland13}, its hierarchical application
\cite{Ma13}, the application of the community detection method
Infomap to the transfer operator \cite{Ser-Giacomi15_1}, and the direct application of the fuzzy C-means algorithm to trajectory data sets \cite{Froyland15}. We demonstrate the applicability and effectiveness of our method through four examples in \Cref{section:results}.

\section{Method}\label{section:method}

The general outline of our method is as follows. To solve the physical
\Cref{prob:MainProblem}, we start with a discrete sample
of the fluid flow and generate an abstract weighted graph, whose
nodes correspond to Lagrangian particles and whose edge weights
are determined according to \Cref{axm:distance}. Next, we apply spectral
clustering to this graph, which is particularly suited to detect
clusters in the graph according to \Cref{axm:coherence} together
with the incoherent background, consistently with \Cref{axm:turbulence}.

\subsection{Input: A trajectory data set}
\label{section: DataSet}

The essential input for our algorithm is a spatio-temporal trajectory
data set, such as particle tracks from a flow experiment, drifter
data from the ocean, or from numerical integration of a differential
equation. The trajectory data set may be sparse or spatially non-uniform at the initial time. Specifically,
we only assume that in a $d$-dimensional configuration space, $n$
trajectory positions $\left\lbrace\mathbf{x}^{i}(t)\right\rbrace_{i=1}^{n}\in\mathbb{R}^{d}$
are available at $m$ discrete times $t_{0}<t_{1}<\ldots<t_{k}<\ldots<t_{m-1}=T$.
This information can be stored in an $n\times m\times d$-dimensional
numerical array, with elements $\mathbf{x}_{k}^{i}\coloneqq\mathbf{x}^{i}(t_{k})\in\mathbb{R}^{d}$.

From this trajectory data, we define the \emph{dynamical distance}
$r_{ij}$ between Lagrangian particles $\mathbf{x}^{i}$
and $\mathbf{x}^{j}$ as
\begin{equation*}
\begin{split}
r_{ij}&\coloneqq\frac{1}{t_{m-1}-t_{0}}\sum_{k=0}^{m-2}\frac{t_{k+1}-t_{k}}{2}\left(\left|\mathbf{x}_{k+1}^{i}-\mathbf{x}_{k+1}^{j}\right|+\left|\mathbf{x}_{k}^{i}-\mathbf{x}_{k}^{j}\right|\right)\\
&\approx\frac{1}{t_{m-1}-t_{0}}\int_{t_{0}}^{t_{m-1}}\left|\mathbf{x}^{i}(t)-\mathbf{x}^{j}(t)\right|\mathrm{d}t.
\end{split}
\end{equation*}
Here $\left|\cdot\right|$ denotes the spatial Euclidean norm, and
hence $r_{ij}$ approximates the $L^{1}$-norm of pairwise trajectory
distances. Since Euclidean coordinate
transformations leave Euclidean distances unchanged, one readily sees
that the pairwise distances are \emph{objective}, i.e., they remain unchanged
in coordinate systems rotating and translating relative to each other \cite{Truesdell04}. Moreover,
it is noteworthy that the pairwise distances remain unchanged under refinements
of the spatial resolution.

\subsection{Similarity graph construction}

\label{section:simliarity}

Next, we convert the spatio-temporal data set with the pairwise distances
$r_{ij}$ into \emph{a similarity graph} $G=(V,E,W)$, which is specified
by the set of its nodes $V=\left\{ v_{1},...,v_{n}\right\} $, the
set of edges $E\subseteq V\times V$ between nodes, and a similarity
matrix $W\in\mathbb{R}^{n\times n}$ which associates weights $w_{ij}$
to the edge $e_{ij}$ between the nodes $v_{i}$ and $v_{j}$.

Specifically, the nodes of $G$ are defined as the Lagrangian particles, i.e.,
$v_{i}=\mathbf{x}^{i}$. The edges between these nodes have the associated weights
\begin{equation}
w_{ij}=1/r_{ij}\qquad\qquad i\neq j,\label{eq:similarities}
\end{equation}
$w_{ij}=1/r_{ij}$ for $i\neq j$, expressing pairwise \emph{similarities}
between distinct Lagrangian particles. Other definitions of similarity
are also possible. In general, converting distance to similarity can be done via any monotonically decreasing function, as long as the distance function $r_{ij}$ is a \emph{metric}, i.e., it satisfies for all points in the space the metric axioms of identity, non-negativity, symmetry, and triangle inequality. This is according to the intuition that the ordering of graph nodes from most dissimilar to least dissimilar should be preserved through the similarity conversion.

Extending the present similarity definition \eqref{eq:similarities}
to the diagonal of $W$ would yield infinitely large quantities. To
regularize $W$, we set the diagonal elements to a large constant
$w_{ii}=K\gg1$, $i=1,\ldots,n$. As we shall see later, the actual
value of $K$ is immaterial in our algorithm.

The entries of $W$ characterize the likelihood of nodes $v_{i}$
and $v_{j}$ to be in the same coherence cluster. By construction,
$W$ is nonnegative ($w_{ij}\geq0$) and symmetric ($W=W^{\top}$,
with the superscript $\top$ referring to matrix transposition).

The \emph{degree} of a node $v_{i}\in V$ is defined as \cite{Chung97}
\[
\deg(v_{i})\coloneqq\sum_{j=1}^{n}w_{ij}.
\]
Subsequently, the \emph{degree matrix} $D$ is defined as the diagonal
matrix with the degrees $\deg(v_{i})$ on the diagonal. For
a subset $A\subset V$ of nodes, we denote its complement in $V$
by $\overline{A}$. We measure the size of $A$ by two different quantities:
\begin{gather*}
\lvert A\rvert\coloneqq\lbrace i;~v_i\in A\rbrace,\\
\vol(A)\coloneqq\sum_{i\in A}\deg(v_{i}).
\end{gather*}
Here, $\lvert A\rvert$ measures the size of $A$ by its number of nodes,
while $\vol(A)$ measures the size of $A$ by summing over the
weights of all edges attached to nodes in $A$.

\subsection{Graph sparsification}\label{section:GraphSparsification}

For large data sets, storing
all entries of the similarity matrix $W$ is prohibitive. For instance,
storing $n=10^{6}$ elements with double precision requires 8 Terabytes
of memory, which clearly exceeds the capacity of today's typical personal
computers \cite{Chen11}.

To address this issue, techniques have been developed to sparsify
$W$ by retaining only elements describing strong enough similarity.
Two widely-used approaches are the \emph{k-nearest neighbors} and
the \emph{$\epsilon$-neighborhood} approaches \cite{Luxburg07}.
In the former, $w_{ij}$ is retained if $v_{j}$ (or $v_{i}$) is
among the $k$ nearest neighbors of $v_{i}$ (or $v_{j}$), $k\ll n$.
In the latter, $w_{ij}$ is retained if it exceeds a specified threshold
$\epsilon$. All other $w_{ij}$ entries are set to zero and hence
require no storage. Other advanced sparsification approaches include
random sampling \cite{Karger99}, sampling in proportion to edge connectivities
\cite{Benczur96}, sampling in proportion to the effective resistance
of an edge \cite{Spielman11}, and sampling using relative neighborhood
graphs \cite{Jaromczyk92,Ahuja82,Gabriel69}.

Here we select the $\epsilon$-neighborhood approach because of its
low computational cost. For the practical determination of nearest
neighbors, a number of efficient packages are available \cite{Garcia08,Muja09}.


\subsection{Spectral clustering}\label{section:SpectralClustering}

With the notation developed so far, our original \Cref{prob:MainProblem}
can be re-formulated as follows.

\begin{problem}{[}Similarity graph clustering{]}\label{prob:similarity}
Given a similarity graph, find a partition of the set of its nodes
into \emph{clusters} such that both of the following hold:
\begin{enumerate}
\item Nodes in the same cluster are similar to each other, which aims to maximize the within-cluster similarities.
\item Nodes in a cluster are dissimilar from those located in other clusters or those not included in any cluster (incoherent background), which aims to minimize the between-cluster similarities.
\end{enumerate}
\end{problem}
These two requirements for clusters implement \Cref{axm:coherence} and \Cref{axm:turbulence}, respectively.
A particularly efficient method to identify
clusters is spectral clustering, which we discuss below (see also
\cite{Luxburg07} for a review).

\subsubsection{Spectral clustering and optimal graph cuts}\label{section:graph_cut}

\begin{figure}

\includegraphics[width=0.3\textwidth]{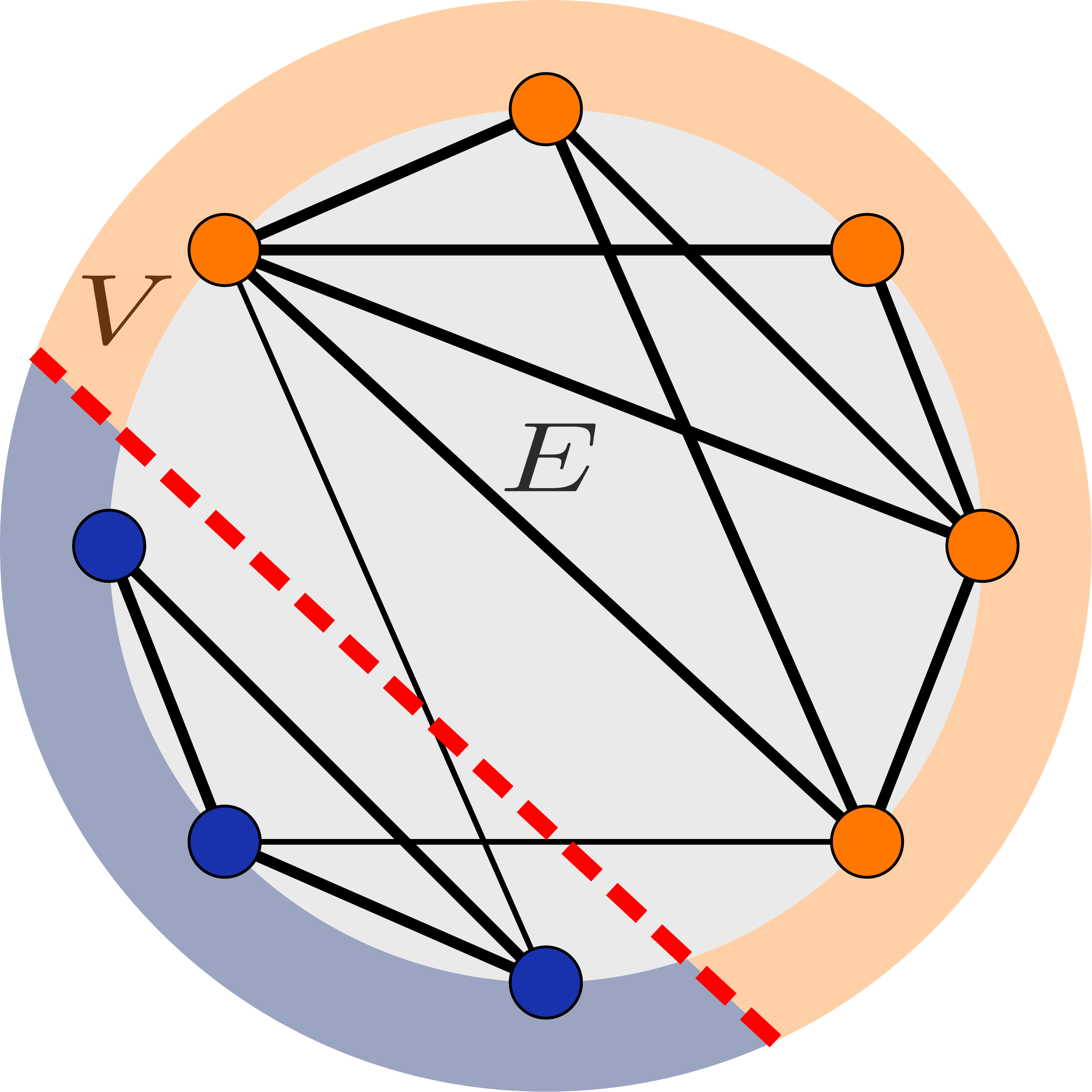}
\caption{Undirected graph partitioning. The dashed line shows the solution
of the problem of finding a decomposition of the graph into two size-balanced
groups with minimal number of edges connecting nodes from distinct
groups.}
\label{fig:UndirectedGraph}
\end{figure}

Given a similarity graph $G=(V,E,W)$, a \emph{graph
cut} is a partition of the set of nodes $V$ into two (or possibly
more) subsets $A$ and $B$. To such a partition, we assign a \emph{weight
cut} $W(A,B)$ defined as the sum of the edge weights between two
sets $A$ and $B$, i.e.,
\[
W(A,B)\coloneqq\sum\limits _{i\in A,j\in B}w_{ij}.
\]
Now, consider a subset of graph nodes with very high within-group
similarity and with weak connections to its complement, such as the
orange set in \cref{fig:UndirectedGraph}.
A graph cut separating this subset from the rest of the graph (such
as the cut indicated by the red dashed line) then yields a much smaller
weight cut $W(A,\overline{A})$ than another graph cut through $A$,
which would necessarily cut some of the strong connections within
$A$.

This suggests the following minimization problem, also known as the
\emph{mincut problem}, as a solution of \Cref{prob:similarity}:
For a given number $k$ of subsets, the mincut problem is to find
a partition $A_{1},...,A_{k}$ of $V$ which minimizes
\begin{equation}
\cut(A_{1},...,A_{k})=\frac{1}{2}\sum_{i=1}^{k}W(A_{i},\overline{A}_{i}).
\end{equation}
For $k=2$, the mincut problem can be solved very efficiently (see,
e.g., \cite{Stoer97}). In practice, however, the solution of the
mincut problem often just separates one individual node (the one with weakest connections)
from the rest of the graph. One way to circumvent this problem is to penalize the
smallness of sets in candidate partitions. The most commonly applied
objective functions that implement this idea are the normalized cut
\cite{Shi00}, or \emph{NCut} for short, RatioCut \cite{Hagen92},
MinMaxCut \cite{Ding01} and Cheeger ratio cut \cite{Cheeger70}.
Notably, not all of these graph cut objective functions have solutions
which satisfy both conditions in \Cref{prob:similarity} (cf.\ \cite{Luxburg07}
for more details).

In this paper, we use the NCut objective function,
whose (approximate) solutions maximize the within-cluster similarity and minimize the between-cluster similarity:
\[
\NCut(A_{i},...,A_{k})=\frac{1}{2}\sum_{i=1}^{k}\frac{\cut(A_{i},\overline{A}_{i})}{\vol(A_{i})},
\]

Introducing the penalizing balancing conditions, however, turns the
originally simple mincut problem into an NP hard problem \cite{Wagner93}.
Spectral clustering is a way to solve relaxed versions of balanced graph
cut problems. 

\subsubsection{Graph Laplacian}

Shi \& Malik \cite{Shi00} showed that the solution of the Ncut problem
can be approximated by solutions of the generalized eigenproblem associated
with the \emph{(unnormalized) graph Laplacian} $L=D-W$, where $D$ is the diagonal degree matrix
of node degrees and $W$ is the similarity matrix defined earlier.

The \emph{generalized eigenvalue problem} for the graph Laplacian
is then defined as
\begin{equation}
Lu=\lambda Du.\label{eq:gen_eigenproblem}
\end{equation}
We refer to its solutions as generalized eigenvectors for short. Generalized
eigenvectors $u$ then offer an alternative representation
of the weighted graph data. As we will see in the next
sections, this change of representation enhances the cluster-properties
in the data, so that clusters can be easily detected in the new
representation. In particular, the simple K-means clustering algorithm
has no difficulties to detect the clusters in this new representation
(see \Cref{section:K-means} regarding K-means clustering). 


It is known from Spectral Graph Theory \cite{Chung97} that the eigenvalues
solving \eqref{eq:gen_eigenproblem} satisfy $0=\lambda_{1}\leq\ldots\leq\lambda_{n}$.
If the underlying graph consists of $k$ disconnected components
(clusters with zero between-cluster similarity), then $\lambda=0$ is a
generalized eigenvalue of multiplicity $k$. In that case, the eigenspace
corresponding to this eigenvalue is spanned by the indicator vectors of the
individual connected components. A perturbation argument implies that
if the between-cluster similarities remain small, then the eigenvectors
of the first $k$ eigenvalues remain close to indicator type \cite{Luxburg07}.
This enables reconstructing the clusters from the first $k$ eigenvectors
obtained from \eqref{eq:gen_eigenproblem}. The main challenge, therefore,
is to extract a meaningful number of clusters directly from the data,
as opposed to postulating its value beforehand.

\subsection{Estimating the number of clusters by eigenspace analysis}\label{section:eigengap}

For a predetermined number $k$, the spectral
clustering algorithm of Shi \& Malik \cite{Shi00} collects the $k$
dominant generalized eigenvectors as \emph{cluster indicators} in a matrix $U=\left(u_{1},\ldots,u_{k}\right)\in\mathbb{R}^{n\times k}$. To retrieve $k$ from the graph data, we adopt here the
eigengap heuristic \cite{Bhatia97} by which
\begin{equation}
k=\arg\min_{i}\left(\max\left(g_{i}\right)\right),\label{eq:kdef}
\end{equation}
where $g_{i}=\lambda_{i+1}-\lambda_{i}$ for $i=1,...,n$. In other
words, $k$ is simply determined as the number of eigenvalues preceding
the largest gap in the eigenvalue sequence. The presence of such a
gap enables us to invoke the perturbation argument of the previous
section, and argue that our graph $G=(V,E,W)$ is a perturbation of
one with $k$ disconnected components.

Expression \eqref{eq:kdef} determines the number of coherent
clusters satisfying the definition given in \Cref{section:SpectralClustering}.
Ultimately, however, we need to partition the graph $G=(V,E,W)$
into $k+1$ clusters to also account for the incoherent cluster surrounding
the coherent clusters, as codified in our \Cref{axm:turbulence}.
We refer to the last, $(k+1)$st cluster arising in this process
as the \emph{noise cluster} or \emph{incoherent cluster} since it
includes nodes that do not belong to any coherent cluster.

Spectral gap arguments were used before in the context of dynamical systems (see \cite{Dellnitz99,Deuflhard00,Froyland03,Budisic12_1,Froyland15_2} for examples). While the number of cluster indicators (leading singular- and eigenvectors) in some of these works (i.e., \cite{Dellnitz99,Deuflhard00}) coincide with the number of coherent structures, in others (i.e., \cite{Froyland03,Budisic12_1,Froyland15_2}) the number of cluster indicators differs from the number of coherent structures (see \cite{Sarkar11} and \Cref{section:TransferOperator} for more details).

\begin{rem}
As discussed, we identify the number of vortices present in a given domain by locating the largest gap in the eigenvalue sequence. This implies that the number of eigenvalues and eigenvectors to be computed should be greater than the maximum number of vortices expected to be present in a domain. In the absence of intuition for the maximum number of vortices, one needs to conduct a full matrix decomposition instead of a partial decomposition. The computational cost of such a decomposition, however, increases dramatically with respect to the number of eigenvalues  to be computed (see \cite{Cai10} for more information).
\end{rem}

\subsection{Retrieving clusters from matrix $U$ by K-means clustering} \label{section:K-means}
As a last step, we employ K-means clustering to convert relaxed continuous spectral vectors, corresponding to $U$'s $k$ columns, into a discrete cluster indicator vector containing the cluster assignment for each node $x^{i}$.

Given the spectral vectors $U \in \mathbb{R}^{n\times k}$ and integer $K$, K-means clustering aims to determine $K$ points in $\mathbb{R}^{k}$, called \emph{centers}, so as to minimize the mean squared distance from each node to its nearest center. In 1957 Stuart Lloyd \cite{Lloyd82} suggested a simple iterative algorithm which efficiently finds a local minimum for this problem. Given any set of $K$ centers, the algorithm proceeds by alternating between the following two steps:

\begin{description}
\item[Assignment]find each node's nearest center and assigns it to the corresponding cluster.
\item[Update] recalculate cluster centers by measuring the mean of all nodes included in each cluster.
\end{description}
These steps repeat until no node is reassigned. Readers not familiar with K-means can read about this algorithm in numerous text books, for example see \cite{Everitt11}. Throughout the paper, we choose the number of cluster centers $K$ equal to $k+1$, where the last, $(k+1)$st cluster corresponds to the incoherent or noise cluster discussed in \Cref{section:eigengap}. The K-means algorithm and its probabilistic counterpart (fuzzy C-means) have been used before to extract coherent structures either directly from a trajectory data set  \cite{Froyland15}, or indirectly from cluster indicators resulted from various spectral dimensionality reduction algorithms \cite{Froyland03,Budisic12_1}.

We summarize our numerical procedure in \Cref{alg:algorithm1}.
\begin{algorithm}
\caption{}
\label{alg:algorithm1} Input: Similarity matrix $W\in\mathbb{R}^{n\times n}$
(cf.\ \Cref{section:simliarity})
\begin{enumerate}
\item Sparsify $W$ by using the NCut algorithm (cf.\ \Cref{section:GraphSparsification}.)
Remove isolated nodes, i.e., nodes with degree zero, from $G=(V,E,W).$
\item Compute the graph Laplacian $L$, and solve the generalized eigenvalue
problem $Lu=\lambda Du$.
\item Identify the number $k$ of coherent clusters as the number of eigenvalues
preceding the largest gap among the increasingly ordered eigenvalues.
Select the first $k$ generalized eigenvectors $u_{1},...,u_{k}$
as coherent cluster indicators.
\item Assemble the matrix $U=\left(u_{1},\ldots,u_{k}\right)$. Each row of $U$ corresponds
to a graph node (excluding the isolated nodes). Apply K-means
to the first $k$ eigenvectors and extract $k+1$ clusters. The last
cluster is the incoherent cluster and corresponds to the mixing region
filling the space between coherent clusters.
\end{enumerate}
Output: Clusters $C_{1},...,C_{k+1}$.
\end{algorithm}
\subsection{Large-scale spectral clustering}\label{section:large_scale}

For large data sets, considerable time
and memory is required to compute and store the similarity matrix $W$ and
the graph Laplacian $L$. The most commonly used approach
to address this issue is graph sparsification, as discussed earlier in
\Cref{section:GraphSparsification}. From the sparse similarity
matrix $W$ so obtained, one determines the corresponding Laplacian
matrix $L$, and calls a sparse eigenvalue solver.

Even after the sparsification of $W$, however, calculating the generalized
eigenvectors of the graph Laplacian $L$ remains challenging with
$O(n^{3})$ worst-case complexity \cite{Chen11}. Several authors
\cite{Chen11,Song08} tried to alleviate the problem by adapting
standard eigenvalue solvers to distributed architecture. Other approaches
are designed to achieve efficiency by finding numerical approximations
to eigenfunction problems \cite{Fowlkes04,Chen06,Liu07}.

Here, we adopt a low-rank matrix approximation
approach. The main idea is to coarse-grain the similarity graph $G=(V,E,W)$,
while keeping as much information as possible from the original graph
and its weights. To this end, we construct a \emph{bipartite} graph $G_{\mathcal{B}}=(V_{\mathcal{B}},E_{\mathcal{B}},W_{\mathcal{B}})$
from the original similarity graph by uniform spatial sampling of $q$ graph nodes, called \emph{supernodes}, from $n$ graph nodes, where $q\ll n$ \cite{Cai14,Liu13}. A bipartite graph is a graph whose
set of nodes $V_{\mathcal{B}}$ admits a partition into two disjoint sets,
$A$ and $B$, such that each edge connects a node in $A$ to one in $B$.
As a result, no two nodes within $A$ and within $B$ are connected by
an edge. Here, we set $A$ as the set of all $n$ original graph nodes,
and $B$ as its subset of $q$ supernodes, considered as \emph{independent copies}.
The weights are now defined as before, such that the square $(n+q)\times (n+q)$ similarity matrix $W_{\mathcal{B}}$ of the bipartite graph can be written as
\begin{equation}
W_{\mathcal{B}} = \begin{pmatrix}0 & Z^{\top}\\ Z & 0 \end{pmatrix}
\end{equation}
where $Z\in\mathbb{R}^{q\times n}$ is a \emph{tight similarity matrix}
containing the edge weights between all nodes and supernodes, i.e.,
between $A$ and $B$. Now, one can pose the Ncut problem to the bipartite
graph whose similarity matrix enjoys a simple block-structure.
As shown by Dhillon \cite{Dhillon01} and Zha et al. \cite{Zha01}, this block-structure breaks the
associated Ncut problem into two parts such that the dominant right
singular vectors of the normalized $q\times n$ tight similarity matrix $\hat{Z}=D_{2}^{-1/2}ZD_{1}^{-1/2}$ play the role of the generalized
eigenvectors of the graph Laplacian in \Cref{section:SpectralClustering}.
Here, $D_{1}$ is an $n\times n$ diagonal matrix whose entries are column
sums of $Z$ and $D_{2}$ is a $q\times q$ diagonal matrix whose entries
are row sums of $Z$ (see \Cref{app:bipartite-clustering} for more details).

\begin{figure}
 
\includegraphics[width=0.2\textwidth]{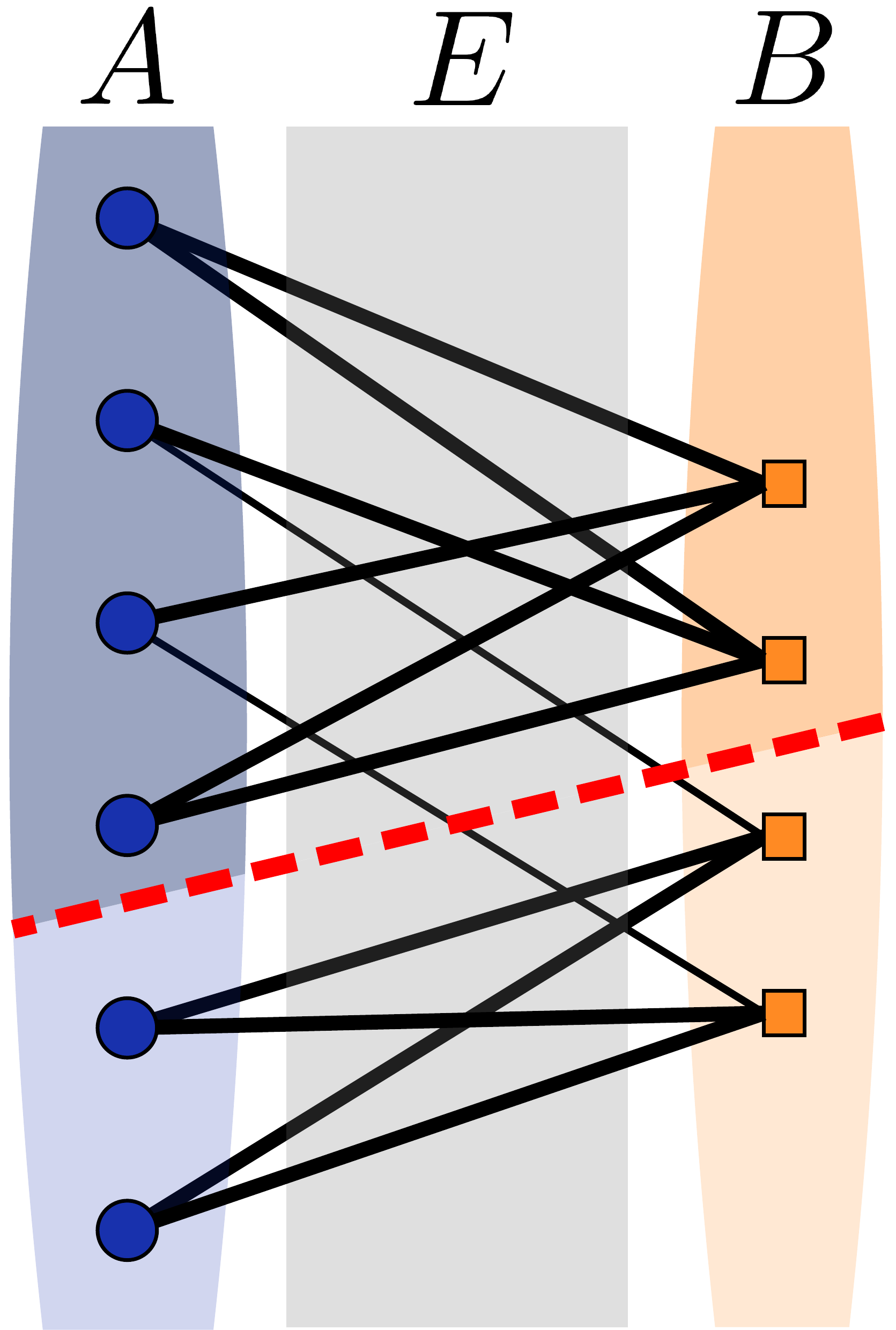}
\caption{Partitioning of a bipartite graph $G_{\mathcal{B}}=(V_{\mathcal{B}},E_{\mathcal{B}},W_{\mathcal{B}})$ whose set of nodes $V_{\mathcal{B}}$ is divided into two disjoint sets $A$ and $B$ such that $V_{\mathcal{B}} = A\cup B$. The dashed line shows the solution of normalized graph cut yielding a simultaneous decomposition of $A$ and $B$.}
\label{fig:BipartiteGraphCut}
\end{figure}

We now summarize our algorithm for large-scale trajectory data sets.
\begin{algorithm}
\caption{}
\label{alg:algorithm2}
\begin{enumerate}
\item Select uniformly $q$ supernodes from $n$ graph nodes.
\item Construct a tight similarity matrix $Z\in\mathbb{R}^{q\times n}$
between all original graph nodes and the supernodes.
\item Given $Z$, form $\hat{Z}=D_{2}^{-1/2}ZD_{1}^{-1/2}$. Compute the singular values and vectors of $\hat{Z}$. Select the first $k$ right singular vectors $u_{1},\ldots,u_{k}$ as cluster indicators for the original graph.
\item Assemble the matrix $U=\left(u_{1},\ldots,u_{k}\right)$. Each row of $U$ corresponds
to a graph node. Apply K-means to the first $k$ right singular vectors and extract $k+1$ clusters. The last cluster is the incoherent cluster and corresponds to the mixing region filling the space between coherent clusters.
\end{enumerate}
Output: Clusters $C_{1},...,C_{k+1}$.
\end{algorithm}
\section{Related previous work}\label{section:previouswork}
\subsection{The transfer-operator approach}\label{section:TransferOperator}
\begin{figure*}
 \includegraphics[width=0.8\textwidth]{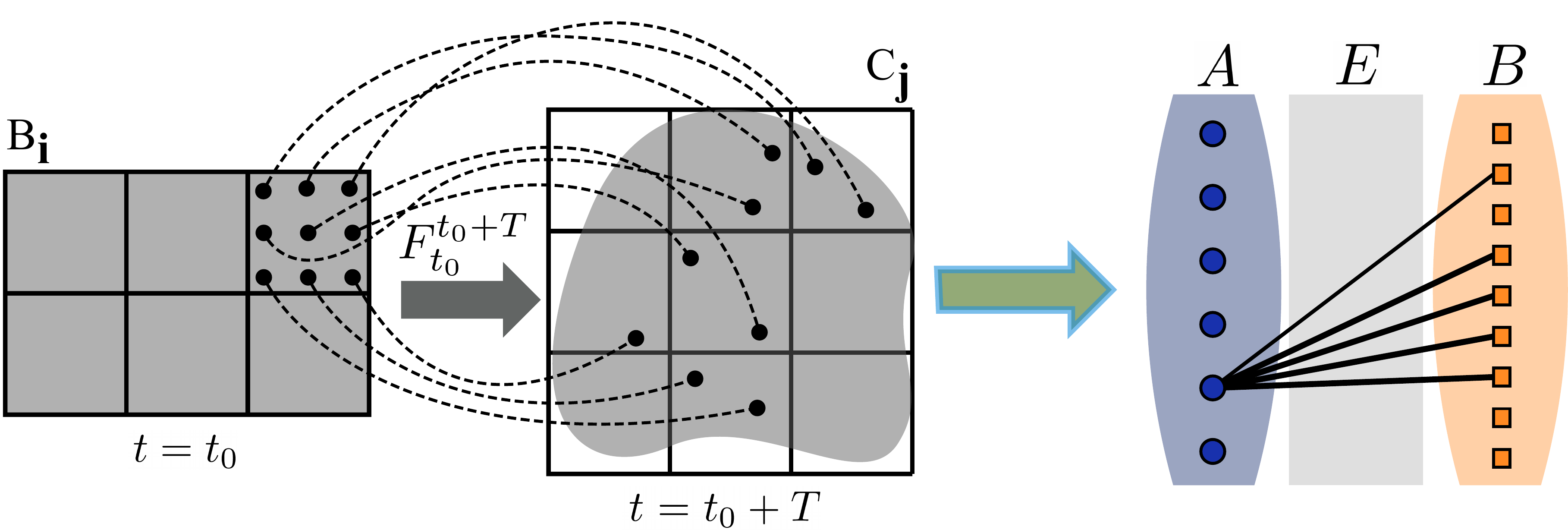}
\caption{Interpreting transition matrix constructed from tracer advection as tight similarity matrix $Z$ of a bipartite graph.}
\label{fig:TransferOperator}
\end{figure*}
In the transfer operator-based approach \cite{Froyland10,Froyland13,Froyland14}
finite-time coherent sets are defined as regions in phase space that
minimally diffuse with the surrounding phase space during a finite
time interval. The method builds on the \emph{Perron-Frobenius operator}
or \emph{transfer operator}, which describes the evolution of material
densities under the flow map.

In practice, the infinite-dimensional transfer operator needs to be
approximated by a finite-dimensional matrix, the \emph{transition
matrix} $P$, which is most commonly obtained from a partition of
the flow domain $\left(B_{i}\right)_{i}$ and the flow image $\left(C_{j}\right)_{j}$
into distinct boxes, and subsequent computation of discrete transition
probabilities: the transition matrix entry $P_{ij}$ is computed as
the number of particles transported from $B_{i}$ to $C_{j}$, normalized
by the total number of particles released from $B_{i}$ (see \cref{fig:TransferOperator}).
This box partitioning is also referred to as Ulam's method, and introduces
(numerical) diffusion at the implementation level \cite{Froyland10}.

In our context, the transition matrix $P$ can be interpreted as the
tight similarity matrix $Z$ of a bipartite graph $G_{\mathcal{B}}$ as follows: define
the first set of nodes $A$ as the collection of initial boxes $B_{i}$,
the second set of nodes $B$ as the collection of final boxes $C_{j}$,
and the edge weights as $Z_{ij}=P_{ij}$, see \cref{fig:TransferOperator}. A similar connection to spectral clustering and graph cuts has been worked out earlier in \cite{Bollt15}. Our presentation here, however, differs from \cite{Bollt15} in that we interpret the graph as a bipartite graph and relate it to the original references \cite{Dhillon01,Zha01}.

\begin{rem} 
The size of the resulting weight matrix depends on the size of the $B_i$'s and $C_j$'s, as well as on the underlying dynamics of the system. For instance, in the presence of chaotic dynamics, particles released at the initial time can scatter in a large domain. This, in return, may require a large number of boxes $C_j$ to cover the final domain, and results in a large number of columns in the subsequent transition matrix. In contrast, the size of the weight matrix of \Cref{alg:algorithm1,alg:algorithm2} depends on the number of tracked particles.
\end{rem}

\begin{figure*}

\subfloat[\label{fig:BickleySingularVector2}]{\includegraphics[width=0.45\textwidth]{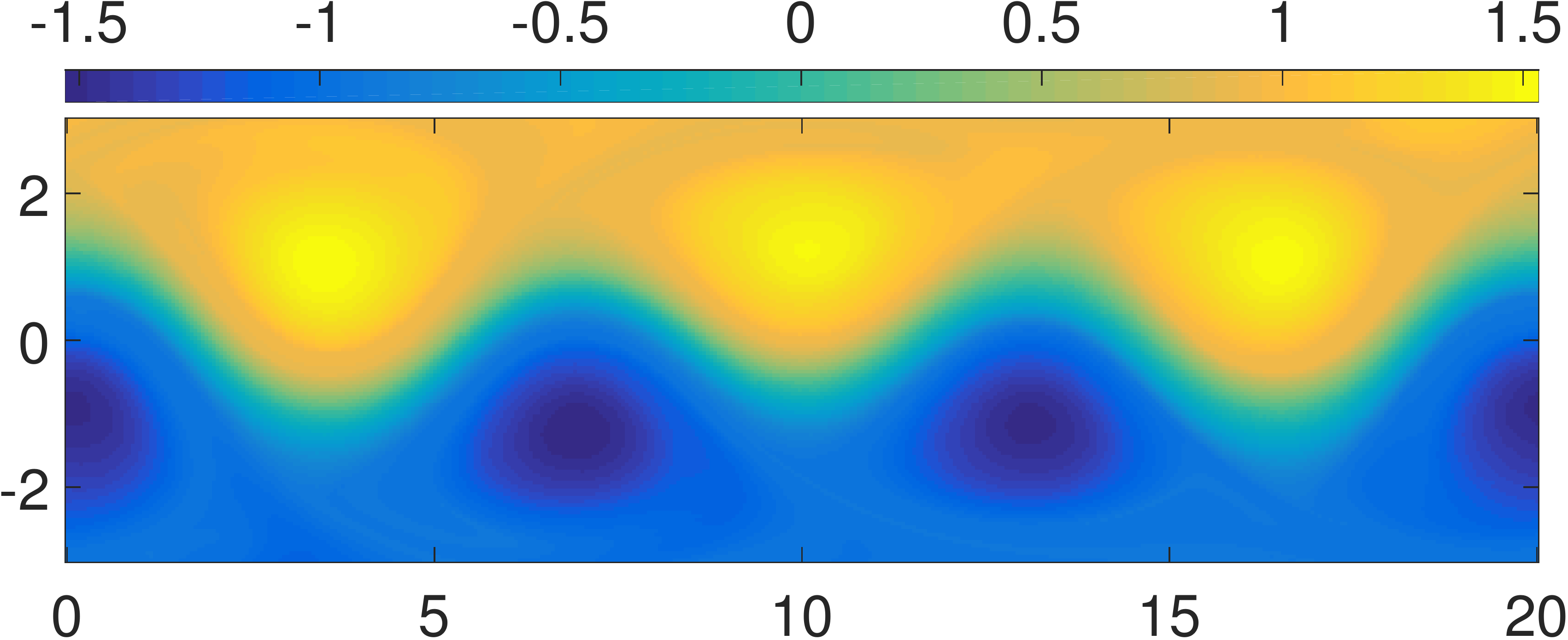}}\qquad
\subfloat[\label{fig:BickleySingularVectorThresh2}]{\includegraphics[width=0.45\textwidth]{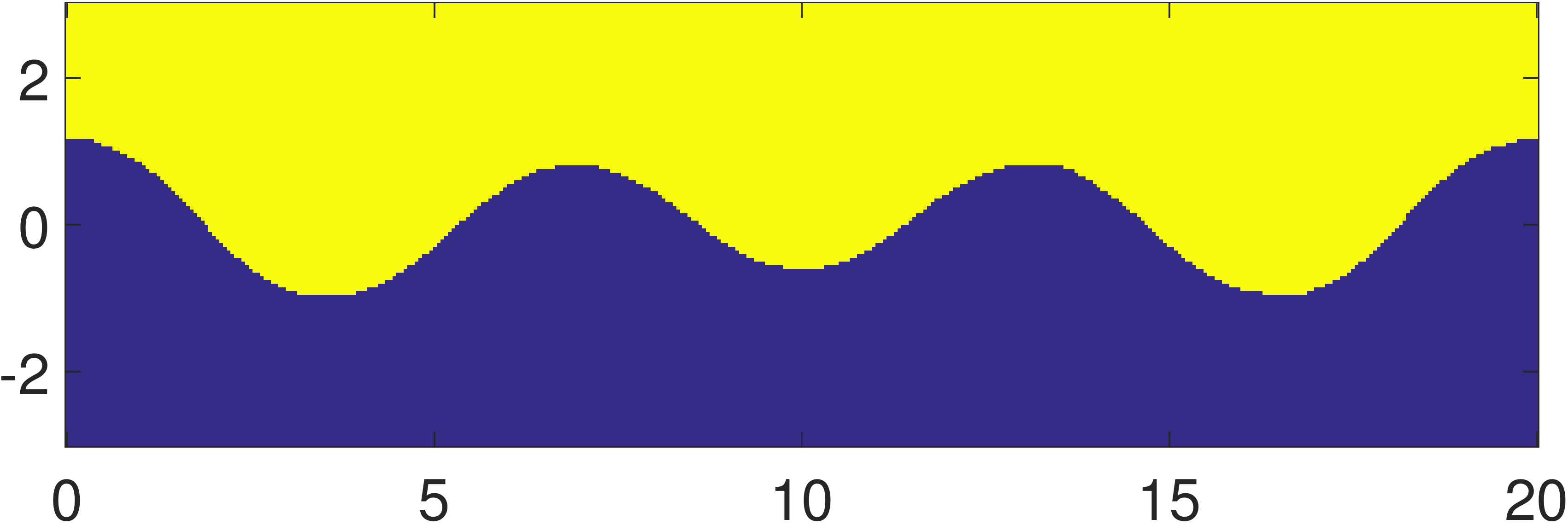}}\\
\subfloat[\label{fig:BickleySingularVector3}]{\includegraphics[width=0.45\textwidth]{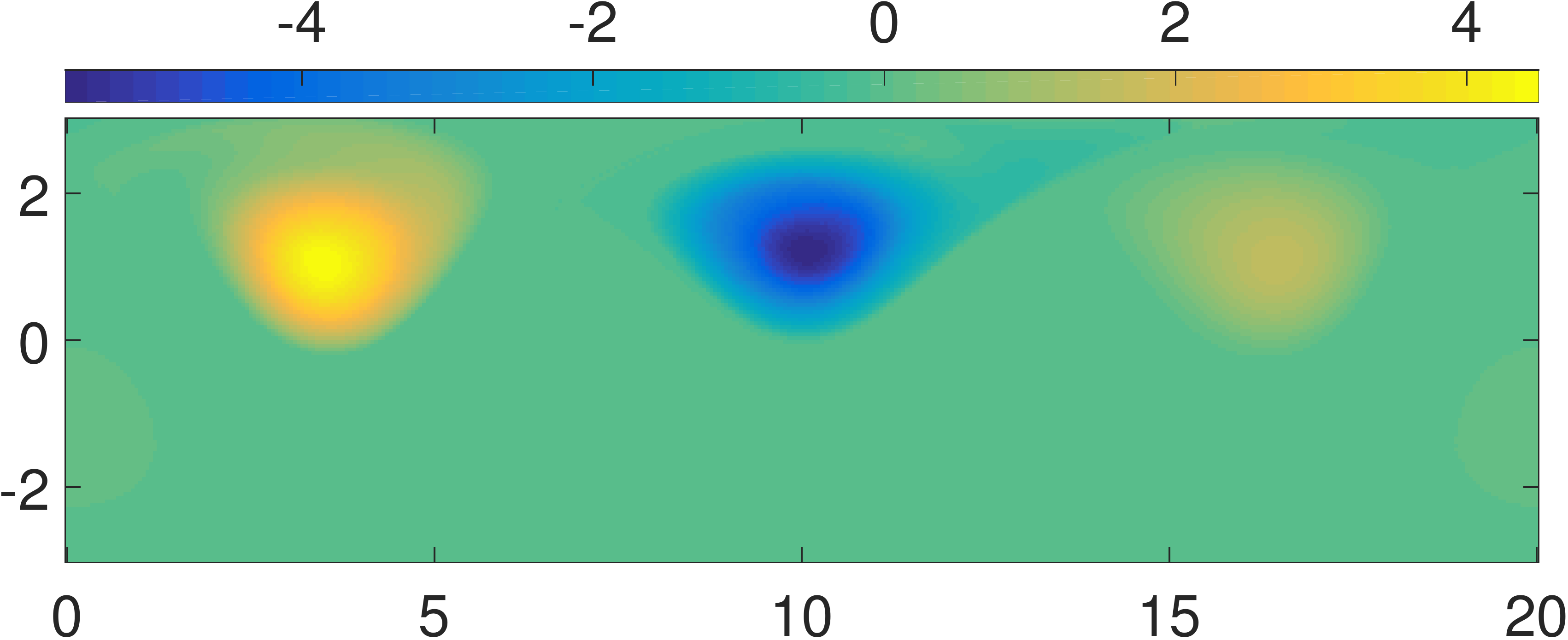}}\qquad
\subfloat[\label{fig:BickleySingularVectorThresh3}]{\includegraphics[width=0.45\textwidth]{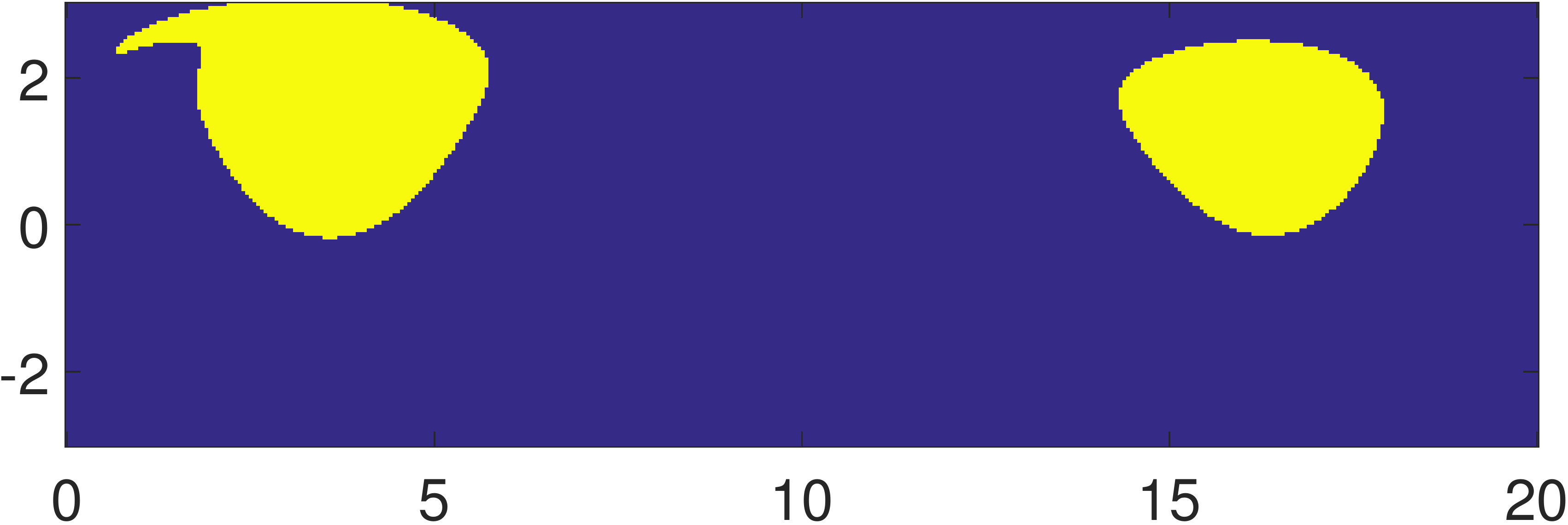}}
\caption{(a,c) Second and third largest (left) singular vectors of the normalized transition matrix for the Bickley jet flow. (b,d) Corresponding coherent sets which are obtained by searching through all possible cuts \cite{Shi00}.
To compute the transition matrix, we subdivided the domain into a grid of $400\times 120$ identical boxes, and released 400 particles in each box. We then advected particles from $t_{0}=0$ to $t=40$ days. }
\label{fig:TransferOperatorComparison}
\end{figure*}

With this bipartite graph construction, the optimization
problem which is underlying the definition of a coherent set in the
transfer-operator setting can be reformulated as a clustering problem.
In a (bipartite) graph cut, such as the one shown in \cref{fig:BipartiteGraphCut},
the weight of the cut can be interpreted as the mass leakage of one set
with its complement.

As discussed in \Cref{app:Ncut}, minimizing the normalized cut for a binary cluster indicator is NP-hard. Relaxation of the binary cluster indicator in the real value domain yields the eigenvector corresponding to the second smallest eigenvalue of $L$ as an approximate cluster indicator \cite{Shi00}. However, in order to obtain a partition of the graph, we need to re-transform the real-valued cluster indicator vector of the relaxed problem into a discrete indicator vector. The simplest way to do this is to use the sign of the eigenvector as a discrete cluster indicator function \cite{Shi00}. Alternatively, one can search for a splitting point such that the resulting partition has the best $\NCut(A,\bar{A})$ value \cite{Shi00}, or apply the line-search algorithm of \cite{Froyland10}. Viewing the transfer operator approach \cite{Froyland10} as a bipartite spectral graph partitioning \cite{Dhillon01,Zha01}, one can similarly recover discrete cluster indicator vectors from real-valued singular vectors.

\Cref{fig:BickleySingularVector2} shows the second largest singular vector of the normalized transition matrix for the Bickley jet model discussed in \Cref{Ex:Bickley}. We obtain the corresponding binary cluster indicator by searching through all possible Ncuts \cite{Shi00}. As shown in \cref{fig:BickleySingularVectorThresh2}, the binary cluster indicator so obtained highlights two coherent sets in which coherent vortices still remain hidden. Therefore, we step in hierarchy of increasingly ordered singular vectors, and search for these vortices in the third singular vector, shown in \cref{fig:BickleySingularVector3}. Similarly, we extract the corresponding discrete-valued indicator vector by examining all possible Ncuts (see \cref{fig:BickleySingularVectorThresh3}). \Cref{fig:BickleySingularVectorThresh3} reveals that the yellow set, as a single entity, is composed of two vortices. The two vortices forming the yellow set have overall small mass exchange with the blue set, implying that the objective function is minimized. The spatial connectedness, however, appears to be missing in this solution as the yellow set is composed of two distant vortices. In other words, a search for sets with minimal mass exchange, without enforcing spatial connectedness, can lead to a union of coherent structures as a solution (see also   \cite{Ser-Giacomi15_1,Deuflhard05} for similar observations). In this case, applying K-means clustering to the collection of leading singular vectors may resolve the issue. However, the number of coherent structures that needs to be extracted, may not be detectable anymore with the eigengap heuristic (see \Cref{section:eigengap}), as each singular vector highlights a combination of coherent structures. In fact, the number of coherent structures generally does not coincide with the number of singular/eigenvalues preceding the largest eigengap (see \cite{Froyland03}, pp. 1852-1853), and therefore needs to be guessed or to be known a priori.

The graph cut approach, in general, does not guarantee that the resulting cluster will form a connected region in the physical space. The connectedness constraint, however, can be enforced indirectly during the construction of the similarity graph. Our method specifically enforces this constraint by measuring and penalizing distances in the spatio-temporal domain. As a result, unlike for the transfer operator method, a union of coherent structures is not a solution for our method. In \Cref{Ex:Bickley}, we will apply our \Cref{alg:algorithm1} to the same Bickley jet flow considered earlier in \cref{fig:TransferOperatorComparison}.

\subsection{Hierarchical partitioning of the transfer-operator}\label{section:hierarchy}

In the spectral clustering community, one distinguishes between two
approaches to detect a specified number of clusters in a given
similarity graph using the graph cut procedure \cite{Shi00,Dhillon01,Zha01,Luxburg07}:
\emph{two-way clustering} and \emph{multi-way clustering}. Our methodology
presented in \Cref{section:method} follows (up to the introduction of
the incoherent cluster) the multi-way clustering approach, in which
$k$ clusters are retrieved from the $k$ dominant eigenvectors at once.

In two-way clustering, the following procedure is ap-
plied recursively to generate multiple clusters: (i) com-
pute the top generalized eigenvector of the unnormalized
graph Laplacian, and (ii) bisect the graph into two sub-
graphs.

In two-way clustering, the following procedure is applied recursively to generate multiple clusters: (i) compute the top generalized eigenvector of the unnormalized graph Laplacian, and (ii) bisect the graph into two sub-graphs. In the transfer-operator context, this procedure has
been put forward in \cite{Ma13} and is stopped when the obtained partitions
no longer satisfy a pre-specified coherence ratio (cf.\ \cite{Ma13} for details). In the clustering analysis community, two-way
clustering is also found to be inefficient due to the fact that separate
eigenvalue problems need to be solved repeatedly \cite{Chan94,Ng02,Shi00}.

\subsection{Application of fuzzy clustering to a trajectory data set}
\label{section:Cmeans}
Recently, Froyland \& Padberg-Gehle \cite{Froyland15} proposed a method based on traditional fuzzy  C-means clustering \cite{Bezdek81,Dunn73} to identify regions of phase space that remain compact over a finite time interval. Specifically, they first build a trajectory data set $X \in \mathbb{R}^{n\times dm}$ whose rows are vectors $(X_{i})_{i= 1,\ldots,n}$ containing concatenated positions of Lagrangian particles in time. Second, they apply the C-means algorithm, with a prespecified number of clusters $K$ and a set of $K$ initial starting points in $\mathbb{R}^{dm}$, to the trajectory data set. The result is a membership value describing the likelihood that a trajectory belongs to a cluster. Thus, each trajectory carries $K$ membership values, showing the degree of belonging to each of the $K$ clusters. Finally, each trajectory is assigned to only one cluster based on the maximum membership value it carries. Those trajectories carrying low membership values for all clusters are occasionally considered to be non-coherent (see \cite{Froyland15} for more details).

Compared with the fuzzy C-mean clustering used in \cite{Froyland15}, the spectral clustering technique considers the \emph{connectedness} of the data, whereas the C-means clustering method considers the \emph{compactness} of the data. Fuzzy C-means algorithm optimizes cluster compactness by assessing the proximity between the uncertain data points assigned to the cluster and the corresponding cluster center. We note that cluster centers are not true trajectories of a dynamical system although they are in the trajectory space \cite{Froyland15}. In contrast, our spectral clustering technique maximizes connectedness inside clusters and disconnectedness between clusters at the same time by measuring pairwise distances between trajectories. 



As opposed to centroid-based clustering algorithms such as K-means or C-means, where the resulting clusters tend to be convex sets \cite{Froyland15,Maimon05,Jain99}, spectral clustering can find any cluster shape, because it has no preference for the
shape of the cluster. This is important as we will show in \Cref{Ex:OceanR2} that vortices with non-convex shapes are the rule rather than the exception considering the known vortex stirring in geophysical flows \cite{Aref84}.

Most clustering methods including centroid-based methods are plagued with the problem of noisy data, i.e., identifying good clusters amongst noise points that just do not belong to any cluster \cite{Dave91}. In some cases, even a few noisy points or outliers may bias the final output of the algorithm \cite{Dave91}. In our specific context, the noise corrsponds to the incoherent or turbulence region itself, where particles do not remain compact. This implies that the turbulence region is not residing in a hypersphere, and consequently cannot be captured by adding an extra cluster to C-means or K-means algorithms (see \cite{Dave91} for more details).

On the other hand, the high dimensionality of the trajectory dataset poses a considerable challenge to K-means or C-means clustering approaches. First, the curse of dimensionality can cause slow convergence for these traditional algorithms, and, second, the existence of redundant subspaces may not allow for the identification of the underlying structure in the data (cf. \cite{Arthur06} and \cite{Cordeiro13}, p. 10).

Similar to many clustering methods, the K-means or C-means algorithms assume that the number of clusters $K$ in the dataset is known beforehand which is not necessarily true in real-world applications. In contrast, the spectral clustering can detect the right number of clusters automatically using techniques such as the eigengap heuristic (cf. \Cref{section:eigengap}).

Finally, the result of K-means or C-means clustering, depends on the initial guess for the cluster centers \cite{Maimon05,Jain99}, and can reach a local minimum of the objective function instead of the desired global minimum \cite{Jain99,Winkler2012}. Often one restarts the procedure a number of times to mitigate the problem. However, when the number of clusters $K$ is large, the number of times to restart K-means or C-means to reach an optimum can be prohibitively high and lead to a substantial increase in runtime (cf. \cite{Winkler2012}).

\section{Results} \label{section:results}
We demonstrate the implementation of \Cref{alg:algorithm1,alg:algorithm2}
on four examples to detect coherent Lagrangian vortices. In the first
example, we consider a periodically forced pendulum for which we can
explicitly confirm our results using an appropriately defined Poincar\'{e}
map. Our second example is one whose temporal complexity is one level
higher: the Bickley jet with quasi-periodic time dependence \cite{Rypina07,Del_Castillo93}.
In the third example, we detect coherent Lagrangian vortices
in a quasigeostrophic ocean surface flow derived from satellite-based
sea-surface height observations \cite{Fu10}. Our last example is a
three-dimensional velocity field, the Arnold-Beltrami-Childress (ABC) flow,
which is an exact solution of Euler's equation \cite{Arnold66}. This is our computationally
most demanding example, where we deploy \Cref{alg:algorithm2} to
reduce the graph size and the associated computational cost. For the
rest of the examples, we use \Cref{alg:algorithm1} with the $\epsilon$-neighborhood
graph sparsification approach described in \Cref{section:GraphSparsification}. We notice that the coherent structures in our first and last examples remain invariant in the phase space, and hence they are in principle detectable in principle by other spectral methods developed specifically for steady flows and maps (see e.g., \cite{Dellnitz99,Deuflhard00,Froyland03,Froyland09,Budisic12_1}).

To implement \Cref{alg:algorithm1,alg:algorithm2} in the forthcoming
examples, we use a variable-order Adams-Bashforth-Moulton solver (ODE113
in MATLAB) to solve the differential equations. The absolute and relative
tolerances of the ODE solver are chosen as $10^{-6}$. In \Cref{Ex:OceanR2},
we obtain the velocity field at any given point by interpolating the
velocity data set using bilinear interpolation.

The dynamic distances $r_{ij}$ can be computed using two approaches that differ in terms of memory consumption, suitability for parallel computation and accuracy. In the first approach, one builds a spatio-temporal trajectory data set by saving trajectory positions over $m$ intermediate times. One then measures pairwise distances using the trapezoidal rule and sparsifies them simultaneously. This can be done effectively using the ExhaustiveSearcher model object in MATLAB or other packages, such as \cite{Garcia08,Muja09}. This approach is memory consuming but highly parallelizable.

In the second approach, one constructs the similarity matrix without building any spatio-temporal trajectory data set. To this end, one measures pairwise distances concurrent with the advection of particles. Specifically, one defines an extra output argument inside the ODE function which measures and cumulates the pairwise distances over a given time interval.
 
Compared with the first approach, the second approach is more accurate and more memory efficient. However, its parallel implementation requires communication between processors, which may make the computation prohibitively slow. For this reason, we only  employ the second approach in our last example, the Arnold-Beltrami-Childress (ABC) flow, and use the first approach otherwise.

\subsection{The periodically forced pendulum}\label{Ex:Pendulum}
\begin{figure*}
\subfloat[\label{fig:PendulumDegree}]{\includegraphics[width=0.28\textwidth]{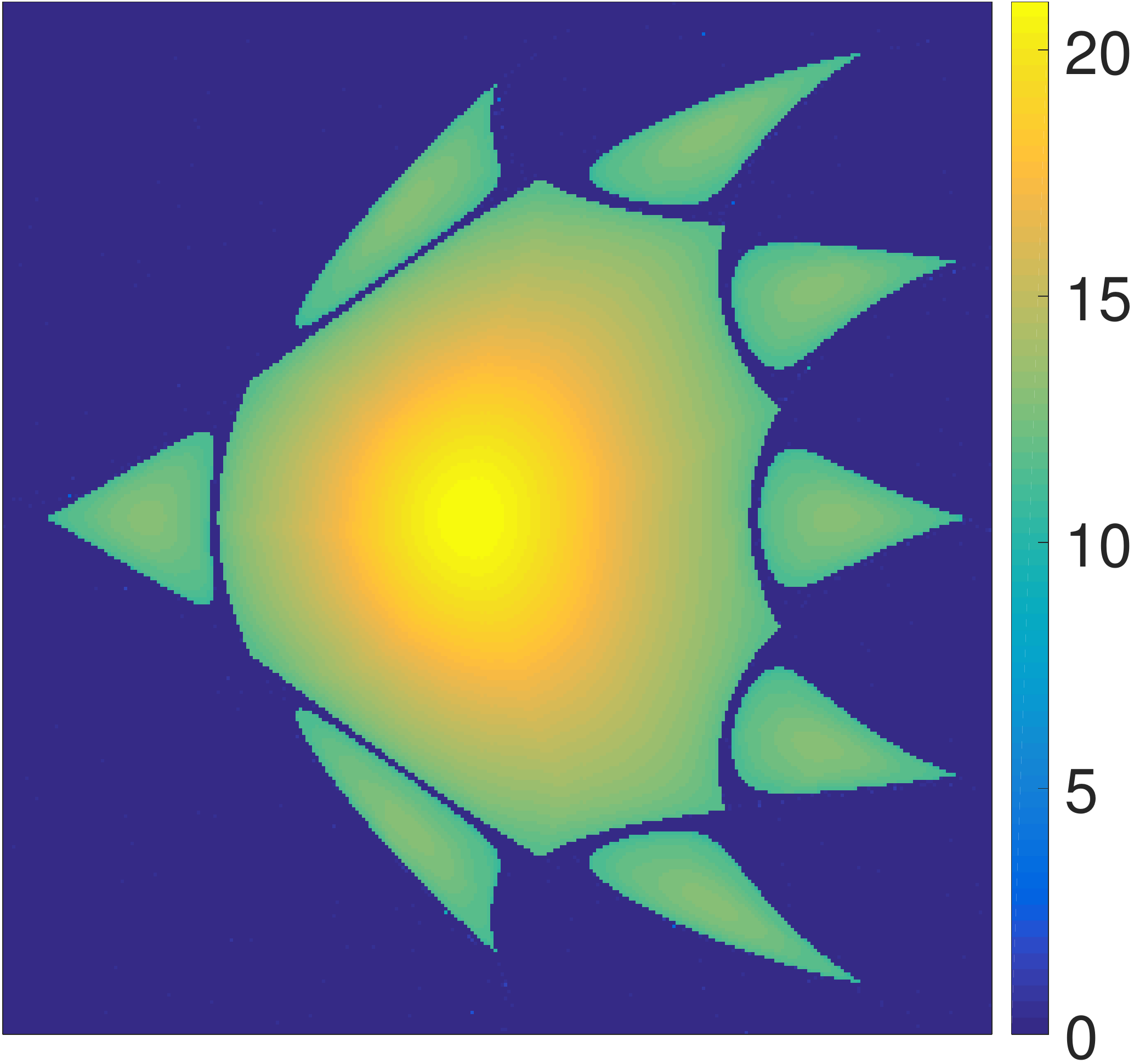}}\qquad
\subfloat[\label{fig:PendulumFTLE}]{\includegraphics[width=0.3\textwidth]{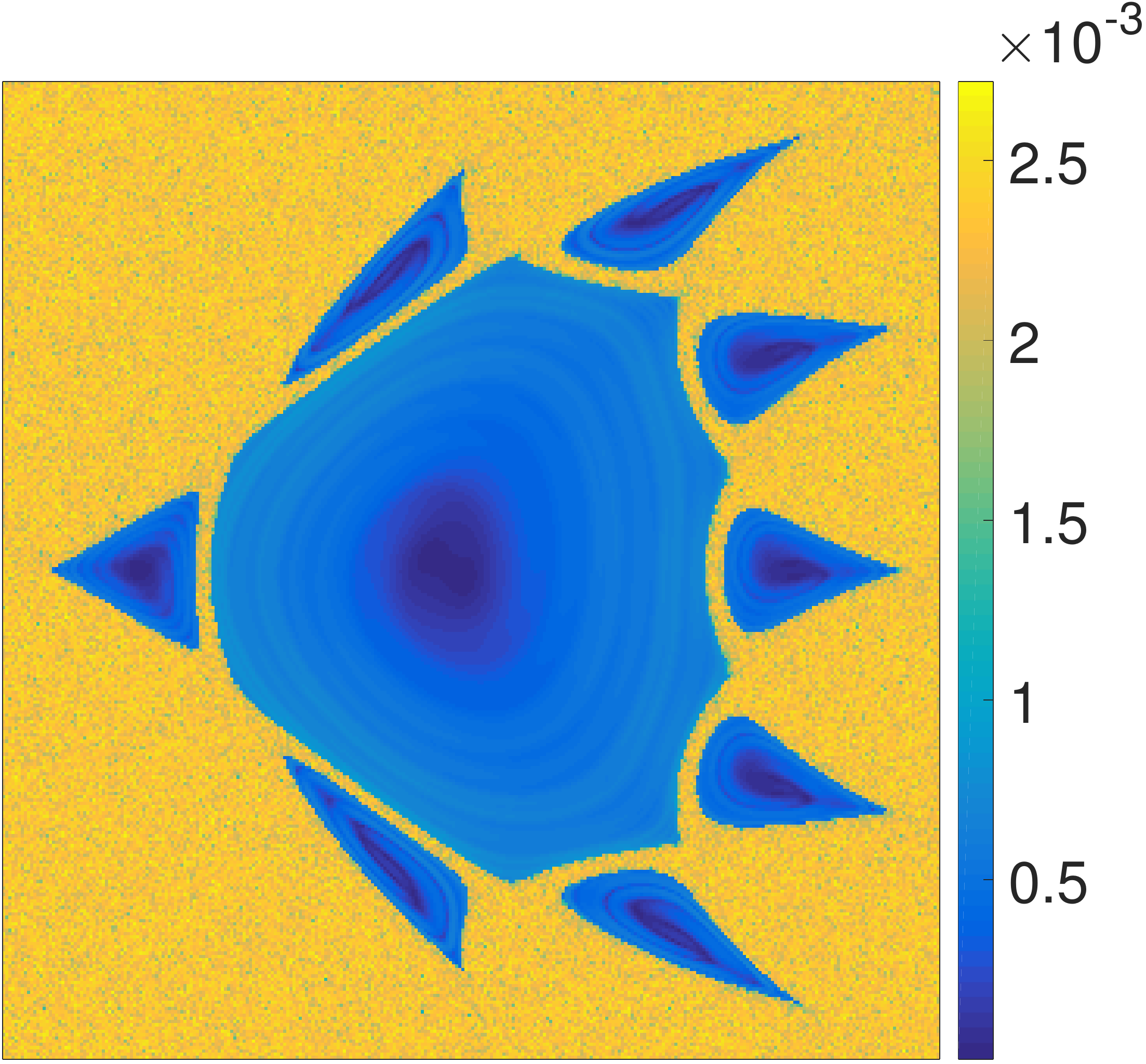}}\qquad
\subfloat[\label{fig:PendulumFSLE}]{\includegraphics[width=0.28\textwidth]{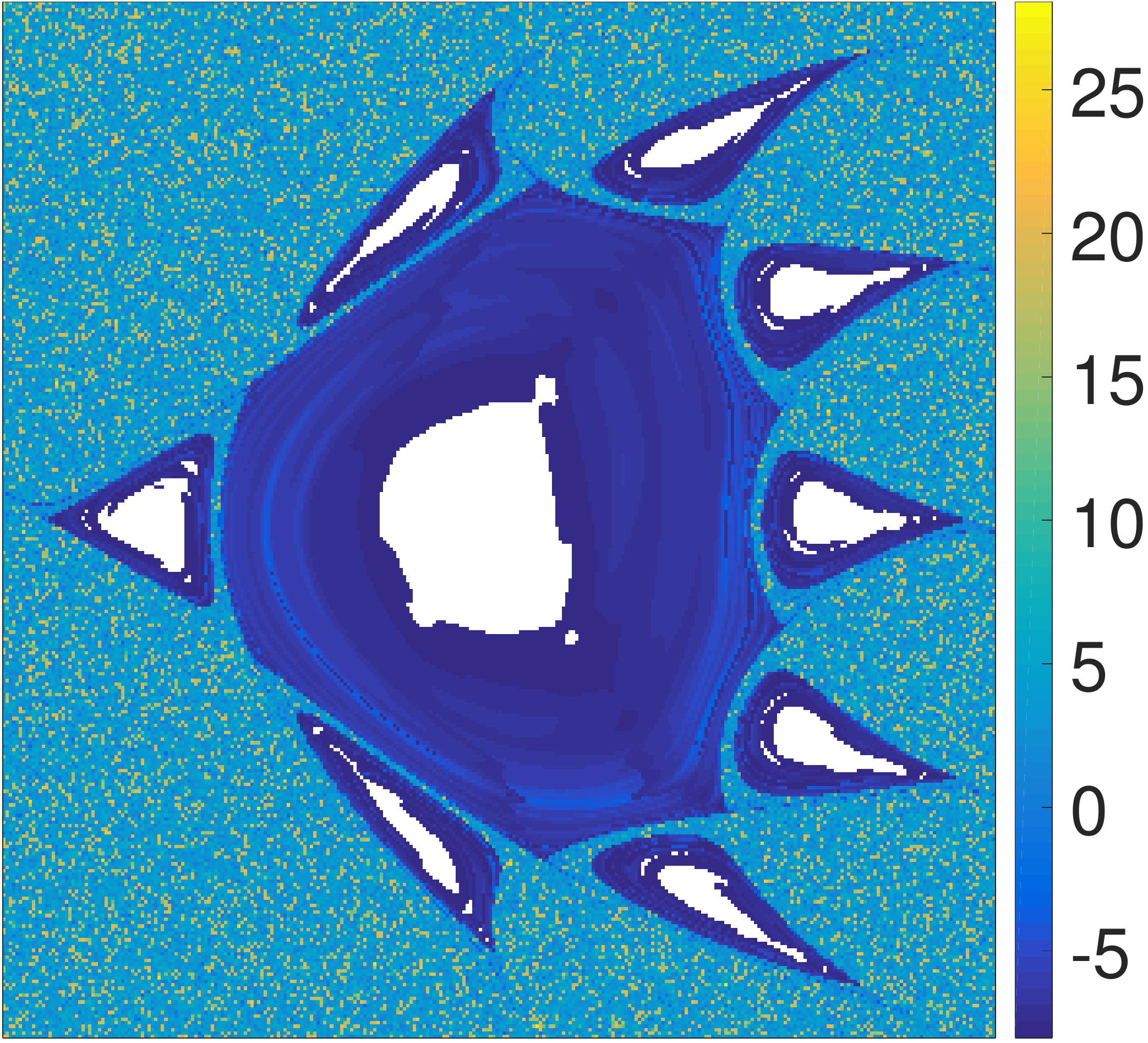}}
\caption{Comparison of three different diagnostic fields for the periodically forced pendulum. The scalar fields are constructed for the same integration time $T = 800\times 2\pi$. (a) Forward-time connectivity field. (b) Forward-time FTLE field. (c) Forward-time FSLE field.}
\label{fig:PendulumComparison}
\end{figure*}

\begin{figure*}
 \subfloat[\label{fig:PendulumEigenvalues}]{\includegraphics[height=0.26\textwidth]{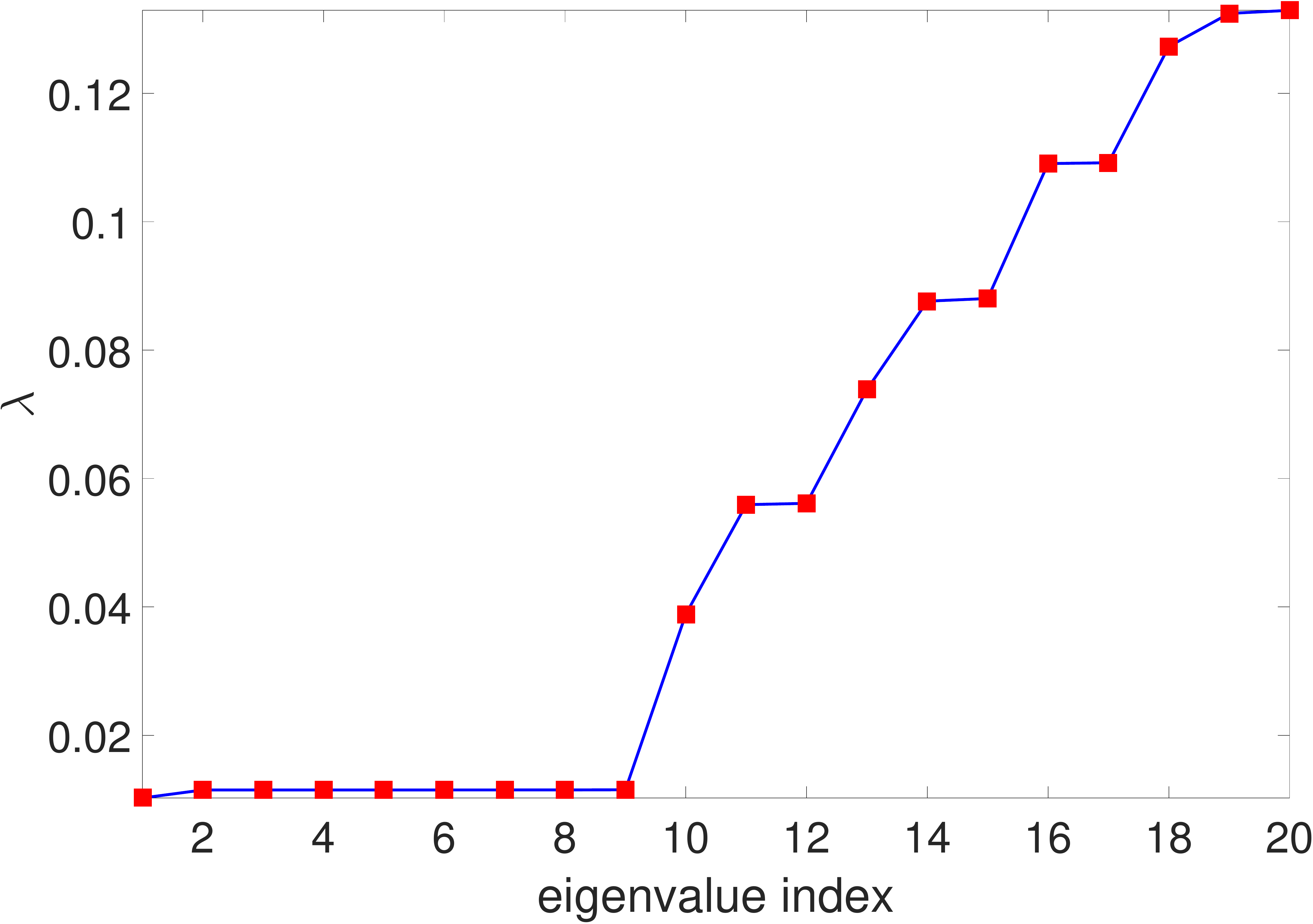}}\;
\subfloat[\label{fig:PendulumEig1}]{\includegraphics[height=0.26\textwidth]{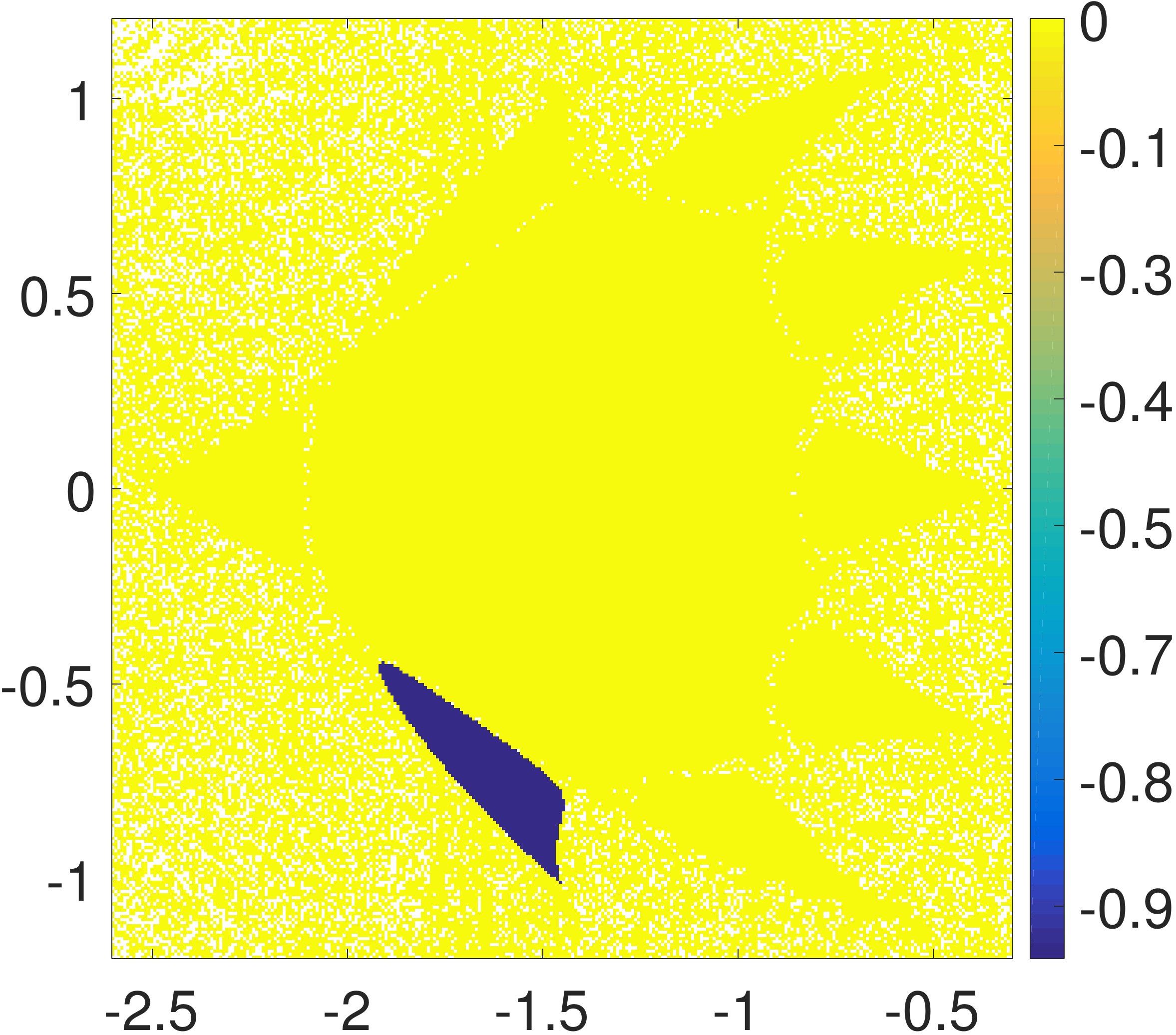}}\;
\subfloat[\label{fig:PendulumEig9}]{\includegraphics[height=0.26\textwidth]{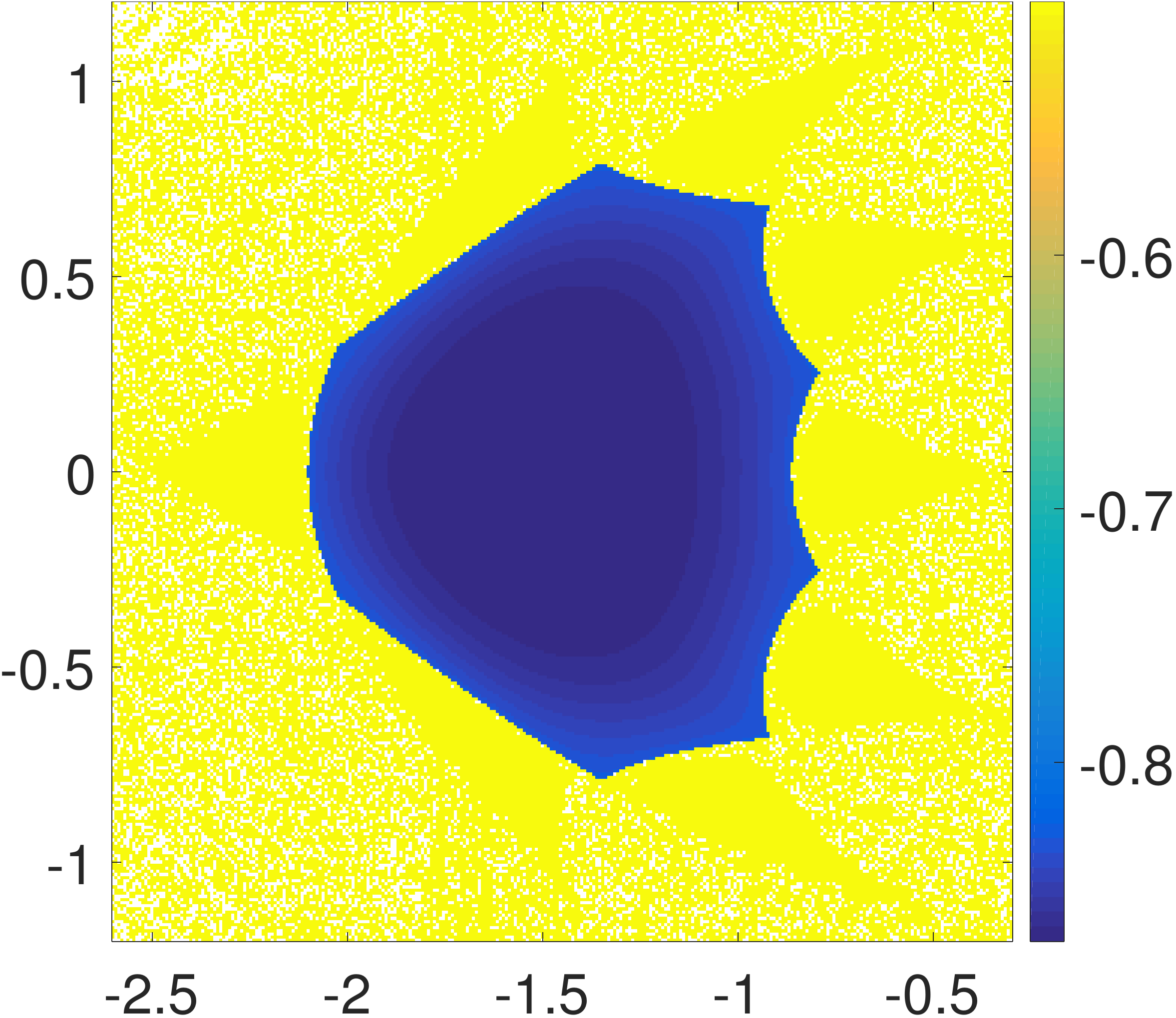}}
\caption{(a) Sorted generalized eigenvalues for graph Laplacian $L$ for the periodically forced pendulum.
(b-c) The first and ninth generalized eigenvectors of graph Laplacian $L$.
Isolated points resulting from the graph sparsification are shown in white.}
\label{fig:parabolic}
\end{figure*}

\begin{figure*}
\subfloat[\label{fig:PendulumClusters}]{\includegraphics[width=0.28\textwidth]{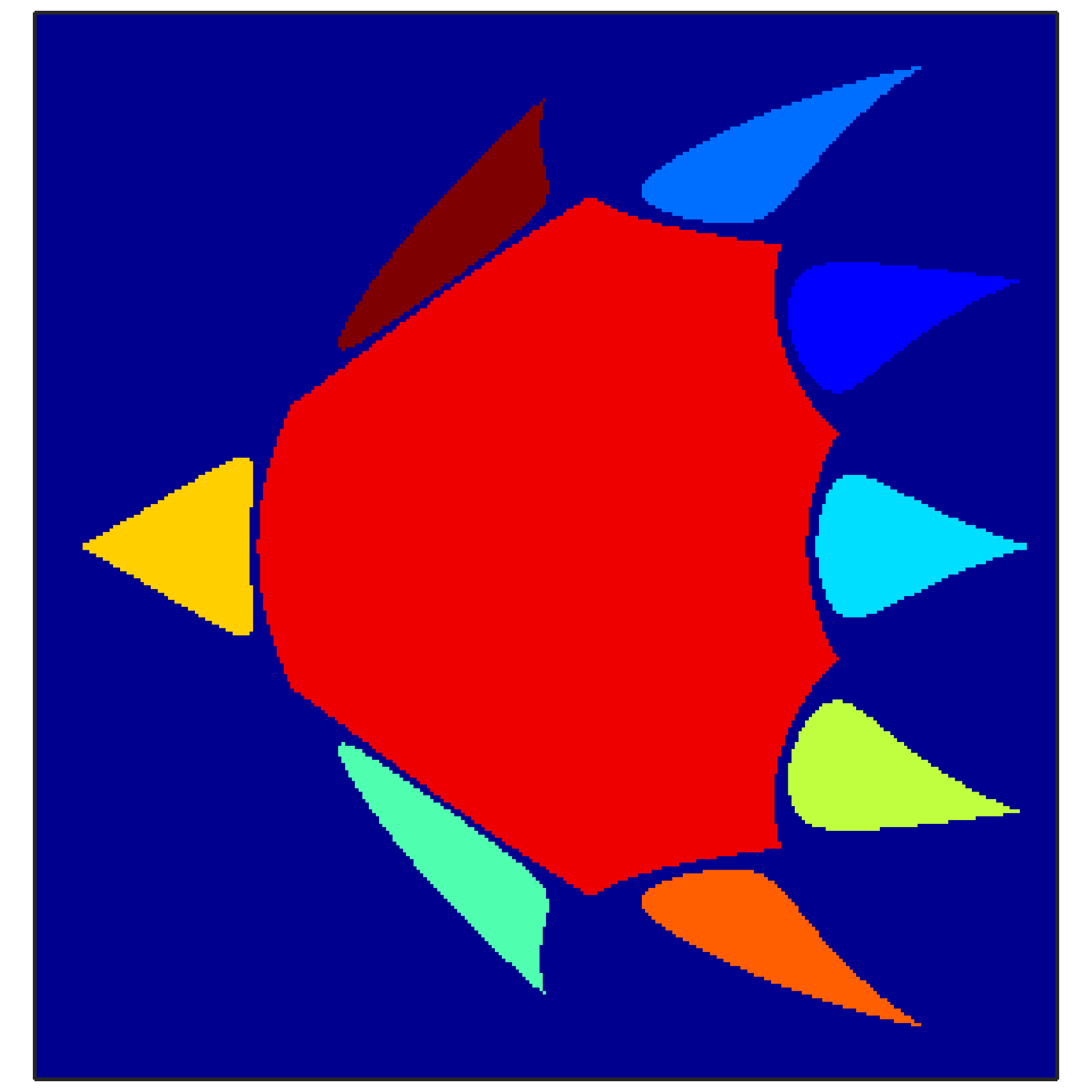}}\qquad
\subfloat[\label{fig:PendulumPmap}]{\includegraphics[width=0.28\textwidth]{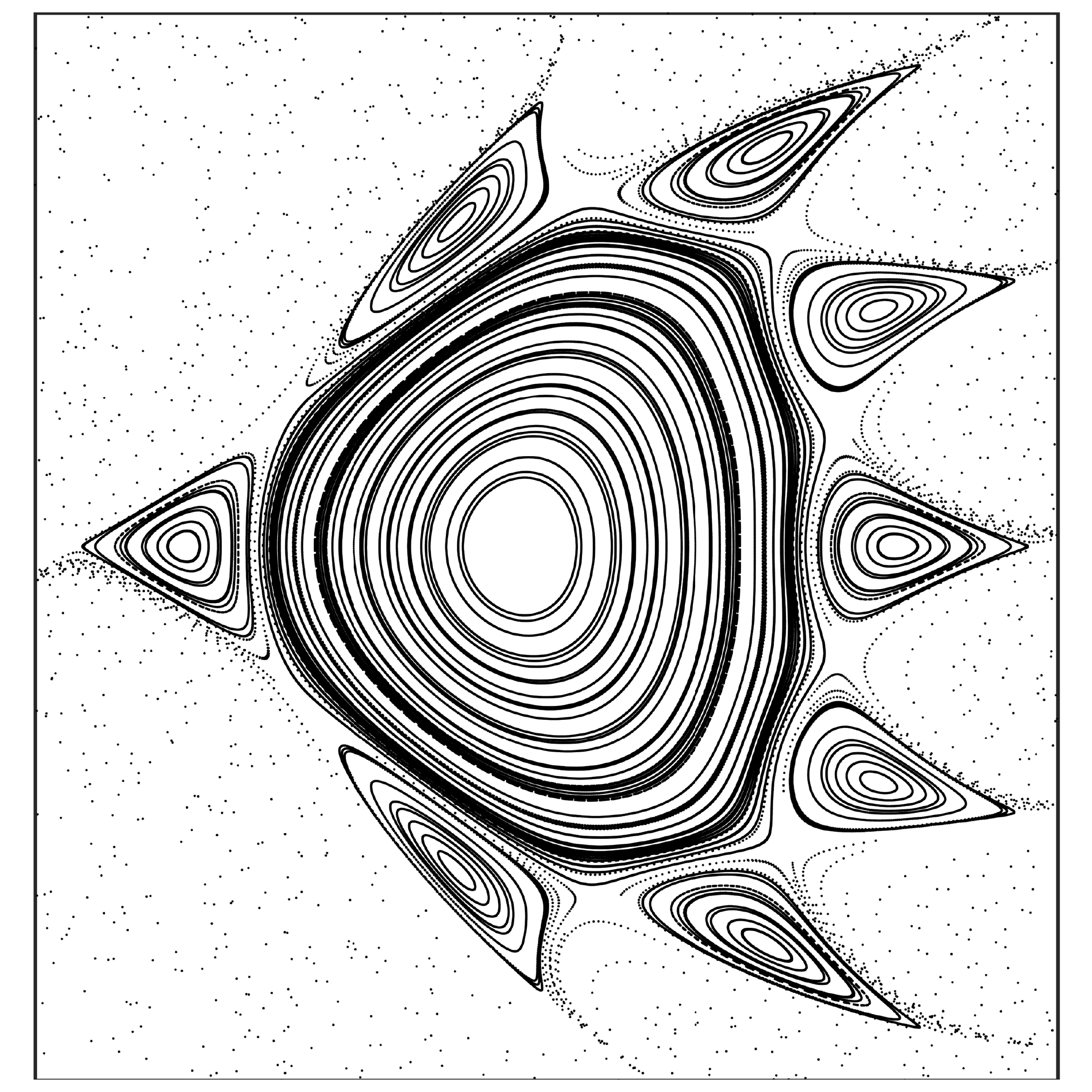}}\qquad
\subfloat[\label{fig:PendulumPmapClusters}]{\includegraphics[width=0.28\textwidth]{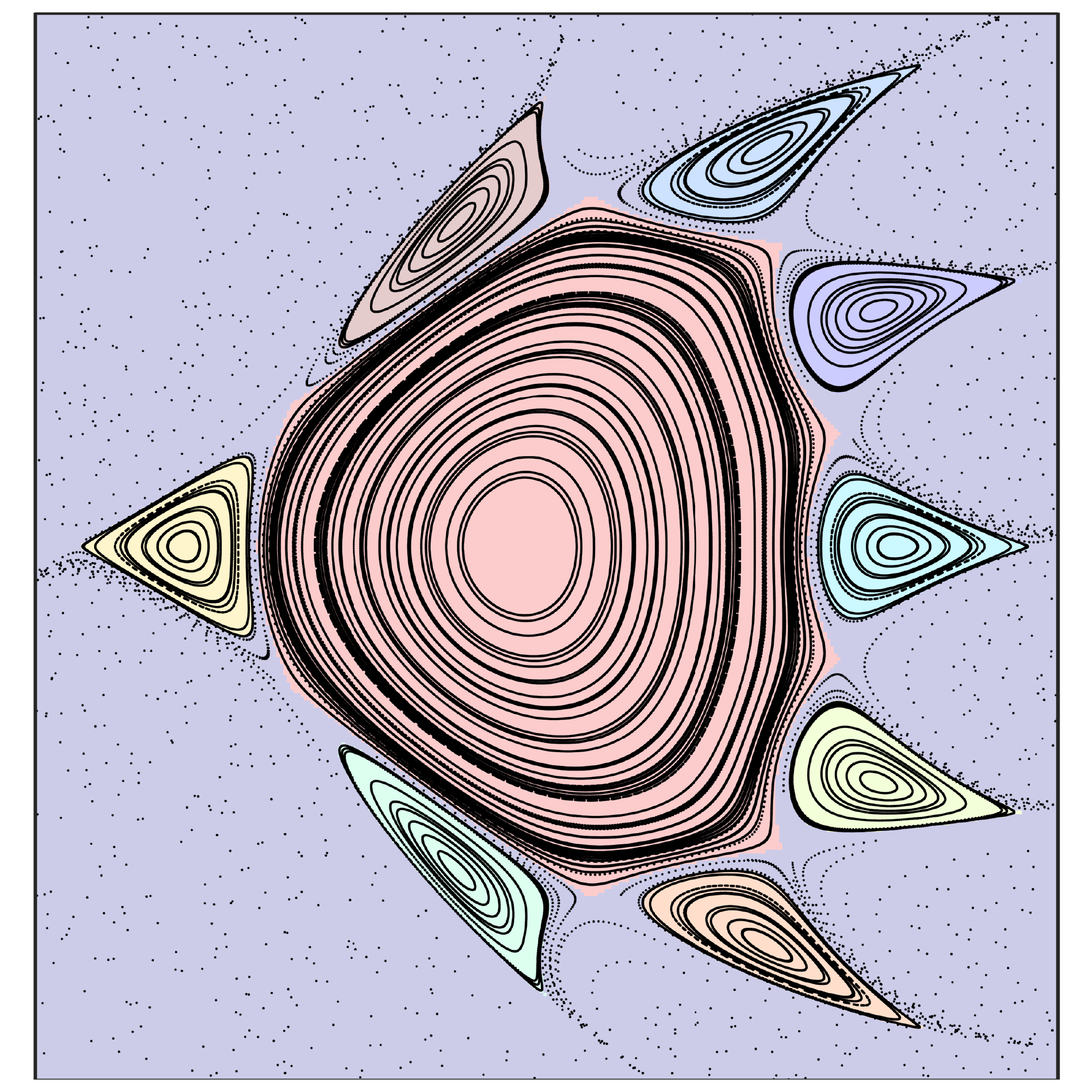}}
\caption{(a) Ten clusters extracted by K-means clustering 
from the first nine generalized eigenvectors of graph Laplacian $L$ for the periodically forced pendulum.
The tenth cluster corresponds to the chaotic sea filling the space
between elliptic regions. (b) $800$ iterations of the Poincar\'{e} map
for the periodically forced pendulum. (c) Computed clusters,
compared with the Poincar\'{e} map computed for the same integration
time (eight hundred iterates).}
\label{fig:pendulum}
\end{figure*}

\begin{figure*}

\subfloat[\label{fig:pendulumRunTime}]{\includegraphics[width=0.48\textwidth]{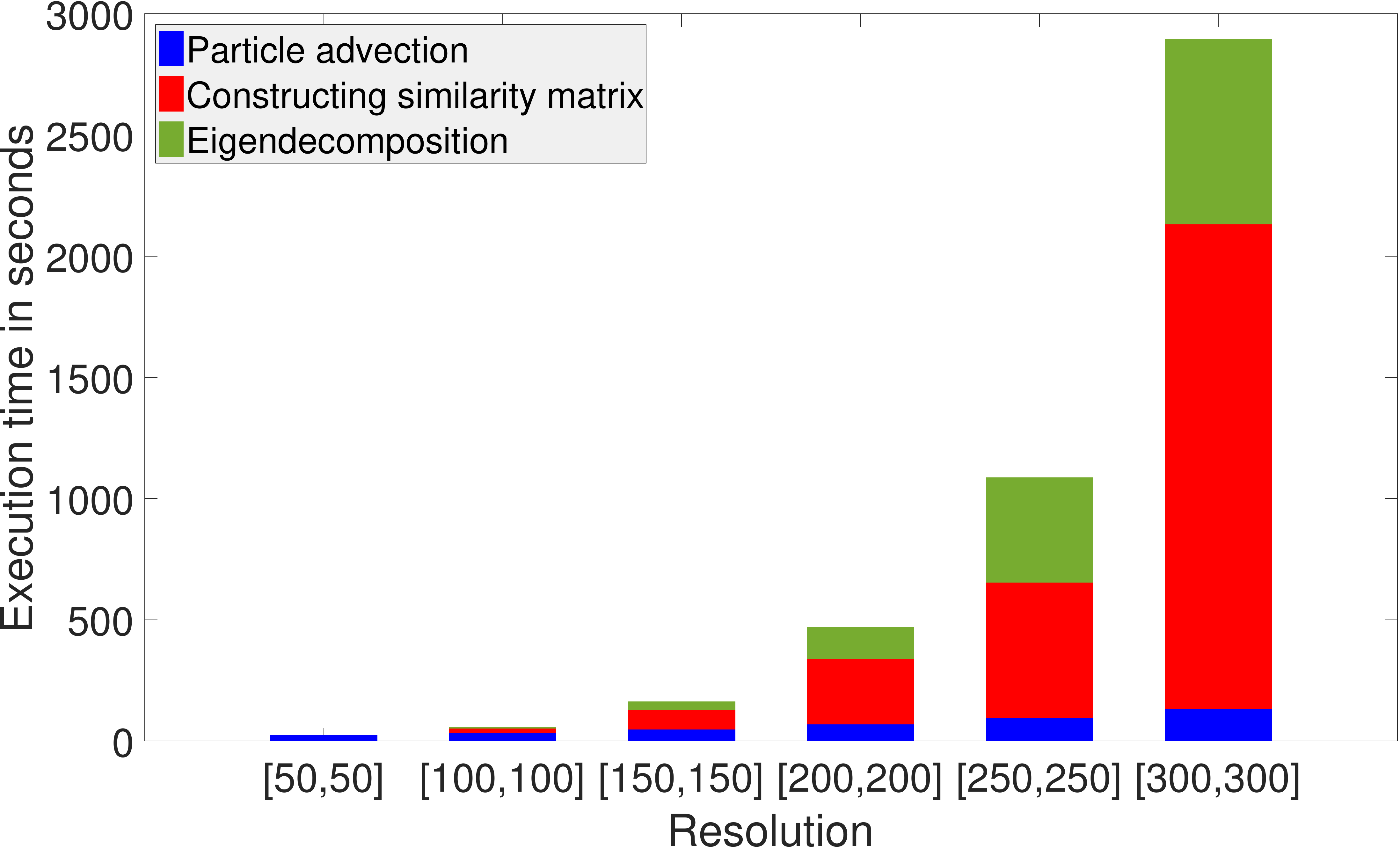}}\qquad
\subfloat[\label{fig:pendulumEpsilonSensitivity}]{\includegraphics[width=0.48\textwidth]{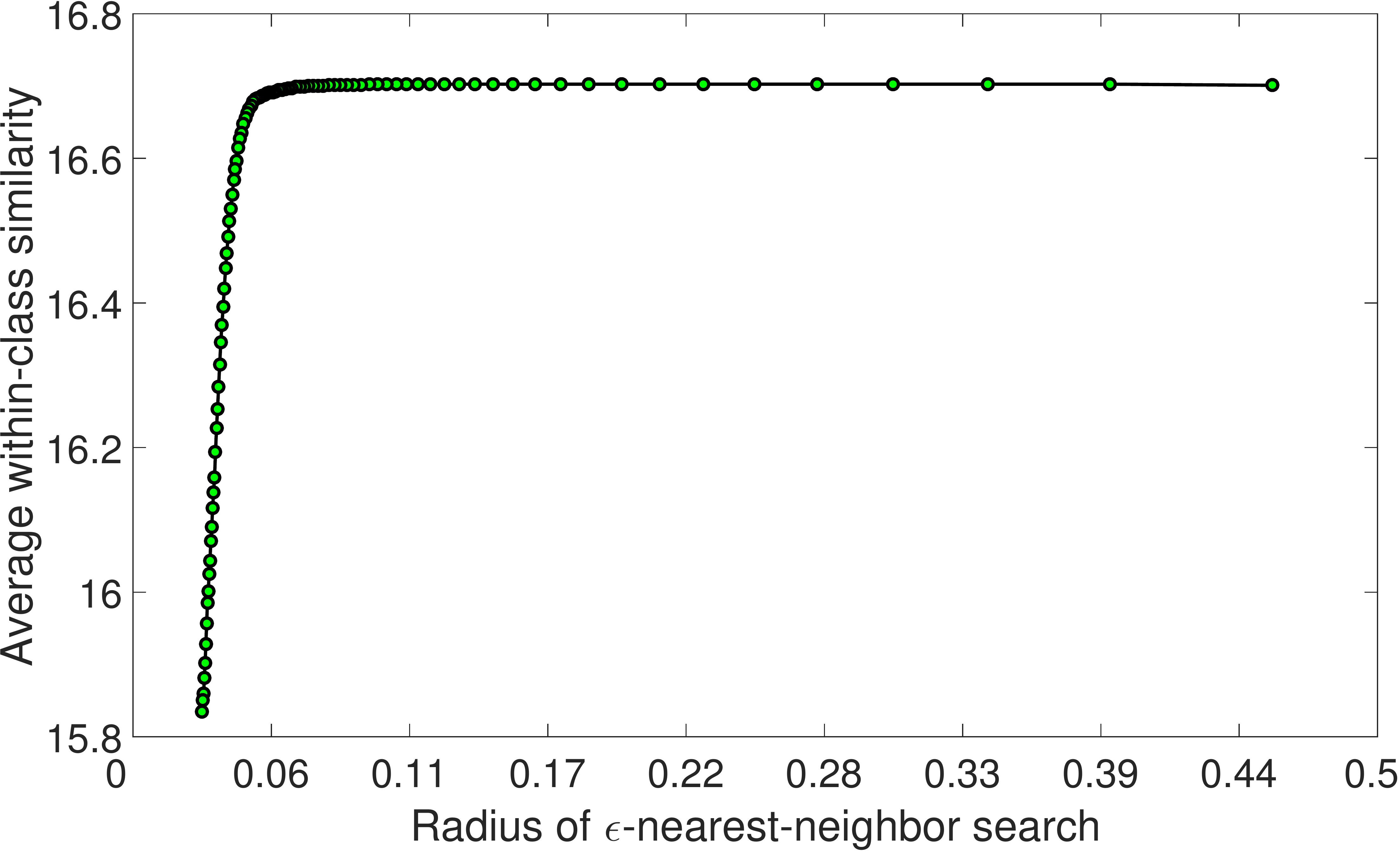}}
\caption{(a) The plot depicts the runtimes of \Cref{alg:algorithm1} for six different resolutions for the periodically forced pendulum. The runtimes represent the average CPU-times for 300 processors used in parallel in these computations. The computations are performed on a supercomputer with $2.7$GHz Intel Xeon CPUs. (b) Clustering sensitivity with respect to the sparsification radius. The plot shows the average within-class similarities (for nine coherent sets) relative to the $\epsilon$-nearest-neighbor radius used to sparsify the pairwise distances $r_{ij}$.}
\label{fig:pendulumComputation}
\end{figure*}
Consider the periodically forced pendulum
\[
\begin{aligned}\dot{x}_{1} & =x_{2}\\
\dot{x}_{2} & =-\sin(x_{1})+\varepsilon\cos(t).
\end{aligned}
\]
For $\varepsilon=0$, the system is integrable with hyperbolic fixed
points at $\left(0,\left(2m-1\right)\pi\right)$, and elliptic fixed
points at $\left(0,2m\pi\right)$, where $m\in\mathbb{Z}$. As is
well known, there are two heteroclinic orbits connecting each successive
pair of hyperbolic fixed points, enclosing an elliptic fixed point,
which is in turn surrounded by periodic orbits. These periodic orbits
appear as closed invariant curves for the Poincar\'{e} map $\mathcal{P}\coloneqq F_{0}^{2\pi}$.
The fixed points of the flow are also fixed points of $\mathcal{P}$.

Kolmogorov-Arnold-Moser (KAM) theory \cite{Arnold89} guarantees the survival
of most closed invariant sets for $\mathcal{P}$ and $0<\varepsilon\ll1$.
Increasing the perturbation strength $\varepsilon$
further leads to the appearance of resonance islands \cite{Arnold07,Birkhoff13} and
to the coexistence of regular and chaotic particle trajectories, as one
would expect in a turbulent fluid flow containing coherent structures.

\Cref{fig:PendulumPmap} shows these surviving invariant sets
(KAM tori and resonance islands) of the Poincar\'{e} map $\mathcal{P}$
obtained for $\varepsilon=0.4$, obtained from 800 iterations
of $\mathcal{P}$. This many iterations are required to obtain continuous-looking
boundaries of the various coherent regions. We would like to capture
the surviving KAM regions as coherent clusters using 
\Cref{alg:algorithm1}.

To construct the pairwise dynamic distances $r_{ij}$ and subsequent
similarity matrix $W$, we advect 90,000 particles, distributed initially
over a uniform grid $\mathcal{G}_{0}^{1}$ of $300\times300$ points,
from $t_{0}=0$ to $t_{1}=800\times2\pi$. The spatial domain ranges from $-2.6$ to $-0.3$ in $x_{1}$ direction and from $-1.2$ to
$1.2$ in $x_{2}$ direction. We output the
trajectory data with $3600$ intermediate points, evenly spaced in time. Moreover, we sparsify edges from the complete graph representing a distance greater than $\epsilon = 0.45$.

\Cref{fig:PendulumDegree} shows the degree of connectivity of graph
nodes, $\deg(v_{i})$, as a scalar field. We refer to this scalar field
here and in our later examples as \textit{connectivity field}. This
field looks generally smoother than other diagnostic fields, such
as the finite-time Lyapunov exponent \cite{Haller00,Haller01} or
finite-size Lyapunov exponent \cite{Aurell97,Artale97} fields (see
\cref{fig:PendulumComparison}). The smoothness of the connectivity
field is the result of two averaging processes which attenuate computational
and in-situ measurement noises. The first averaging process happens
as we integrate Euclidean distances between graph nodes over time.
The second averaging takes place once we compute $d_{i}$, i.e., when summing
the edge weights connected to a node $v_{i}$.

\Cref{fig:PendulumEigenvalues} shows the first 20 generalized eigenvalues
as a function of their indices. We can see that the first nine eigenvalues
are very close to 1, while the tenth has an appreciable difference,
creating the largest gap in the eigenvalue plot. This eigengap implies
that the first nine eigenvectors are cluster indicators from which
coherent structures should be extracted. For example,  \cref{fig:PendulumEig1,fig:PendulumEig9}
show the first and ninth generalized eigenvector of the graph Laplacian
$L$.

Finally, \cref{fig:PendulumClusters} shows the ten clusters extracted by
the K-means algorithm from the first nine generalized
eigenvectors of graph Laplacian $L$. The tenth cluster corresponds to the
chaotic background filling the space between the coherent clusters.
In \cref{fig:PendulumPmapClusters}, the extracted clusters are
superimposed on the Poincar\'{e} map, showing close agreement with the
Lagrangian vortices of this example, i.e., the elliptic islands.

\Cref{fig:pendulumRunTime} shows the execution times for three major steps of \Cref{alg:algorithm1} as a function of increasing spatial resolution of the graph nodes. The main computational bottleneck, as shown in the figure, is computing the pairwise distances and subsequently the similarity matrix W. For this purpose, we utilized parallel computing techniques with 300 CPUs, with each processor just computing a few rows/columns of the sparse similarity matrix. \Cref{fig:pendulumRunTime} shows the averaged CPU-times spent on each processor on carrying out the particle advection, sparse similarity matrix construction and eigen-decomposition.

Figure \ref{fig:pendulumEpsilonSensitivity} shows the sensitivity of the clustering results to the choice of the neighborhood radius used to sparsify the pairwise distances $r_{ij}$. In particular, the figure shows how the averaged within-class similarities of coherent sets change with respect to the choice of neighborhood radius. Figure \ref{fig:pendulumEpsilonSensitivity} suggests the existence of a critical radius below which the size and shape of clusters can change. This critical radius simply corresponds to a distance where even strong edges within coherent sets are affected by graph sparsification. It is important to choose the sparsification radius such that strong edges will be maintained. As a rule of thumb, we  set the sparsification radius such that only $5\%$-$10\%$ of the elements in the similarity matrix $W$ will be kept. To estimate such a radius, one can compute the pairwise distances for a subsample of the original graph (e.g., 40 nodes) and choose the sparsification radius accordingly.
\subsection{Quasiperiodic Bickley jet}\label{Ex:Bickley}
\begin{figure}
\includegraphics[height=0.3\textwidth]{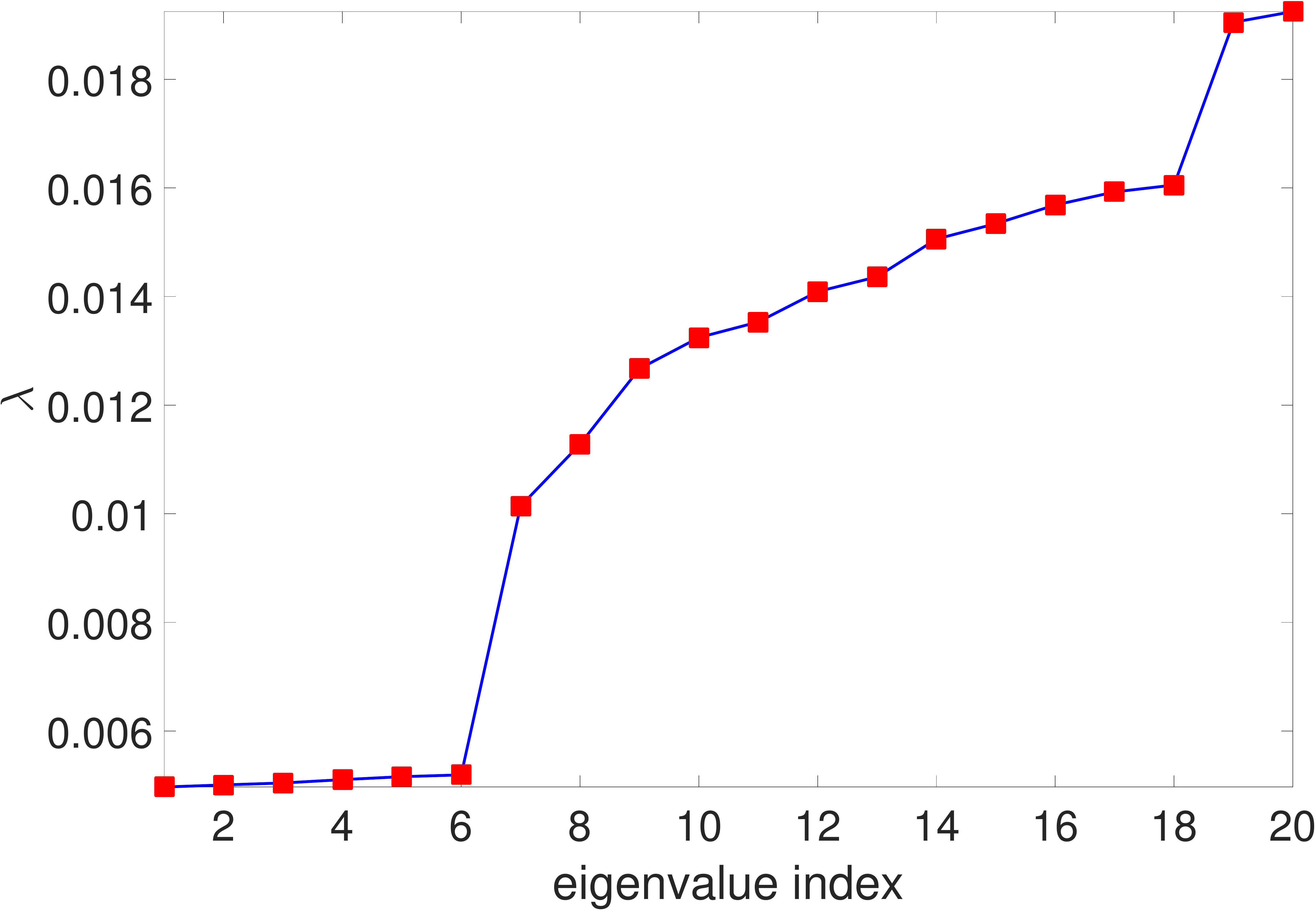}
\caption{Sorted generalized eigenvalues for the graph Laplacian $L$ for the quasiperiodic Bickley jet flow.}
\label{fig:BickleyEigenvalues}
\end{figure}

\begin{figure*}

\subfloat[\label{fig:BickleyEig1}]{\includegraphics[width=0.45\textwidth]{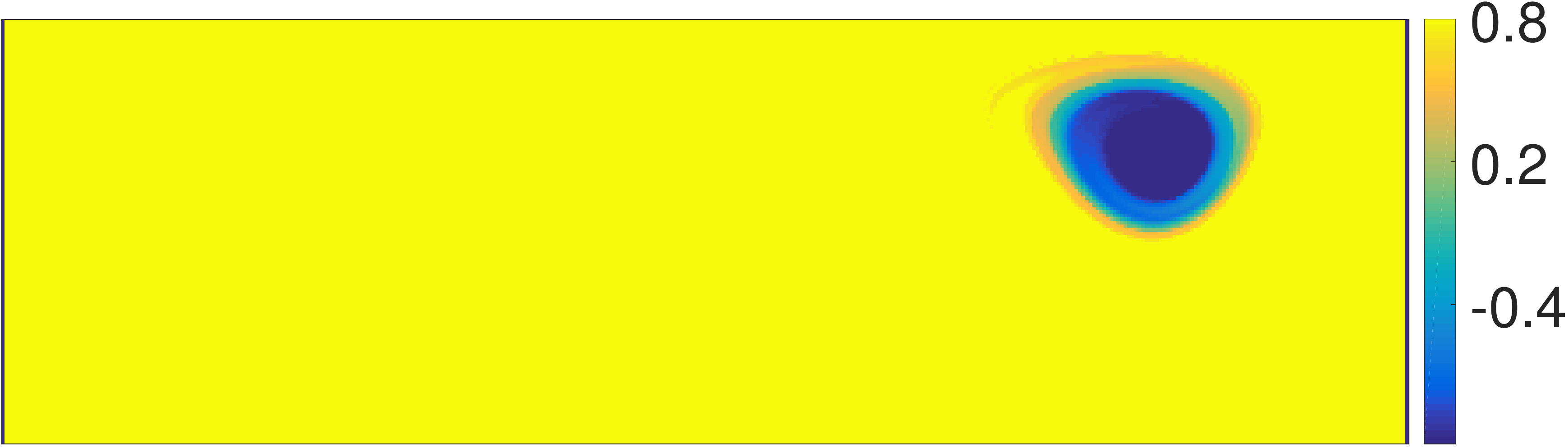}}\quad
\subfloat[\label{fig:BickleyEig2}]{\includegraphics[width=0.45\textwidth]{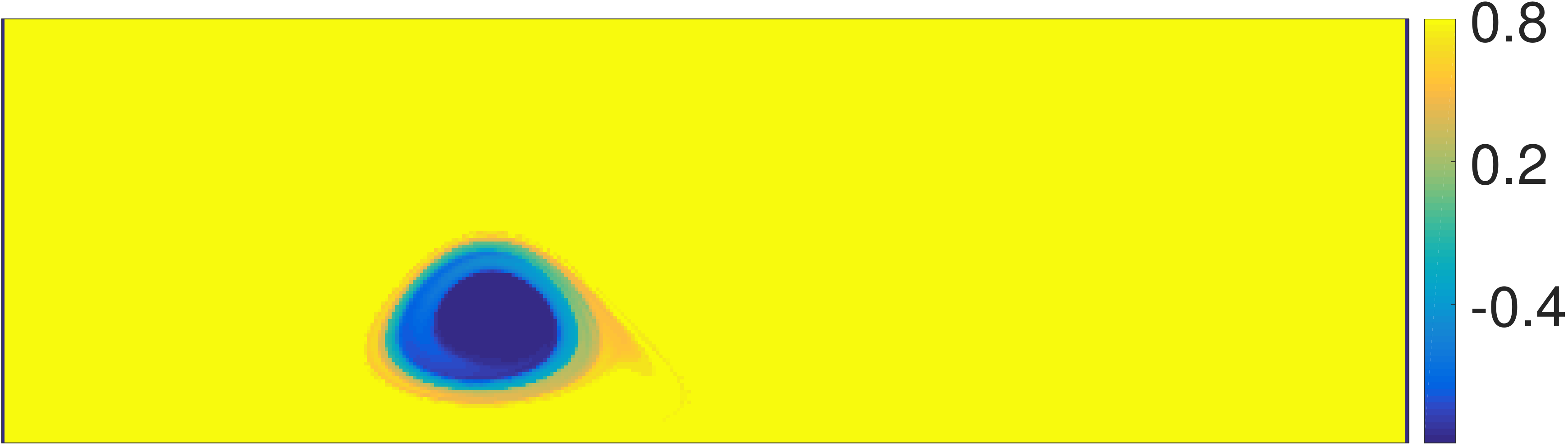}}\\
\subfloat[\label{fig:BickleyEig3}]{\includegraphics[width=0.45\textwidth]{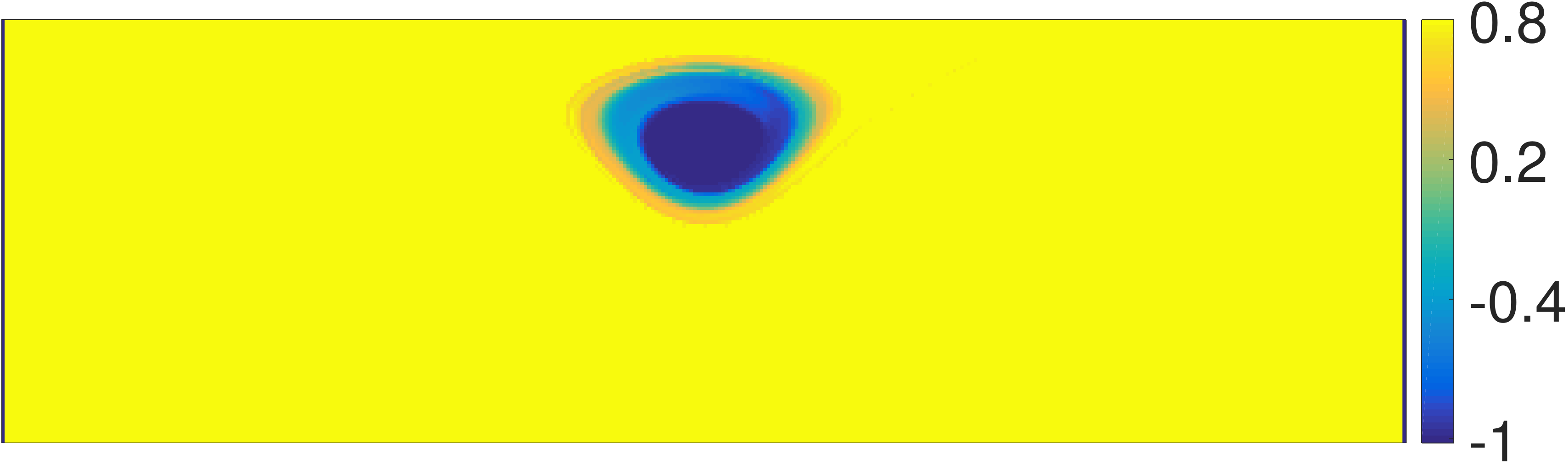}}\quad
\subfloat[\label{fig:BickleyEig4}]{\includegraphics[width=0.45\textwidth]{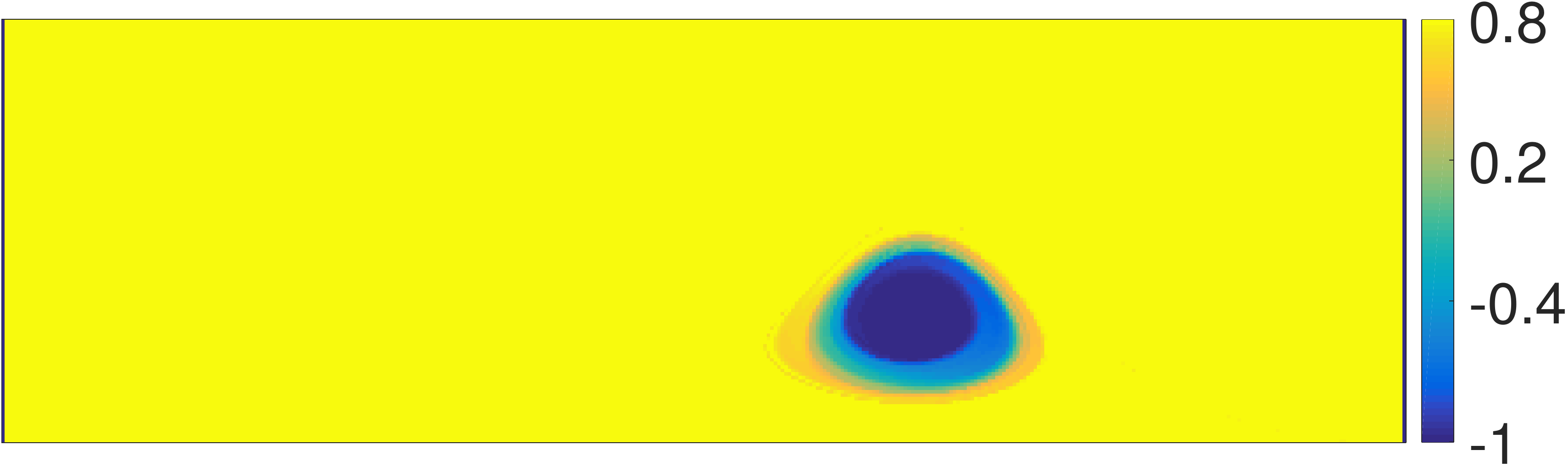}}\\
\subfloat[\label{fig:BickleyEig5}]{\includegraphics[width=0.45\textwidth]{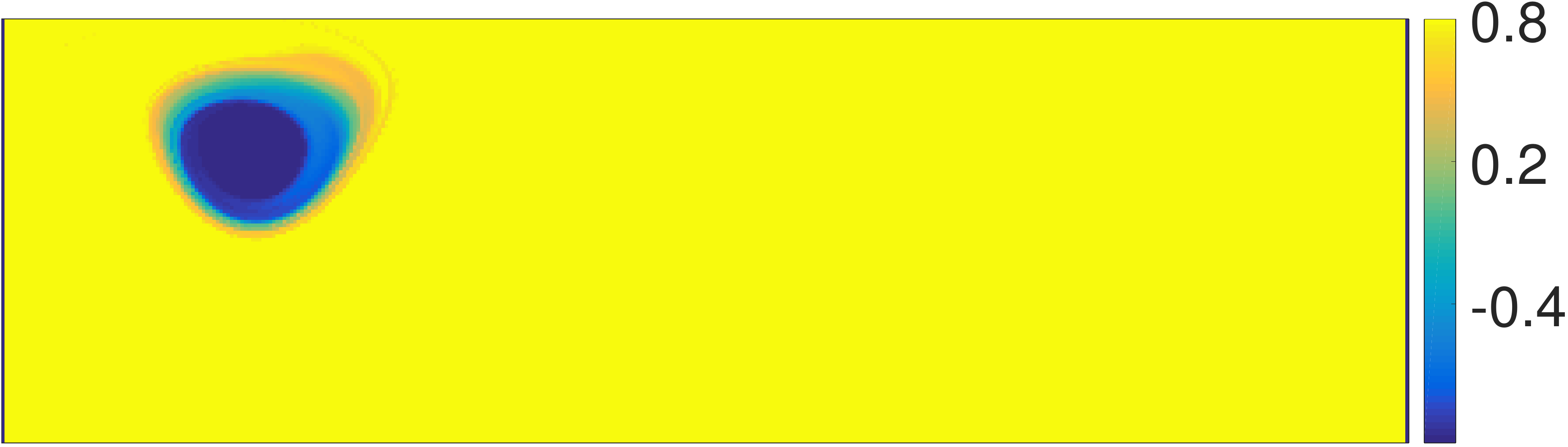}}\quad
\subfloat[\label{fig:BickleyEig6}]{\includegraphics[width=0.45\textwidth]{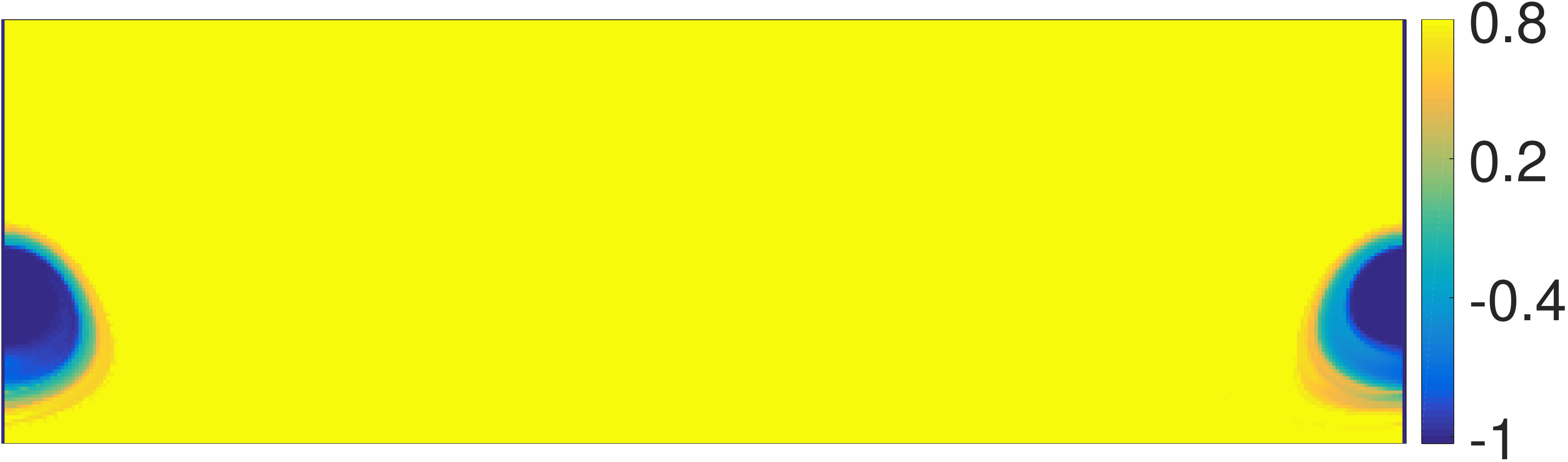}}\\
\caption{The leading generalized eigenvectors of the graph Laplacian $L$ for the Bickley jet flow.}
\label{fig:Bickley_eigenvectors}
\end{figure*}

\begin{figure*}
\subfloat[\label{fig:BickleyClusterst0}]{\includegraphics[width=0.45\textwidth]{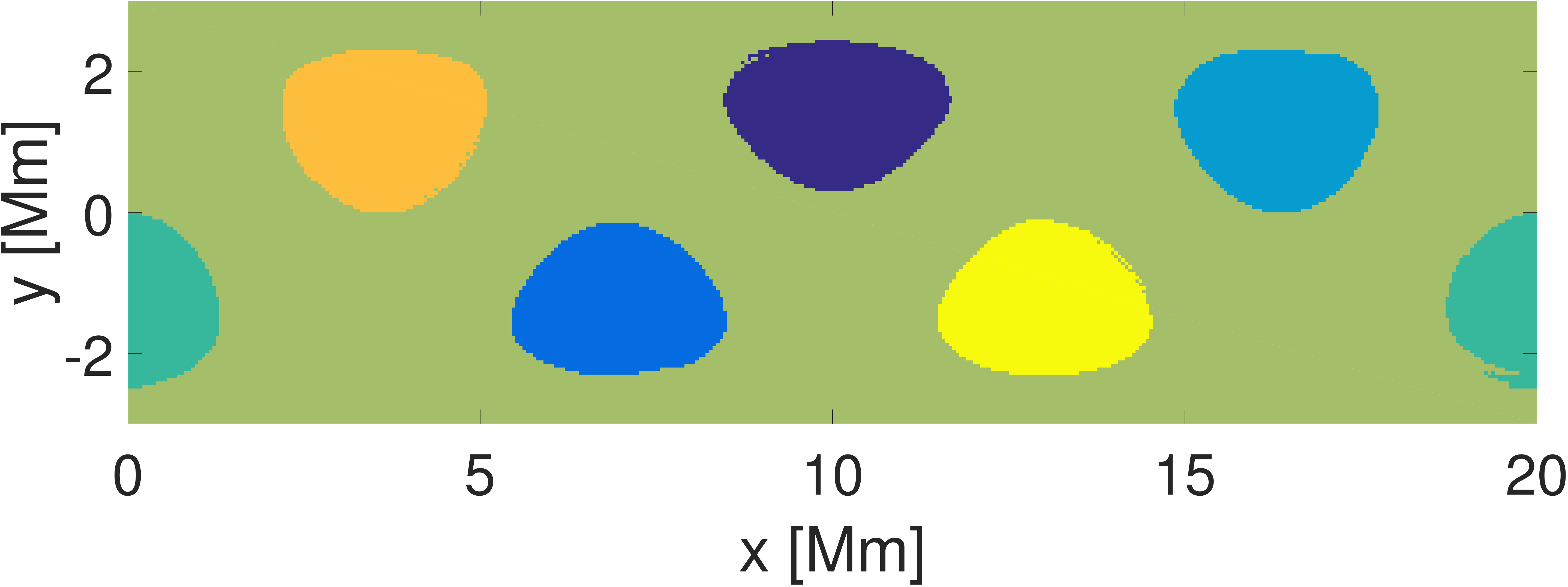}}\qquad
\subfloat[\label{fig:BickleyClustersT}]{\includegraphics[width=0.45\textwidth]{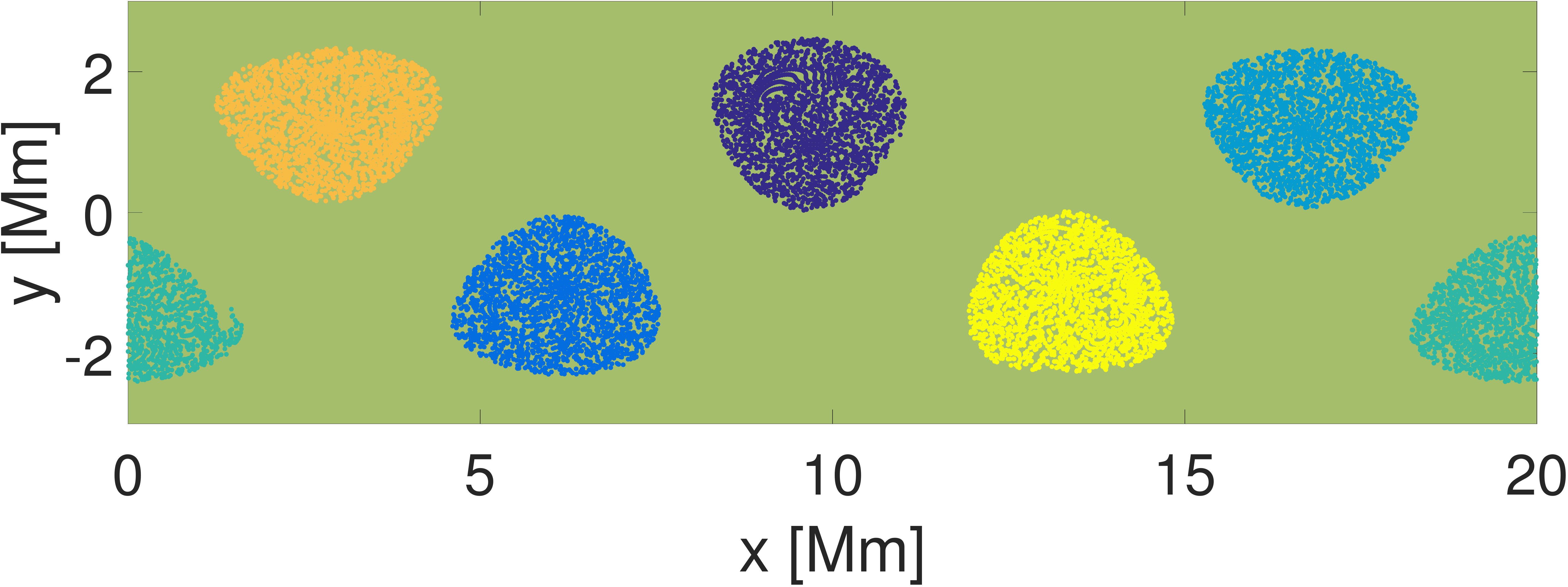}}
\caption{(a) Seven clusters extracted by K-means clustering from the first
six generalized eigenvectors of graph Laplacian $L$ at initial time, $t_{0}=0$. The seventh cluster corresponds to the mixing region filling the space between the coherent clusters. (b) The same clusters advected passively to the final time, $t = 40$ day.  The complete advection sequence over 40 days is illustrated in the online supplemental movie M1 \cite{url_1}.}
\label{fig:BickleyClusters}
\end{figure*}
Next, we consider the Bickley jet, an idealized model
of a meandering zonal jet flanked above and below by counter rotating
vortices \cite{Del_Castillo93,Rypina07}. This model consists of a
steady background flow subject to a time-dependent perturbation. The
time-dependent Hamiltonian for this model reads as
\begin{align*}
\psi(x,y,t) & =\psi_{0}(y)+\psi_{1}(x,y,t),\\
\psi_{0}(y) & = -U_{0}L_{0}\tanh(\frac{y}{L_{0}}),\\
\psi_{1}(x,y,t) & = U_{0}L_{0}\sech^{2}(\frac{y}{L_{0}})\Re\left[\sum_{n=1}^{3}f_{n}(t)\exp(ik_{n}x)\right],
\end{align*}
where $\psi_{0}$ is the steady background flow and $\psi_{1}$ is
the perturbation. The constants $U_{0}$ and $L_{0}$ are characteristic
velocity and characteristic length scale, respectively. For the following
analysis, we apply the set of parameters used in \cite{Rypina07}:
\[
U_{0} = 62.66\;\text{ms}^{-1},\; L_{0} = 1770\;\text{km},\;k_{n}=2n/r_{0},
\]
where $r_{0}=6371$ km is the mean radius of the earth.

For $f_{n}(t)=\varepsilon_{n}\exp(-ik_{n}c_{n}t)$, the time-dependent
part of the Hamiltonian consists of three Rossby waves with wave numbers
$k_{n}$ traveling at speeds $c_{n}$. The amplitude of each Rossby
wave is determined by the parameters $\varepsilon_{n}$. Specifically,
the parameter values used are: $c_1=0.1446U_0$, $c_{2}=0.205U_{0}$, $c_{3}=0.461U_{0}$, $l_{y}=1.77\times10^{6}$, $\varepsilon_{1}=0.0075$, $\varepsilon_{2}=0.15$, $\varepsilon_{3}=0.3$,
$l_{x}=6.371\times10^{6}\pi$, $k_{n}=2n\pi/l_{x}$.

To construct the dynamic distances $r_{ij}$ and the similarity
matrix $W$, we advect 48000 particles, distributed initially over
a uniform grid of $400\times120$ points, from
$t_{0}=0$ to $t=40$ days. The spatial domain $U$ ranges from $0$
to $20$ in $x$ direction and from $-3$ to $3$ in $y$ direction. We output the trajectory data with $600$ intermediate points, evenly spaced in time. Moreover, we sparsify edges from the complete graph representing a distance greater than $\epsilon = 3$.

In \cref{fig:BickleyEigenvalues}, we show the first 20 generalized
eigenvalues of the graph Laplacian $L$ with respect to their indices.
We can observe that the largest eigengap is between the sixth
and seventh generalized eigenvalues, signaling the presence of six
coherent clusters in the domain. Hence, we extract seven clusters
from the first six generalized eigenvectors shown in \cref{fig:BickleyEig1,fig:BickleyEig2,fig:BickleyEig3,fig:BickleyEig4,fig:BickleyEig5,fig:BickleyEig6}). The last cluster, as described earlier in \Cref{section:eigengap}, corresponds
to the incoherent region filling the space between the coherent vortices. The 
observed fuzziness of the vortex boundary region is due to the fact that
coherent and incoherent motion is--on the chosen time interval--not as distinguished as in the forced 
pendulum example considered in the previous section. After all, this 
distinction is retrieved from the trajectory data, as opposed to being
imposed externally through some threshold, for instance. Interestingly,
this dynamic distinction is very clear in the ocean example considered 
in the next section, which results in very pronounced cluster indicators.

\Cref{fig:BickleyClusterst0} shows the identified clusters at the
initial time, and \cref{fig:BickleyClustersT} shows them at the final
time, confirming the coherence of extracted vortices over the 40-day period.
The complete advection sequence over 40 days is available
in the online supplemental movie M1 \cite{url_1}.
\subsection{An ocean surface data set}\label{Ex:OceanR2}
\begin{figure*}
 \subfloat[\label{fig:OceanR2Degree}]{\includegraphics[height=0.3\textwidth]{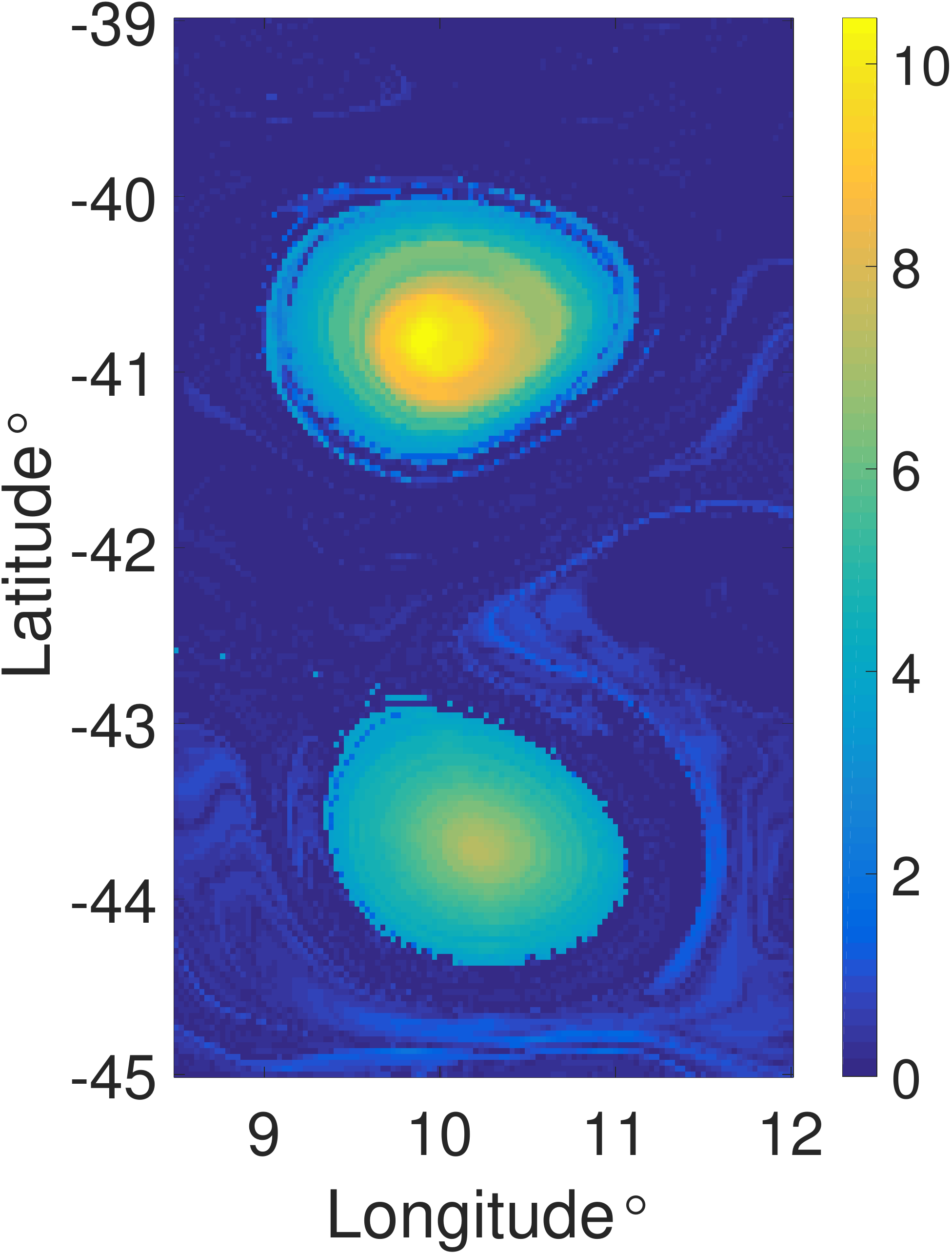}}\quad
\subfloat[\label{fig:OceanR2FTLE}]{\includegraphics[height=0.3\textwidth]{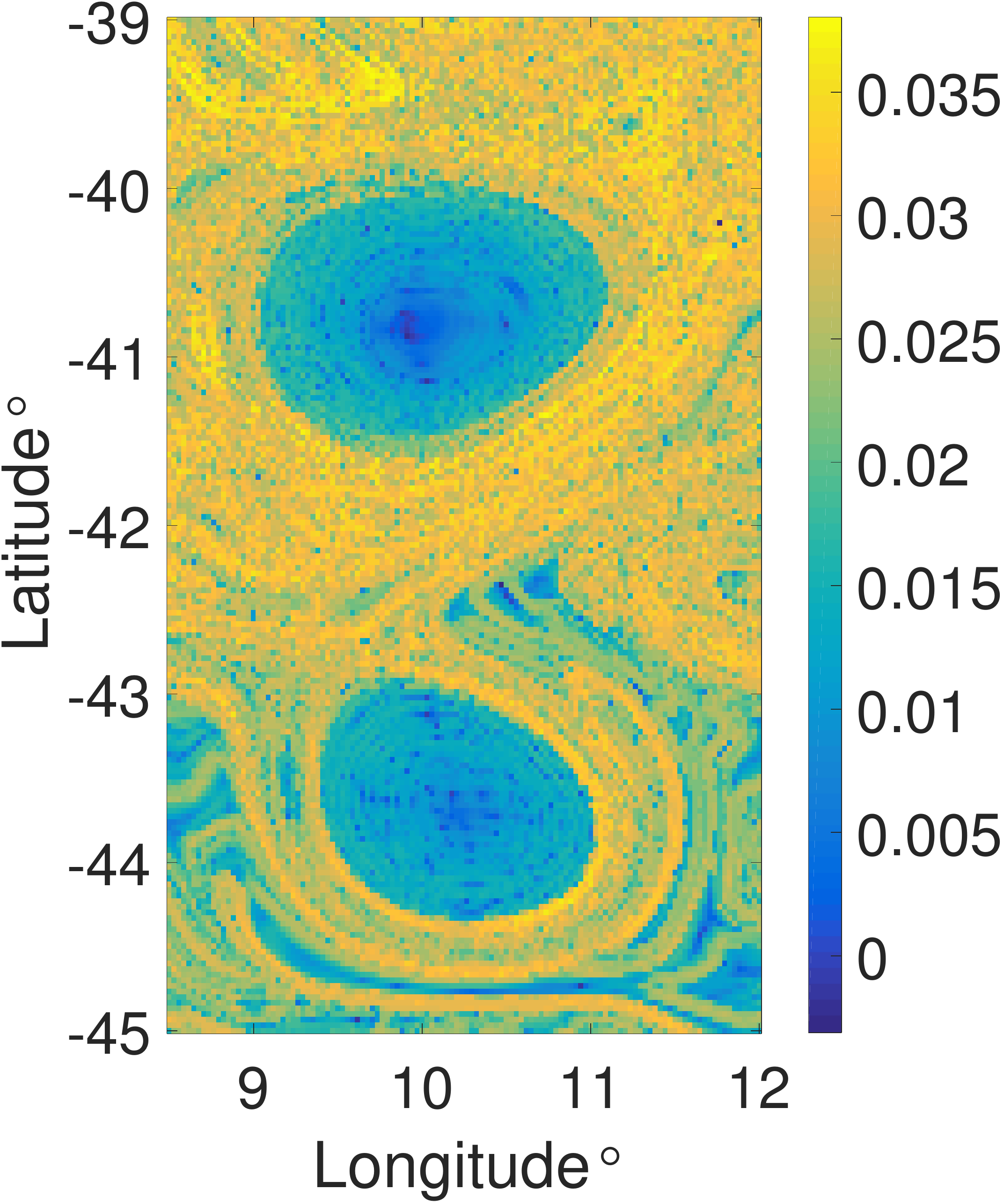}}\quad
\subfloat[\label{fig:OceanR2FSLE}]{\includegraphics[height=0.3\textwidth]{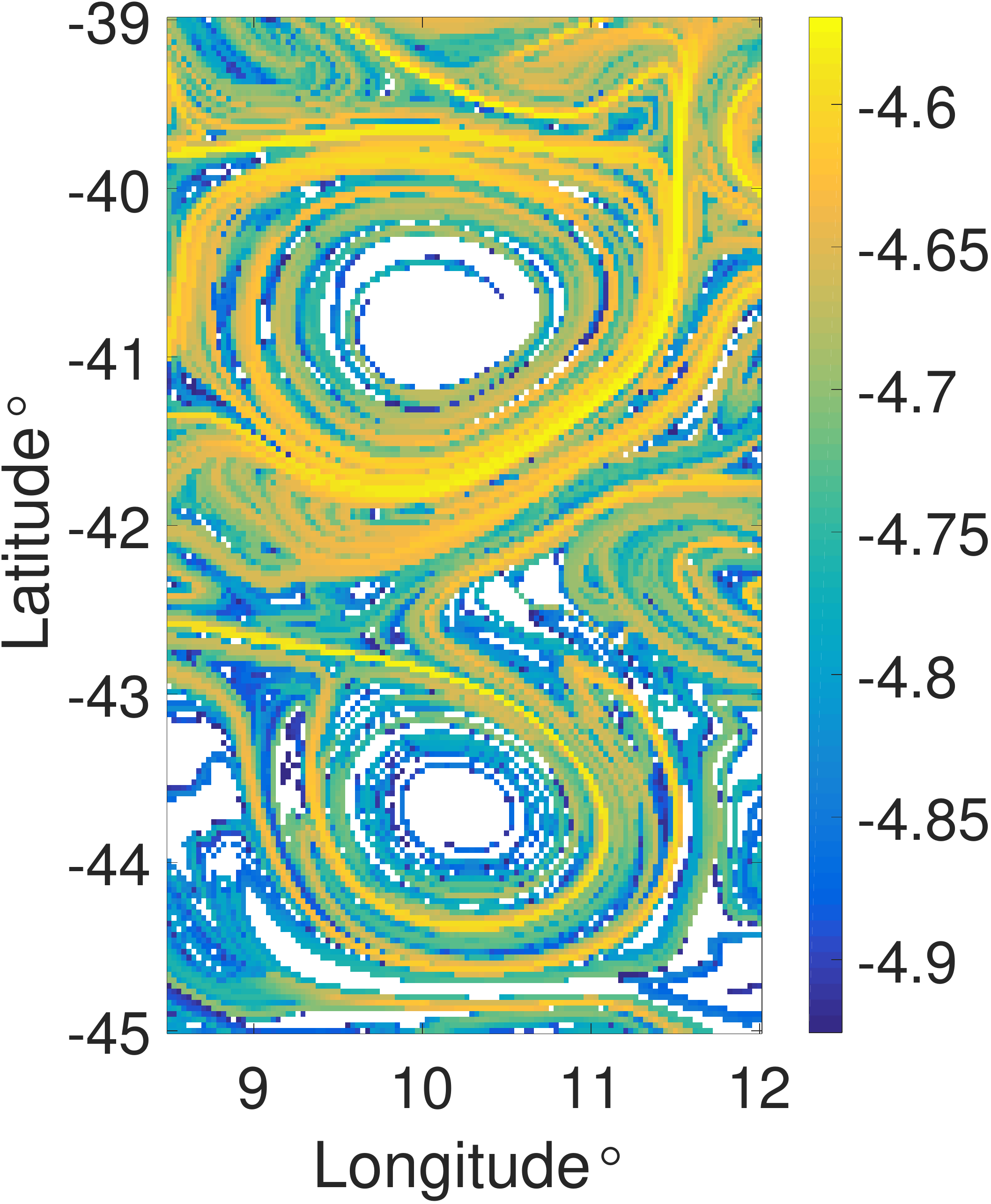}}
\caption{Comparison of three different diagnostic fields for the ocean data set. The scalar fields are constructed for the same integration time $T = 168$ days. (a) Forward-time connectivity field. (b) Forward-time FTLE field. (c) Forward-time FSLE field.}
\label{fig:OceanR2Comparison}
\end{figure*}

\begin{figure*}

\subfloat[\label{fig:OceanR2Eigenvalues}]{\includegraphics[height=0.3\textwidth]{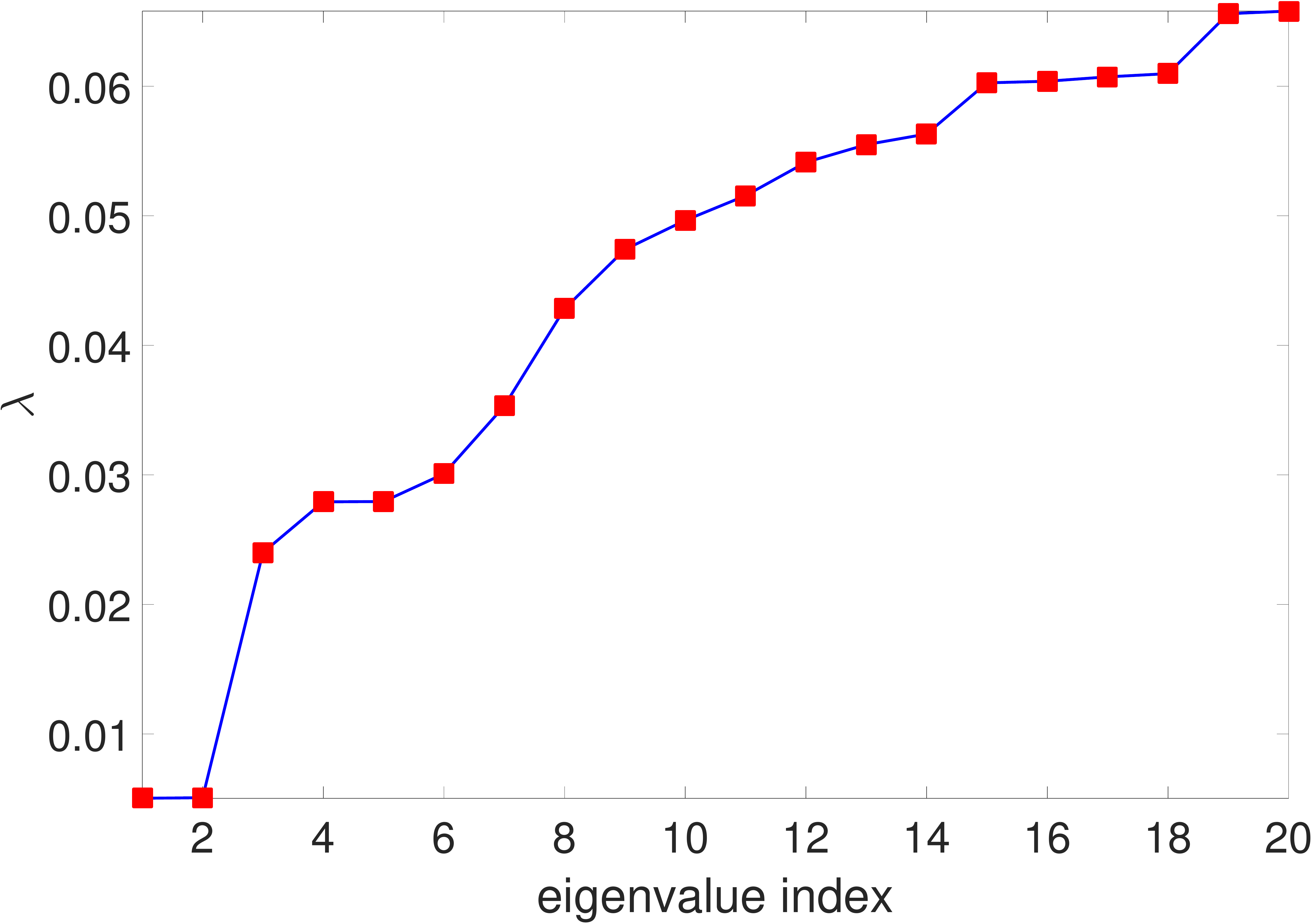}}\qquad
\subfloat[\label{fig:OceanR2Eig1}]{\includegraphics[height=0.3\textwidth]{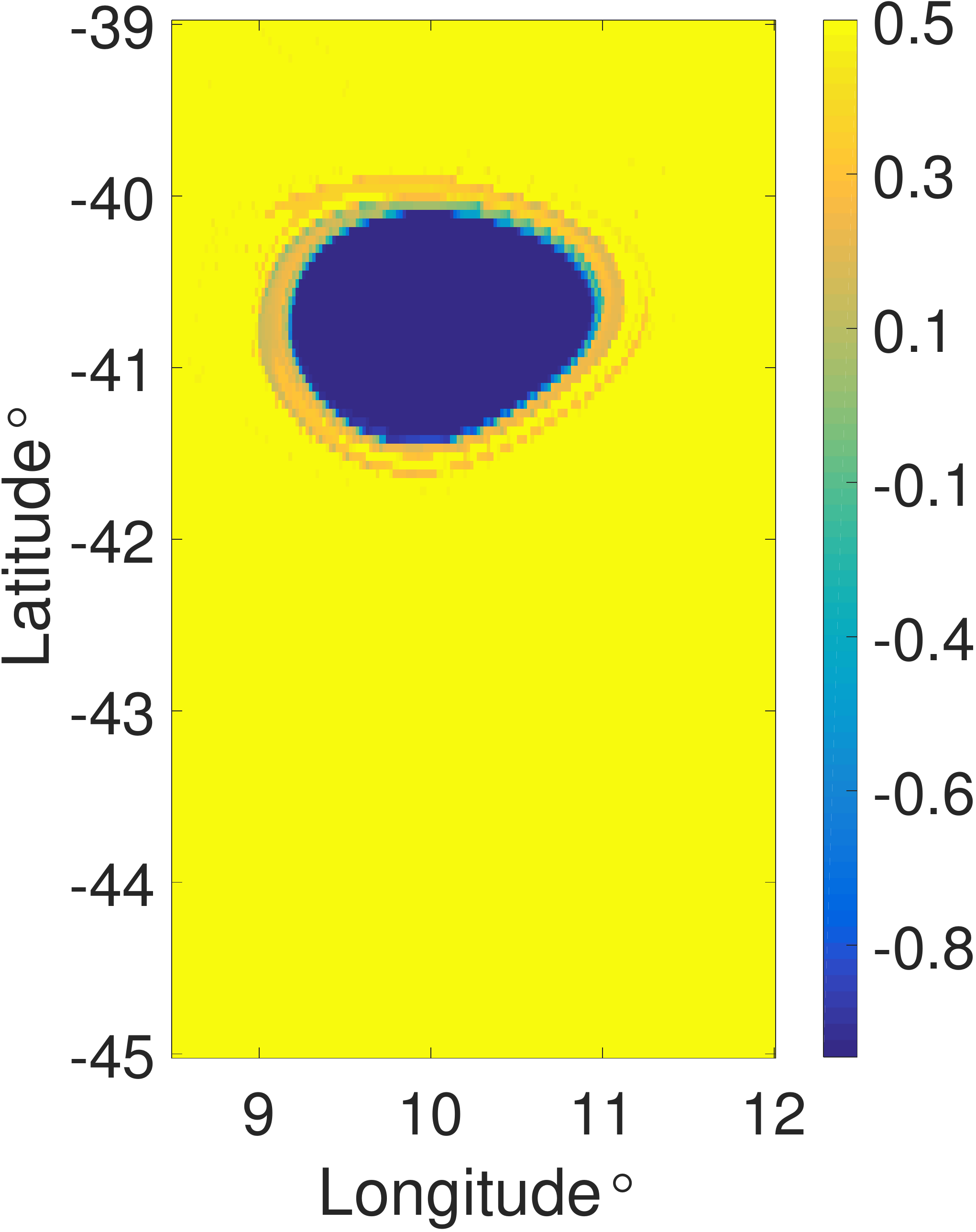}}\qquad
\subfloat[\label{fig:OceanR2Eig2}]{\includegraphics[height=0.3\textwidth]{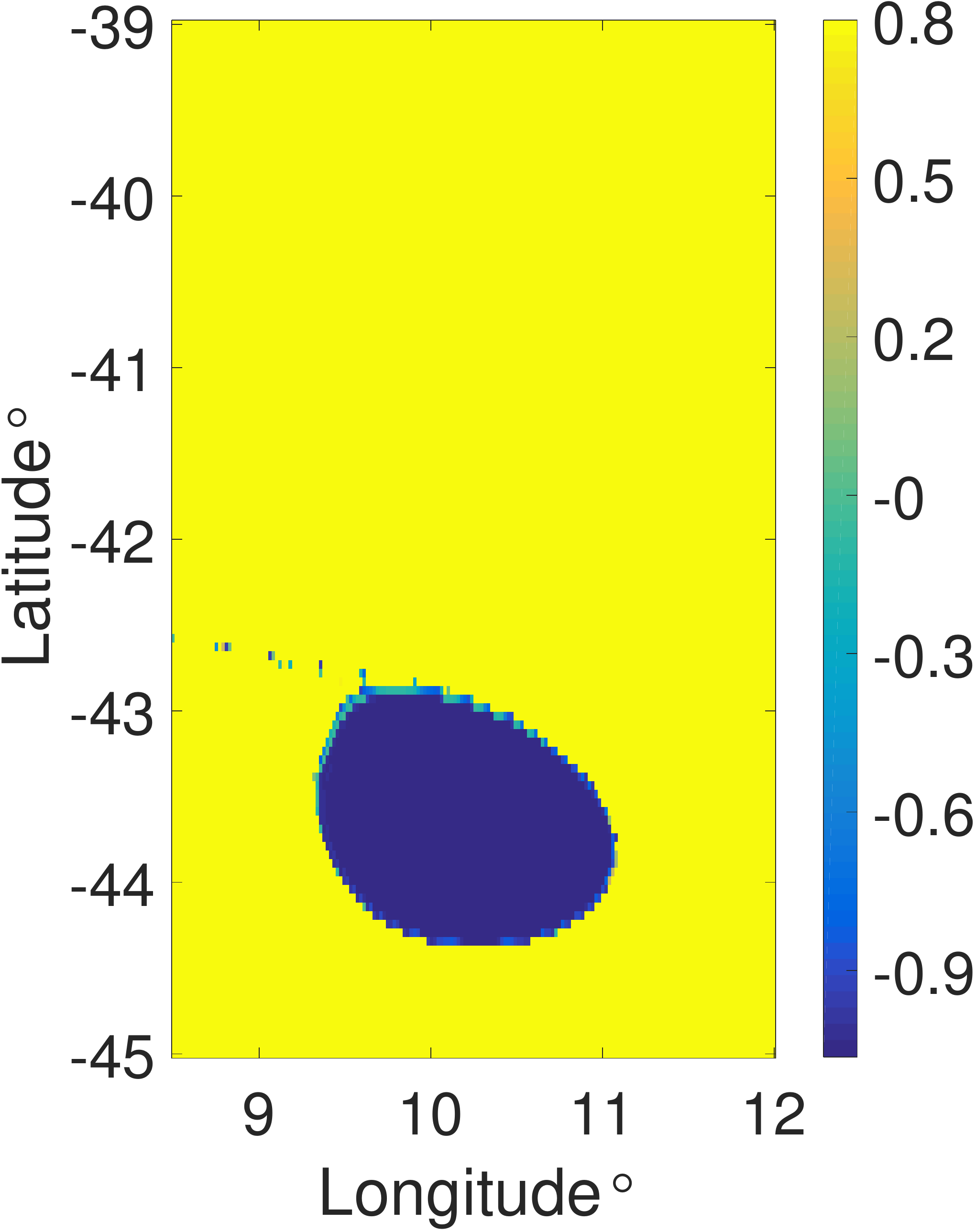}}
\caption{(a) Sorted generalized eigenvalues for the graph Laplacian $L$ for the ocean data set.
(b-c) The first two generalized eigenvectors.}
\label{fig:OceanR2Comparion}
\end{figure*}

\begin{figure*}
\subfloat[\label{fig:OceanR2Clusterst0}]{\includegraphics[height=0.27\textwidth]{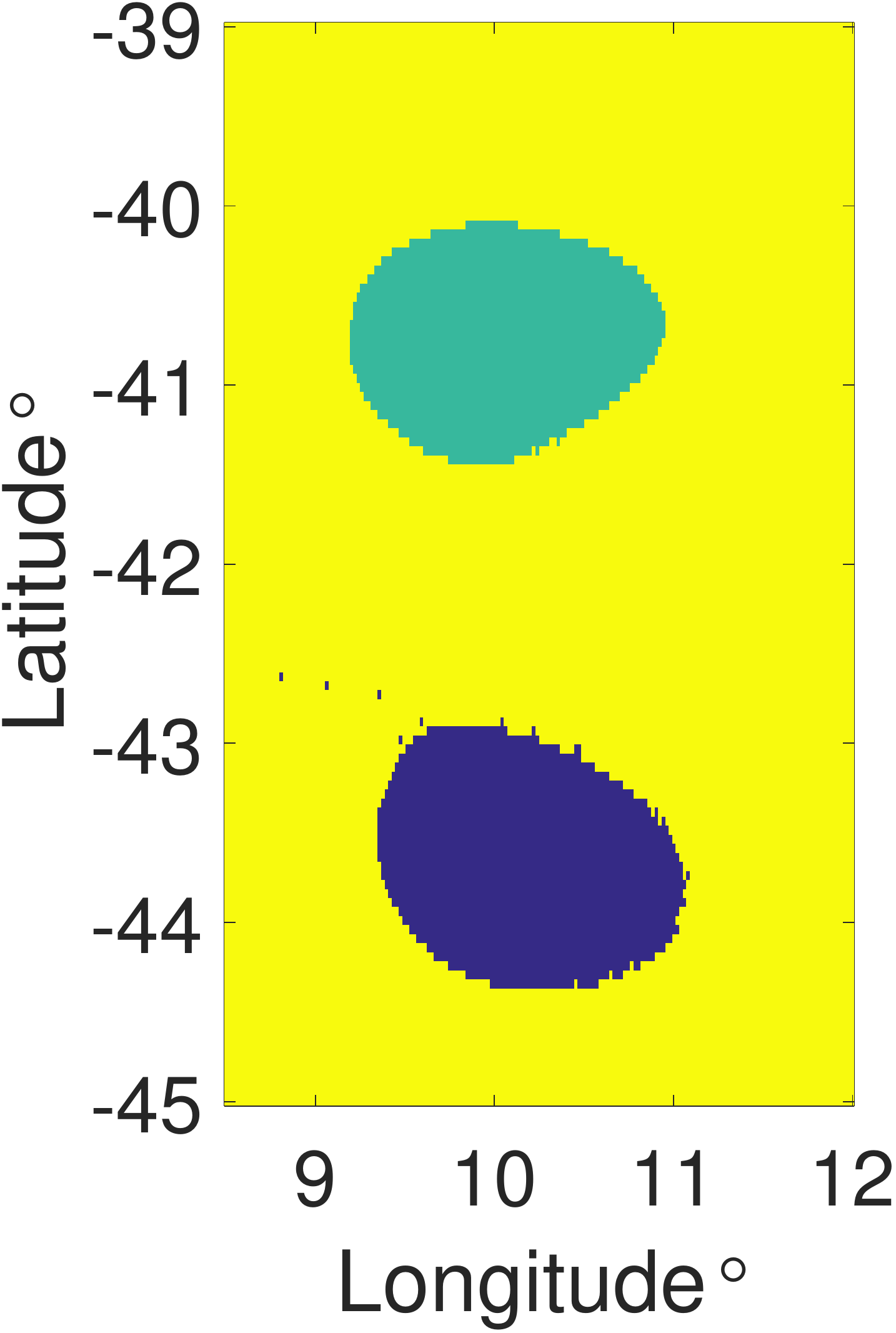}}\qquad
\subfloat[\label{fig:OceanR2ClustersT}]{\includegraphics[height=0.27\textwidth]{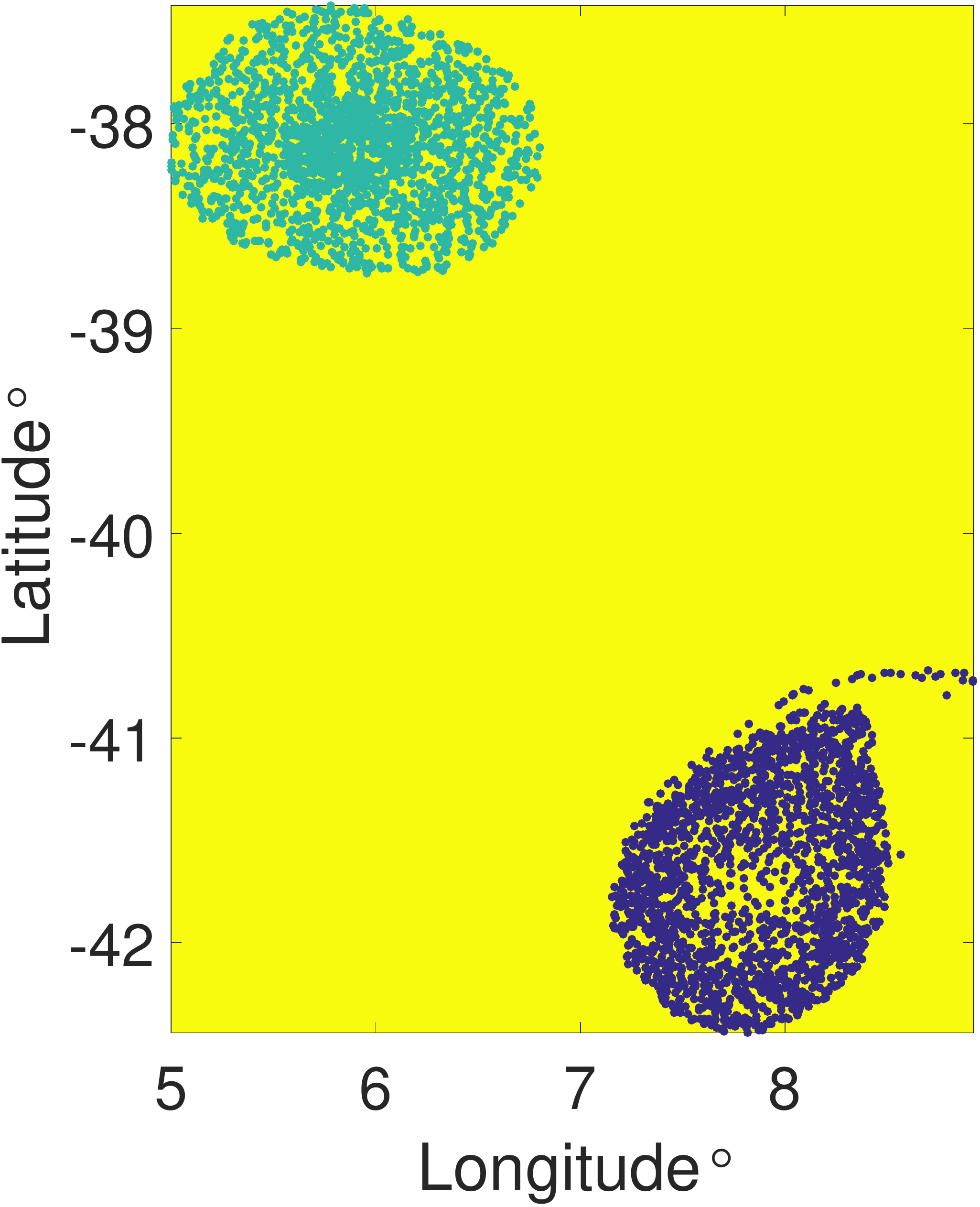}}\qquad
\subfloat[\label{fig:OceanR2Cluster2Zoom}]{\includegraphics[height=0.23\textwidth]{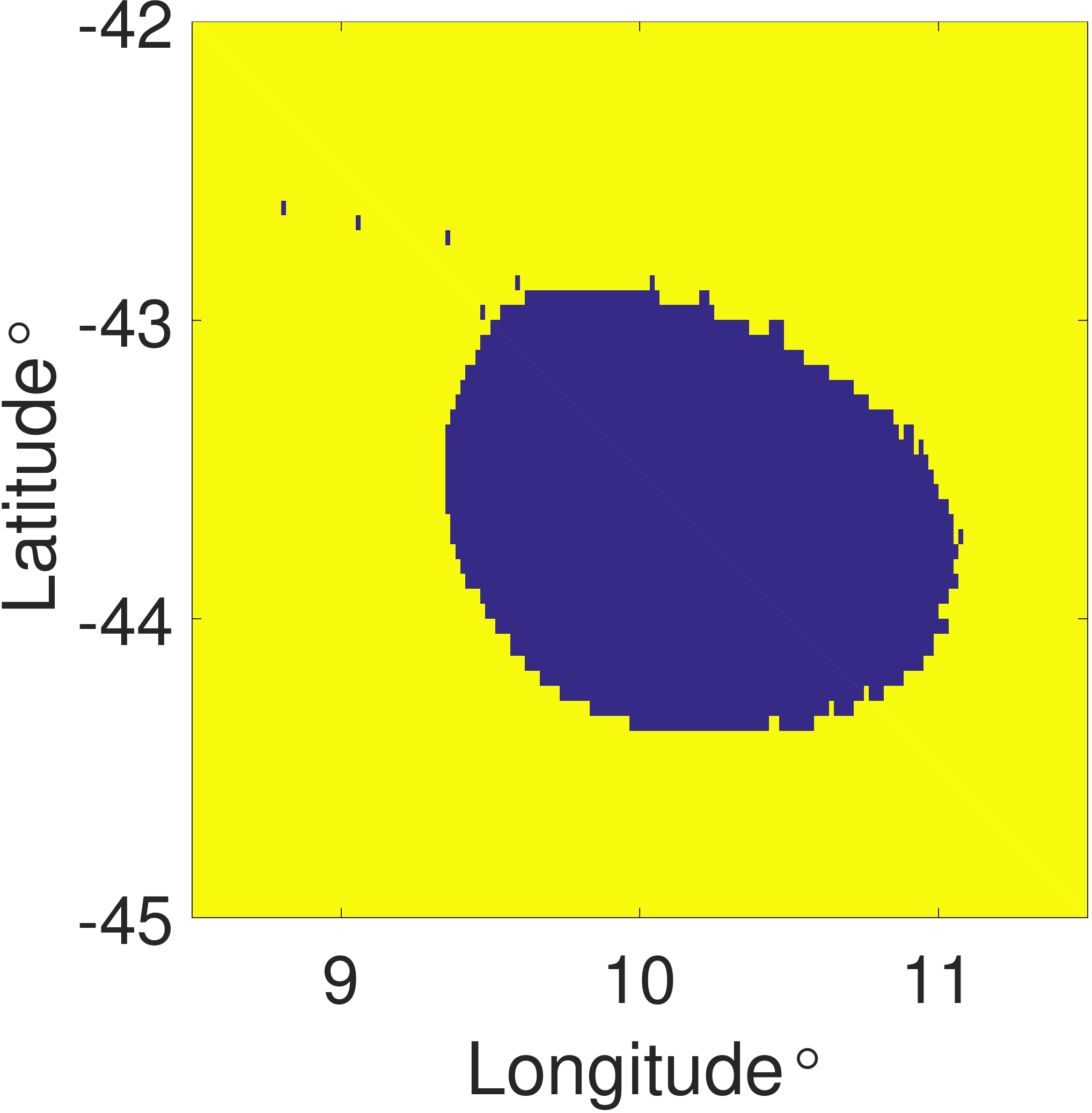}}\qquad
\subfloat[\label{fig:OceanR2Cluster3HighRes}]{\includegraphics[height=0.23\textwidth]{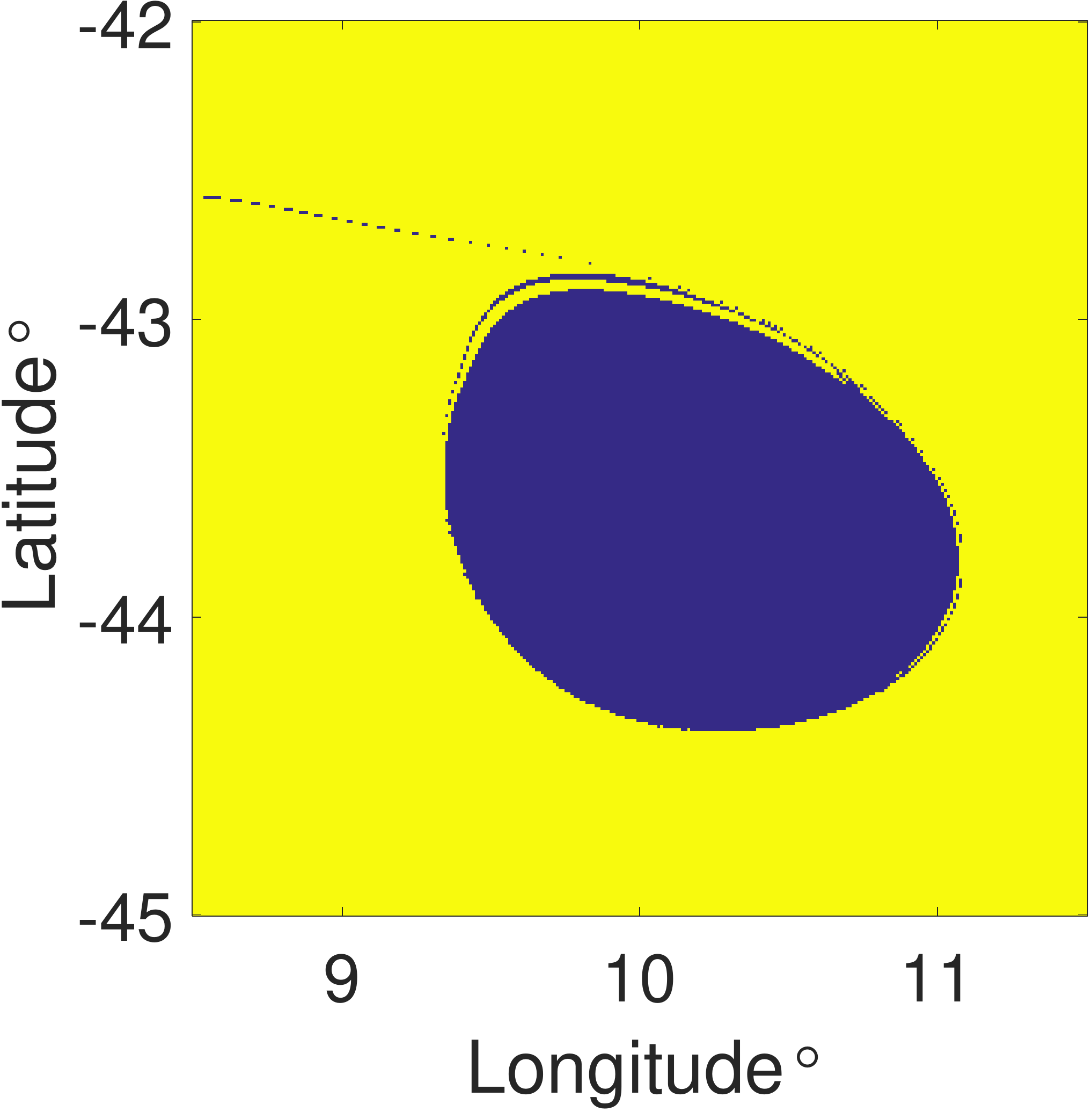}}
\caption{(a) Coherent vortices at initial time $t_{0}=$ 11 January 2006.
(b) Advected image of the vortices at the final time $t=$
28 June 2006. (c) Magnification of the blue cluster shown in the first
panel. The figure shows some isolated points located far from the
cluster core. (d) Corresponding cluster obtained from a higher resolution
computation, revealing that the previously
detected isolated points are part of a narrow fingering emanating
from the cluster core. The complete advection sequence over 168 days
is illustrated in the online supplemental movies M2 and M3 \cite{url_2,url_3}.}
\label{fig:OceanR2}
\end{figure*}

\begin{SCfigure}
\includegraphics[height=0.3\textwidth]{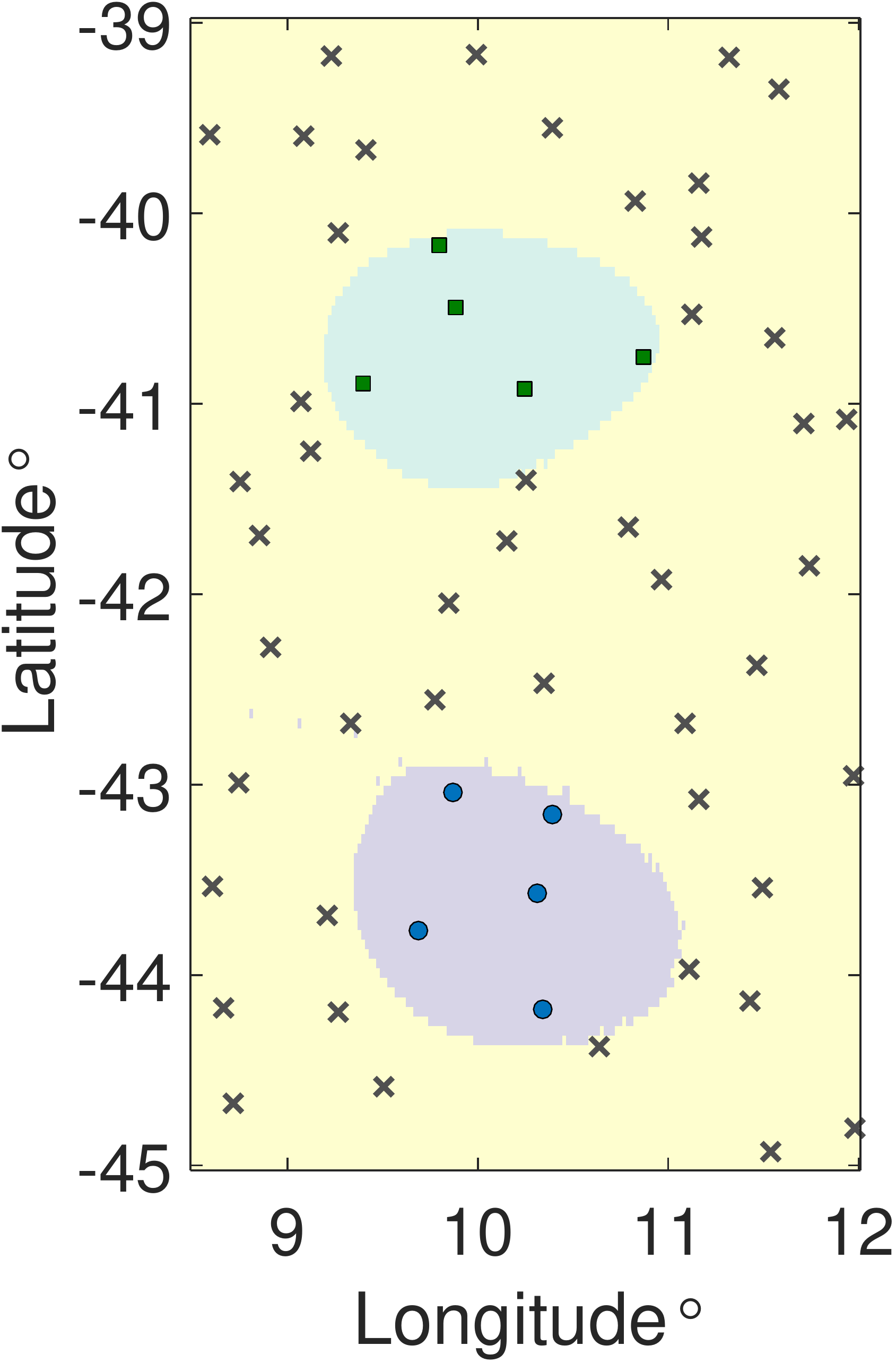}
\caption{Coherent vortices are captured at initial time $t_{0}=$ 11 January 2006, with a sparse trajectory data set. \Cref{fig:OceanR2Clusterst0} is shown in the background for comparison.}
\label{fig:OceanR2_sparse}
\end{SCfigure}
Next, we apply \Cref{alg:algorithm1} to a two-dimensional unsteady
velocity data set obtained from AVISO satellite
altimetry measurements \cite{LeTraon98}. The domain of the
data set is the Agulhas leakage in the Southern Ocean, characterized
by large coherent eddies that pinch off from the Agulhas current of
the Indian Ocean.

Here, we show how our coherent Lagrangian vortex detection principle
uncovers the material eddies over integration time of 168 days, ranging from
$t_{0}=$ 11 January 2006 to $t=$ 28 June 2006. The South Atlantic
ocean region in question is bounded by longitudes $[8.5^{\circ}\text{E},12^{\circ}\text{E}]$
and latitudes $[45^{\circ}\text{S},39^{\circ}\text{S}]$. The region in question is chosen away from the coast so that particle positions will be available for the entire integration time. Otherwise, one has to discard those particles hitting obstacles or the coast at some intermediate times from the computation. We compute
the pairwise accumulative distances over a uniform grid
of $120\times180$ points using a trajectory data set composed of $600$ evenly spaced intermediate times. We sparsify edges from the complete graph representing a distance greater than $\epsilon = 1$.

\Cref{fig:OceanR2Comparison} compares the connectivity field with the
FTLE and FSLE fields. Note that we view the connectivity
field as a simple visualization tool from which one may diagnose the
existence of coherent structures before taking the eigendecomposition step.

In \cref{fig:OceanR2Eigenvalues}, we show the first 20 generalized
eigenvalues of the graph Laplacian $L$. We can observe that the largest
eigengap exists between the second
and third generalized eigenvalues, signaling the presence of two coherent
clusters in the domain, which are indicated by the corresponding generalized
eigenvectors (see \cref{fig:OceanR2Eig1,fig:OceanR2Eig2}).

\Cref{fig:OceanR2Clusterst0} show the coherent
vortices extracted from the first two generalized eigenvectors of
graph Laplacian $L$ at initial time $t_{0}=$ 11 January 2006 and
final time $t=$ 28 June 2006 respectively. In \cref{fig:OceanR2ClustersT}, we confirm the coherence of extracted vortices by advecting them to the final time $t=$ 28 June 2006.

Interestingly, the coherent cluster shown in blue contains isolated points located far away from the cluster core (see \cref{fig:OceanR2Cluster2Zoom}). The presence of isolated points in a given cluster, however, seems to be unphysical due to the continuity of fluid flows. To investigate the true nature of these isolated points, we repeat our computation with a higher resolution, over a uniform grid of $300\times300$ points, ranging from $[8.5^{\circ}\text{E},12^{\circ}\text{E}]$ in longitudes and from $[45^{\circ}\text{S},39^{\circ}\text{S}]$ in latitudes (see \cref{fig:OceanR2Cluster3HighRes}). The higher resolution computation reveals that the previously detected isolated points are part of a narrow fingering emanating from the core of the blue cluster. This is in line with the known vortex stirring reported by several authors (see \cite{Aref84}, for example).

Despite the strange fingering-type appearance, the cluster remains
highly coherent over the extraction period of 168 days. The complete
advection sequence over 168 days is illustrated in the online supplemental
movies M2 and M3 \cite{url_2,url_3}.

This example underlines that a Lagrangian vortical region can have
an instantaneously non-convex geometry. It may also, over time, absorb an 
initial finger-type protrusion and form a convex circular boundary in the end. 
This illustrates that while requiring convexity \cite{Pratt14,Haller16}, 
lack of filamentation \cite{Haller13}, or shape coherence \cite{Ma14} of 
the vortex boundary may yield boundaries meeting high coherence requirements, 
they will not necessarily identify the largest set of trajectories forming a
coherent cluster.

Finally, we repeat our computation with a sparse trajectory data set, composed of $57$ particles distributed non-uniformly on an unstructured grid. Here, we select the number of intermediate times $m$, and sparsification distance $\epsilon$ similar to our earlier computation. \Cref{fig:OceanR2_sparse} shows the clustering result, with \cref{fig:OceanR2Clusterst0} shown in the background for comparison.
\subsection{The ABC flow}
\begin{figure}
\subfloat[\label{fig:ABCcoherentIncoherentClusters}]{\includegraphics[height=0.3\textwidth]{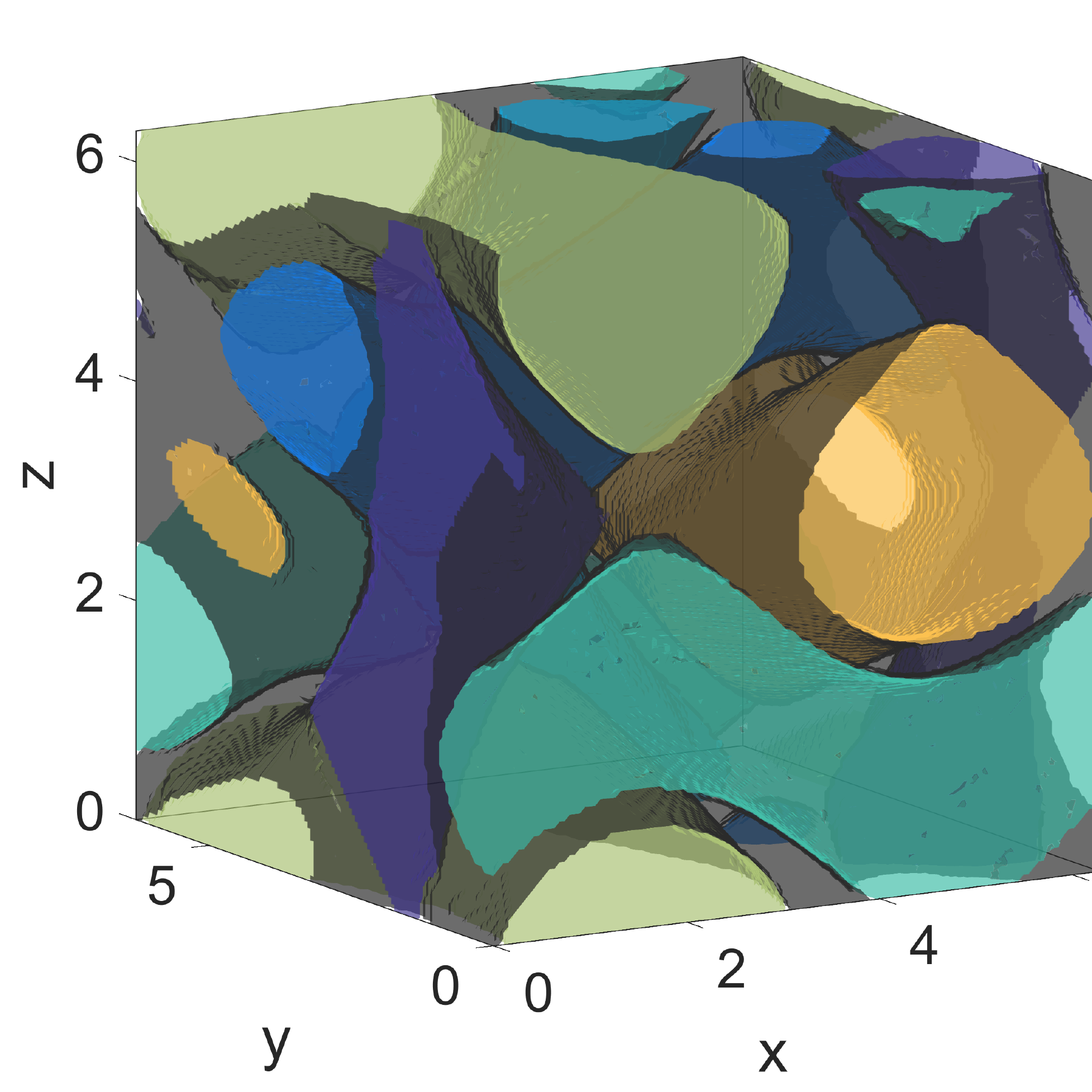}}\\
\subfloat[\label{fig:ABCincoherentclusters}]{\includegraphics[height=0.3\textwidth]{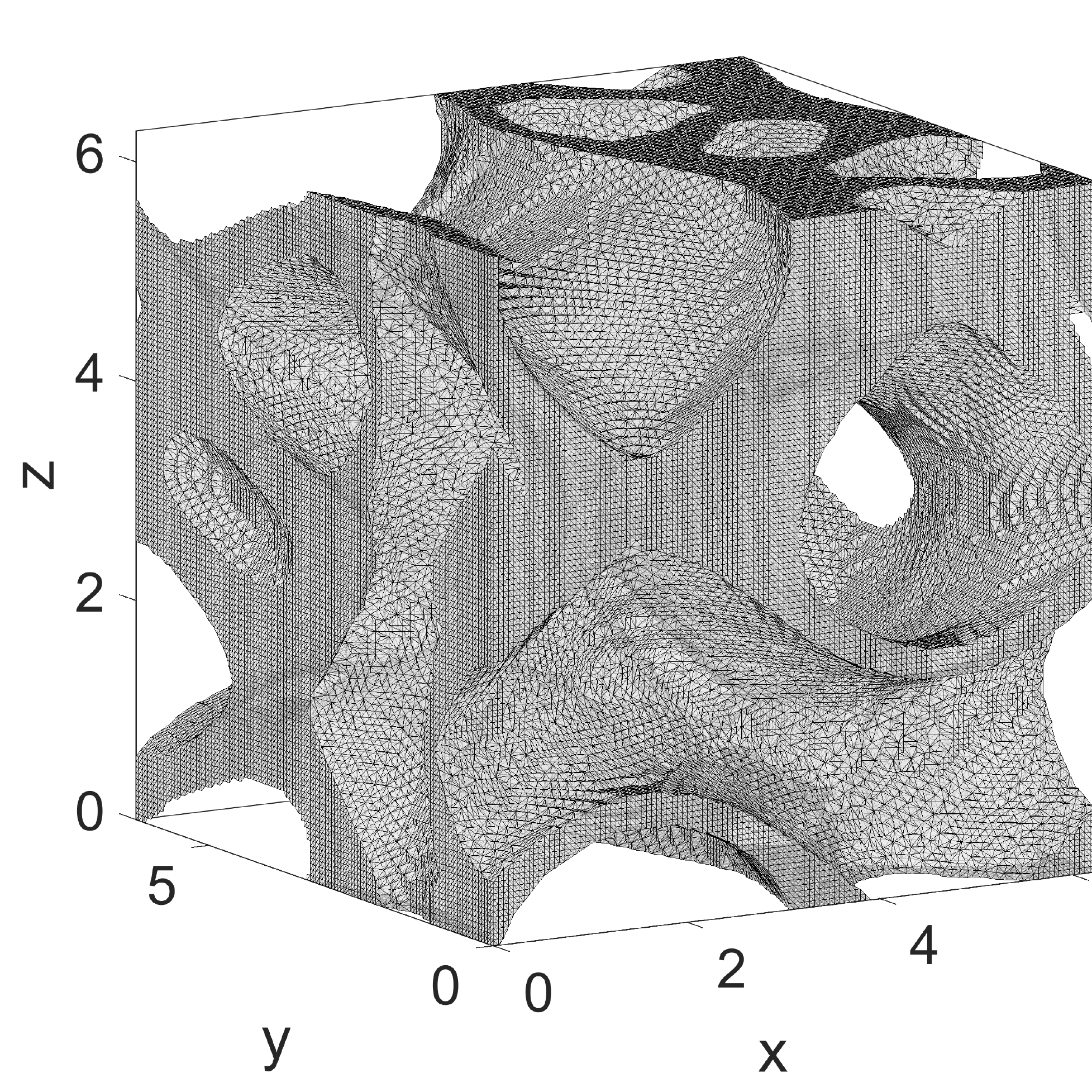}}\\
 \subfloat[\label{fig:ABCclusters136}]{\includegraphics[width=0.27\textwidth]{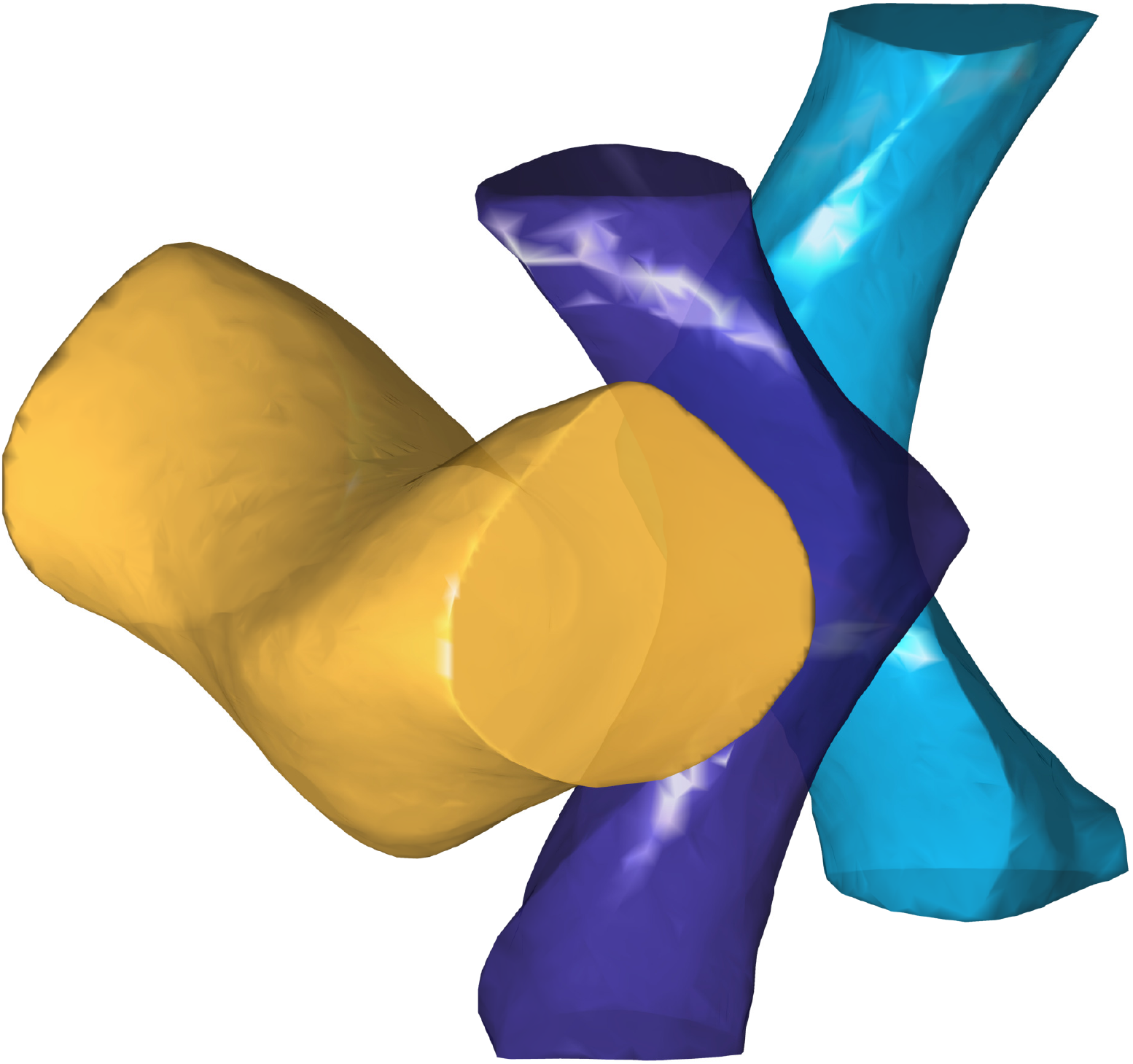}}\quad
\subfloat[\label{fig:ABCclusters245}]{\includegraphics[width=0.27\textwidth]{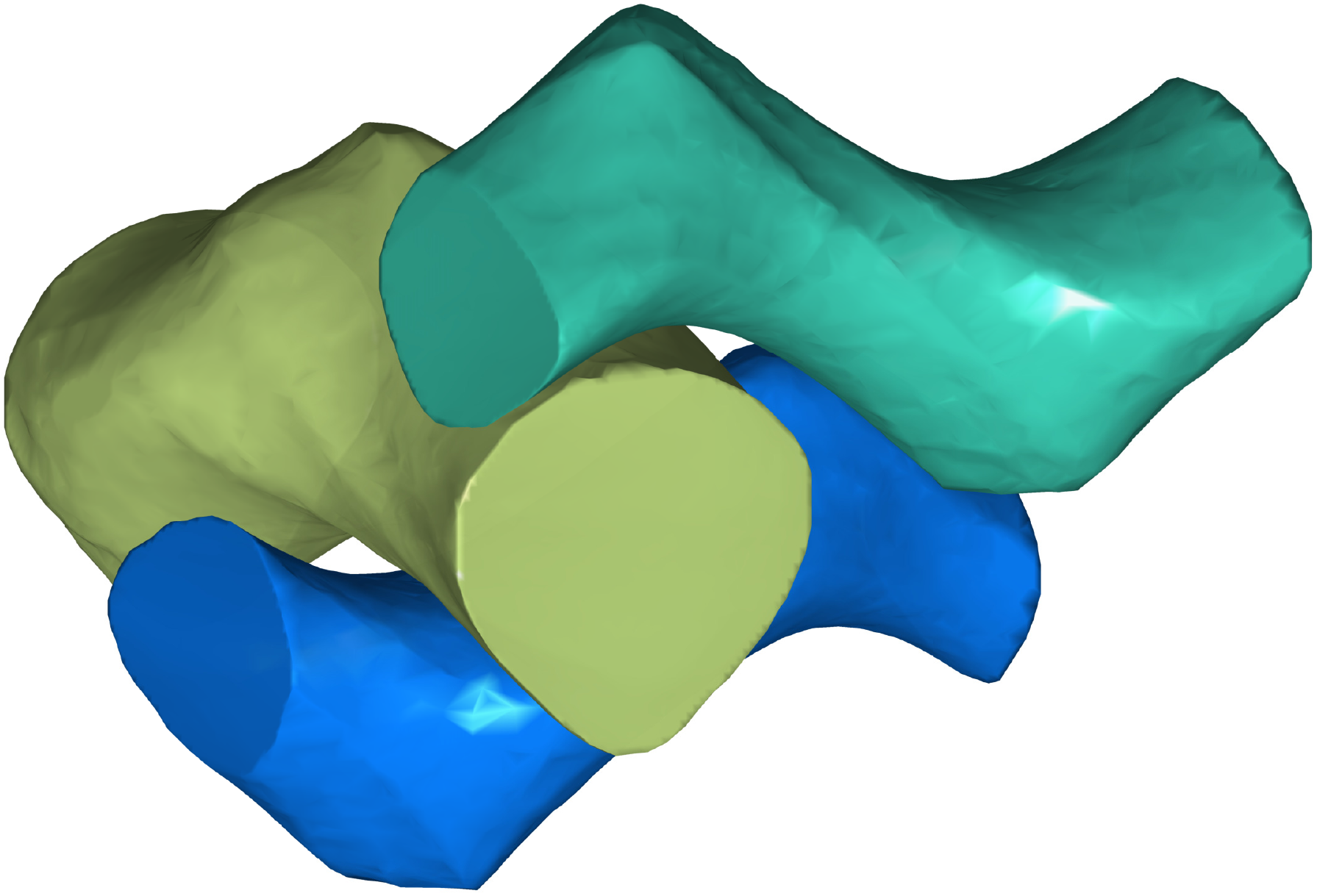}}
\caption{(a) Seven clusters extracted by K-means clustering ($k=7$) from the
first six eigenvectors of $L$. The first six clusters correspond
to six coherent vortices that were identified earlier in \cite{Dombre86}.
The chaotic sea between coherent vortices is the seventh cluster and
appears as the void between them. (b) The seventh cluster that appears
as the chaotic sea between coherent vortices. (c)-(d) 3D vortices are 
reconstructed by putting together the coherent cluster pieces.}
\label{fig:ABCclusters}
\end{figure}

\begin{figure}
 
\includegraphics[width=0.44\textwidth]{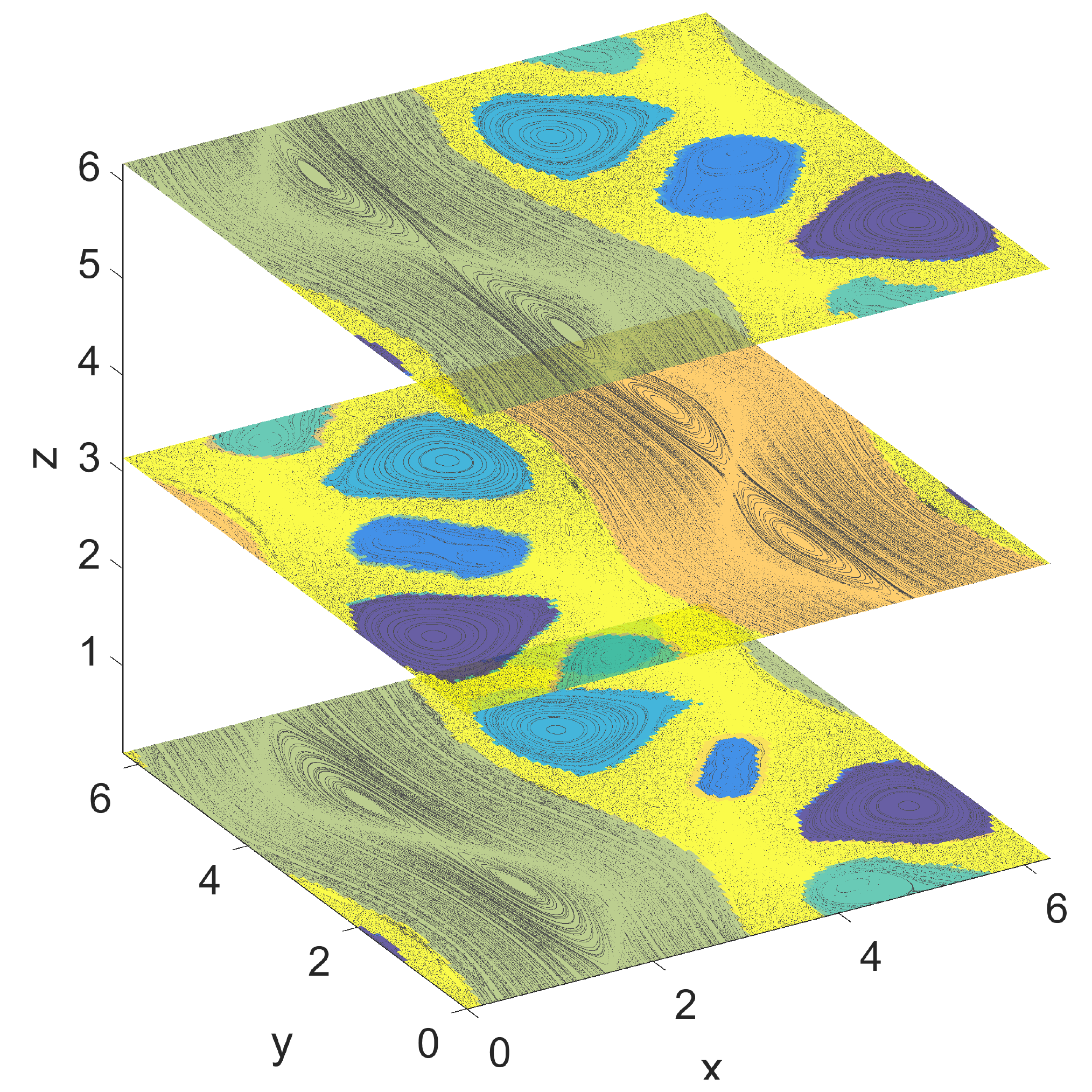}
\caption{Coherent vortices extracted by \Cref{alg:algorithm2}
are compared with the Poincar\'{e} map constructed for integration time
$T=3000$.}
\label{fig:ABCclustersPmap}
\end{figure}
As a last example, we consider the steady Arnold-Beltrami-Childress
(ABC) flow \cite{Arnold66}
\[
\begin{aligned}\dot{x} & =A\sin z+C\cos y,\\
\dot{y} & =B\sin x+A\cos z,\\
\dot{z} & =C\sin y+B\cos x,
\end{aligned}
\]
an exact solution of Euler's equation. We select the parameter values
$A=\sqrt{3}$, $B=\sqrt{2}$, and $C=1$. This well-studied set of
parameter values \cite{Blazevski14,Dombre86,Froyland09,Budisic12_1} yields six coherent vortices.

We construct a high resolution graph by selecting a uniform grid of $120\times120\times120$ points over the spatial domain ranging from $0$ to $2\pi$ in $x$, $y$, and $z$ directions.

Next, we subsample the phase space uniformly on a coarser grid by selecting  $q=1000$ supernodes out of the $120^{3}$ nodes
of the original graph, and construct the tight similarity matrix $Z\in\mathbb{R}^{q\times n}$, expressing similarity between the $q$ supernodes and the $n$ nodes of the original graph. To construct the tight similarity matrix $Z$, we measure the dynamic distances in the lifted system, where trajectories can flow out of the $2\pi$ cube. Having the similarity matrix $Z$ in hand, we compute the dominant singular values and singular vectors of $\hat{Z}=D_{2}^{-1/2}ZD_{1}^{-1/2}$. The left singular eigenvectors are cluster indicators for the reduced graph built upon $q$ supernodes, while the right singular vectors are cluster indicators for the original graph.

As the last step, we retrieve seven clusters from six cluster indicators using the K-means algorithm. The last cluster, as before, shows the incoherent region filling the space between the coherent clusters or vortices. \Cref{fig:ABCcoherentIncoherentClusters} shows the six coherent clusters which are separated by the incoherent cluster. The six clusters capture the six known coherent structures of the ABC flow identified earlier in \cite{Dombre86}.


Due to the existence of the spatial periodic boundary condition, the coherent vortices are broken into pieces in the initial cubic domain. In \cref{fig:ABCclusters136,fig:ABCclusters245},
we put together the pieces of six coherent vortices, and show their
full cylindrical geometry. The colors used in \cref{fig:ABCclusters136,fig:ABCclusters245}
are consistent with those in \cref{fig:ABCcoherentIncoherentClusters}. 
In \cref{fig:ABCclustersPmap}, the clusters are superimposed on the Poincar\'{e} map showing close agreement between the results of the two approaches.
\section{Conclusion}
We have developed here an approach to locate coherent structures based 
on spectral graph theory. To identify coherent structures, we measure 
the pairwise Euclidean distance between Lagrangian trajectories, and 
construct an undirected weighted graph describing the spatio-temporal 
evolution of fluid flows. We then identify coherent vortices as clusters 
of Lagrangian particles remaining close under the flow using two different 
algorithms. In the first algorithm, we used Shi \& Malik \cite{Shi00} 
normalized cut to identify coherent vortices whose nodes on graph have 
large internal (external) (in-)coherence. We demonstrate the effectiveness
of the corresponding \Cref{alg:algorithm1} to detect Lagrangian coherent 
vortices in periodic, quasiperiodic, and unsteady two-dimensional flows. 
This includes the determination of the a priori unknown number of present 
vortices in a given domain using the eigengap heuristic.

In \Cref{alg:algorithm2}, we apply a recently developed graph sub-sampling 
technique \cite{Cai14,Liu13} to handle the memory bottleneck associated 
with large-scale graphs. We apply \Cref{alg:algorithm2} in our last example, 
the 3D steady ABC flow, where we succeeded to combine high sampling resolution with computational efficiency. 

An advantage of our approach is that it requires a relatively low number of Lagrangian trajectories as input, making it suitable for the analysis of low-resolution trajectory data sets (see also \cite{Froyland15,Williams15,Froyland15_2} for similar approaches designed for low numbers of Lagrangian trajectories). Moreover, our method is taking advantage of trajectories' intermediate 
positions, i.e., information that comes in most cases without additional computational cost, e.g., in time resolved trajectory data sets or numerical integration of velocity data sets/vector fields (see also \cite{Froyland15}).

Moreover, we argue that in fluid-like flows coherence-related phenomena 
can only be conceived in the presence of an incoherent background, which 
prohibits the partitioning of the fluid domain into purely coherent sets 
or regions. Here, we introduced the definition of incoherent cluster and 
partitioned the fluid domain into coherent and incoherent clusters, an idea 
that appears to be missing in other similar approaches \cite{Ma14,Froyland15,Ser-Giacomi15_1}.

Finally, we chose spectral clustering as a tool of choice due to its 
solid mathematical foundation and its performance. However, other clustering 
algorithms such as density-based clustering approaches \cite{Ester96} that 
can incorporate the definition of noise or incoherent cluster may be used 
alternatively. Incorporating other clustering algorithms, and comparing their 
performance for the purpose of Lagrangian coherent vortex identification 
remains a viable future research direction. Moreover, further work is needed 
to connect graph properties with physical or mechanical quantities characterizing 
the fluid motion, beyond the heuristic and numerical arguments given in \Cref{section:introduction,section:results}.

\section*{Acknowledgments}

The altimeter products were produced by Ssalto/Duacs and distributed
by Aviso, with support from CNES (\href{http://www.aviso.altimetry.fr/duacs/}{http://www.aviso.altimetry.fr/duacs/}).

\appendix

\section{Approximating Ncut}\label{app:Ncut}

In this section, we recall how the NCut problem can be solved for
the case $k=2$, which partitions the graph into two disjoint sets.
We follow closely the arguments of \cite{Shi00,Luxburg07}.

Our goal is to solve the optimization problem
\begin{equation}
\min\limits _{A\in V}\text{NCut}(A,\bar{A}).\label{eq:Ncut_k_2}
\end{equation}
First, we rewrite the problem in a more convenient form. Given a subset
$A\subset V$ we define the cluster indicator vector $f=(f_{1},...,f_{n})^{\top}\in\mathbb{R}^{n}$
with entries
\begin{equation}
f_{i}=\begin{cases}
\sqrt{\frac{\text{vol}(\bar{A})}{\text{vol}(A)}}, & \textnormal{if }v_{i}\in A,\\
-\sqrt{\frac{\text{vol}(A)}{\text{vol}(\bar{A})}}, & \textnormal{if }v_{i}\in\bar{A}.
\end{cases}\label{eq:Ncut_indicator}
\end{equation}
Now, Eq.\ \eqref{eq:Ncut_k_2} can be conveniently rewritten using
the graph Laplacian $L$ as
\[
\min_{A}f^{\top}Lf\quad\text{subject to}\;f\;\text{as in}\;\eqref{eq:Ncut_indicator},\;Df\bot1,\;f^{\top}Df=\text{vol}(V).\label{eq:Ncut_NP_hard}
\]
This is a Rayleigh quotient, and minimizing it is of complexity NP-hard,
since we have constrained $f$ to take on only discrete values as
described in \eqref{eq:Ncut_indicator}. We relax the problem by allowing
$f$ to take arbitrary real values (\textit{$l_{2}$-relaxation}),
to obtain:
\[
\min_{f\in R^{n}}f^{\top}Lf\quad\text{subject to}\;Df\bot1,\;f^{\top}Df=\text{vol}(V).
\]
After substitution of $g\coloneqq D^{1/2}f$, the problem converts to
\[
\min_{g\in R^{n}}g^{\top}D^{-1/2}LD^{-1/2}g\quad\text{subject to}\;g\bot D^{1/2}1,\;{\left\Vert g\right\Vert }^{2}=\text{vol}(V),
\]
to which the standard Rayleigh-Ritz theorem applies, such that its
solution $g$ is given by the second eigenvector of $D^{-1/2}LD^{-1/2}$.
Re-substituting $f=D^{-1/2}g$, we see that $f$ is the second generalized
eigenvector of $Lu=\lambda Du$.

Similarly, we can decompose the graph into $k$ partitions by using cluster
indicator vectors $h_{j}=(h_{1,j},...,h_{n,j})^{\top}$

\begin{equation}
h_{i,j}=\begin{cases}
\frac{1}{\sqrt{\text{vol}(A_{j})}}, & \textnormal{if }v_{i}\in A_{j},\\
0, & \textnormal{otherwise,}
\end{cases},\qquad i=1,...,n,\,j=1,...,k.\label{eq:Ncut_indicator_matrix}
\end{equation}
Then we set the matrix $H\in R^{n\times k}$ as the matrix containing
those $k$ cluster indicator vectors as columns. Observe that the columns
in $H$ are orthonormal to each other, that is $H^{\top}H=I$, and
$h_{i}^{\top}Lh_{i}=\text{cut}(A_{i},\bar{A}_{i})/\text{vol}(A_{i})$.
So we can write the problem of minimizing NCut as
\[
\min_{A_{1},...,A_{k}}\Tr(H^{\top}LH)\quad\text{subject to }H^{\top}DH=I,\;H\text{ as in }\eqref{eq:Ncut_indicator_matrix}
\]
Relaxing the discreteness condition and substituting $T=D^{1/2}H$
we obtain the relaxed problem
\[
\min_{T\in\mathbb{R}^{n\times k}}\Tr(T^{\top}D^{-1/2}LD^{-1/2}T)\quad\text{subject to }T^{\top}T=I.
\]
Again, this is the standard trace minimization problem, which is solved
by the matrix $T$ composed of the first $k$ eigenvectors of $D^{-1/2}LD^{-1/2}$
as columns. Re-substituting $H=D^{-1/2}T$, we see that the solution
$H$ consists of the first $k$ generalized eigenvectors of $Lu=\lambda Du$.
This yields the normalized spectral clustering algorithm according
to \cite{Shi00}.

\section{Bipartite spectral graph partitioning}

\label{app:bipartite-clustering}

In this section, we briefly recall how spectral clustering is
applied to bipartite graphs. This specification is also referred
to as \emph{spectral co-clustering} \cite{Dhillon01,Zha01}, and is
presented here in the sub-sampling terminology introduced in \Cref{section:large_scale}
It applies, however, verbatim to the bipartite transfer-operator graph.

Let $Z\in\mathbb{R}^{q\times n}$ be a tight similarity matrix between the $n$
graph nodes and the $q$ supernodes. To explicitly capture the node-supernode
relationship, we consider a bipartite graph $G_{\mathcal{B}}=(V_{\mathcal{B}},E_{\mathcal{B}},W_{\mathcal{B}})$ whose nodes 
can be divided into two disjoint sets $A$ and $B$ such that internal edges all 
have zero weights, i.e.,
$w_{ij}^{\mathcal{B}}=0$ if $v_{i}^{\mathcal{B}},v_{j}^{\mathcal{B}}\in A$ or $v_{i}^{\mathcal{B}},v_{j}^{\mathcal{B}}\in B$. The similarity matrix of the 
whole bipartite graph $W_{\mathcal{B}}$ then reads as
\begin{equation}
W_{\mathcal{B}}=\begin{pmatrix}0 & Z^{\top}\\
Z & 0
\end{pmatrix}\label{eq:bipartite_affinity2}
\end{equation}

To partition the bipartite graph, the optimization task can be formalized
as a generalized eigenvalue problem with suitable relaxation, see
\Cref{app:Ncut},

\begin{equation}
L_{\mathcal{B}}q=(D_{\mathcal{B}}-W_{\mathcal{B}})q=\lambda D_{\mathcal{B}}q\label{eq:bipartite_opt}
\end{equation}
where $D_{\mathcal{B}}$ is the degree matrix of $W_{\mathcal{B}}$.

Substituting \eqref{eq:bipartite_affinity2} in \eqref{eq:bipartite_opt}, we
get
\begin{equation}\label{eq:block_gen_eigenproblem}
\begin{pmatrix}0 & Z^{\top}\\
Z & 0
\end{pmatrix}\begin{pmatrix}q_{1}\\
q_{2}
\end{pmatrix}=(1-\lambda)\begin{pmatrix}D_{1} & 0\\
0 & D_{2}
\end{pmatrix}\begin{pmatrix}q_{1}\\
q_{2}
\end{pmatrix},
\end{equation}
where $D_{1}$ is an $n\times n$ diagonal matrix whose entries are
column sums of $Z$ and $D_{2}$ is an $q\times q$ diagonal matrix
whose entries are row sums of $Z$. Breaking the block matrix form into
parts, Eq.\ \eqref{eq:block_gen_eigenproblem} can be rewritten as:
\begin{align*}
Z^{\top}q_{2} & =(1-\lambda)D_{1}q_{1},\\
Zq_{1} & =(1-\lambda)D_{2}q_{2}.
\end{align*}
Let $b=D_{1}^{1/2}q_{1}$ and $a=D_{2}^{1/2}q_{2}$, and after variable
substitution, we have
\[
\begin{aligned}D_{1}^{-1/2}Z^{\top}D_{2}^{-1/2}a=(1-\lambda)b,\\
D_{2}^{-1/2}ZD_{1}^{-1/2}b=(1-\lambda)a.
\end{aligned}
\]
These equations define the SVD of the normalized
matrix $\hat{Z}=D_{2}^{-1/2}ZD_{1}^{-1/2}$. Particularly, $a$ and
$b$ are the left and right singular vectors and $1-\lambda$ is the
corresponding singular value \cite{Zha01}.

\newpage
\bibliography{Hadjighasem_clustering_lib2}

\end{document}